\documentclass[10pt,twocolumn]{aastex631}

\usepackage{chngcntr}

\usepackage{soul}
\usepackage{lipsum}
\newcommand{\ms}{$\rm m\,s^{-1}$}
\newcommand{\kms}{$\rm km\,s^{-1}$}

\newcommand{\vsini}{\textit{v}\,\textnormal{sin}\,\textit{i}}
\newcommand{\logg}{log\,$g$}

\newcommand{\teff}{$T_{\rm eff}$}
\newcommand{\porb}{$P_{\rm orb}$}
\newcommand{\prot}{$P_{\rm rot}$}

\received{April 20, 2024}
\revised{June 08, 2024}
\accepted{June 17, 2024}
\submitjournal{ApJ}

\graphicspath{{./}{figures/}}
\shorttitle{TTS \teff{} variability}
\shortauthors{Tang et al.}

\defcitealias{boyajian2012}{B12}
\defcitealias{mann2015}{M15}

\begin{document}

\title{
    Measuring the Spot Variability of T Tauri Stars Using Near-IR Atomic Fe and Molecular OH Lines
    }

%-----------------------------------

\author[0000-0003-4247-1401]{Shih-Yun Tang}
    \affiliation{Department of Physics and Astronomy, Rice University, 6100 Main Street, Houston, TX 77005, USA}
    \affiliation{Lowell Observatory, 1400 West Mars Hill Road, Flagstaff, AZ 86001, USA}
    \email{sytang@rice.edu}

\author[0000-0002-8828-6386]{Christopher M. Johns-Krull}
    \affiliation{Department of Physics and Astronomy, Rice University, 6100 Main Street, Houston, TX 77005, USA}

\author[0000-0001-7998-226X]{L. Prato}
    \affiliation{Lowell Observatory, 1400 West Mars Hill Road, Flagstaff, AZ 86001, USA}

\author[0000-0002-0848-6960]{Asa G. Stahl}
    \affiliation{Department of Physics and Astronomy, Rice University, 6100 Main Street, Houston, TX 77005, USA}

%-----------------------------------
%-----------------------------------

\begin{abstract}
As part of the Young Exoplanets Spectroscopic Survey (YESS), this study explores the spot variability of 13 T Tauri Stars (TTSs) in the near-infrared $H$ band, using spectra from the Immersion GRating INfrared Spectrometer (IGRINS). By analyzing effective temperature (\teff{}) sensitive lines of atomic Fe\,\textsc{i} at $\sim$1.56259 \micron{} and $\sim$1.56362 \micron{}, and molecular OH at $\sim$1.56310 and $\sim$1.56317 \micron{}, we develop an empirical equivalent width ratio (EWR) relationship for \teff{} in the range of 3400–5000 K. This relationship allows for precise relative \teff{} estimates to within tens of Kelvin and demonstrates compatibility with solar metallicity target models. However, discrepancies between observational data and model predictions limit the extension of the \teff{}--EWR relationship to a broader parameter space. Our study reveals that both classical and weak-line TTSs can exhibit \teff{} variations exceeding 150 K over a span of two years. The detection of a quarter-phase delay between the EWR and radial velocity phase curves in TTSs indicates spot-driven signals. A phase delay of 0.06 $\pm$ 0.13 for CI\,Tau, however, suggests additional dynamics, potentially caused by planetary interaction, inferred from a posited 1:1 commensurability between the rotation period and orbital period. Moreover, a positive correlation between \teff{} variation amplitude and stellar inclination angle supports the existence of high-latitude spots on TTSs, further enriching our understanding of stellar surface activity in young stars.
\end{abstract}

%-----------------------------------
%-----------------------------------


%-----------------------------------
\section{Introduction}\label{sec:intro}

To comprehend how planets come into being, studying them while they are still forming is crucial. Yet searches for planets around young active stars, like T Tauri Stars (TTSs), often face challenges related to stellar activity \citep[e.g., cool spot(s) on the stellar surface,][]{prato2008,dittmann2009,donati2014}. The more we know about these surface features, the better chance we have of finding planets around young active stars.

One way for spots to hinder radial velocity (RV) planet searches is by generating periodic signals that mimic the RV variations induced by a planet. For simplicity, we refer to these spot-induced apparent RV shifts as ``spot-RV signals". As star spots are cooler areas on the stellar disk, with local convective flux suppressed by strong magnetic fields, their nonaxisymmetric distribution will cause observed spectral lines to become asymmetric. Because spots co-rotate with the star, the periodic variation in the distortion of the absorption lines can be mistaken for a periodic RV signal, which can then be interpreted as a planet-induced signal \citep[e.g.,][]{saar1997,crockett2012}. 

Purely data-driven methods for modeling the spot-RV signal, like Gaussian Process Regression (GPR) \citep[e.g.,][]{garnett2023}, have lately seen wide use \citep[e.g.,][]{haywood2014,benatti2020,tran2023a}; however, prior knowledge of the stellar rotation period and spot(s) lifetime is essential for the GPR to trace the activity signal properly. Knowing these stellar properties is even more critical for applying GPR to young stellar systems where the spot-RV signals from the extreme stellar activity can be several times larger than those induced by planets \citep[e.g.,][]{cale2021,sikora2023}. Young systems in which a planet's orbital period (\porb) is approximately equivalent to the rotation period of the host star (\prot) \citep{lanza2022} are even more challenging to characterize. An approximate 1:1 ratio between \prot{}:\porb{} has been found in mature solar-type systems like CoRoT-4 \citep[\prot $\sim$ \porb $\sim$9.2 d,][]{lanza2009,bonomo2017} and $\tau$\,Boo \citep[\prot $\sim$ \porb $\sim$3.31 d,][]{baliunas1997,brogi2012,borsa2015}, and may not be surprising to find in young stars \citep{dawson2018}. 

A star's \prot{} can often be determined from analyzing its lightcurve; however, for disk-bearing classical TTS (CTTS), the rotation period is often more difficult to determine compared with that of a weak-line TTS (WTTS) as the CTTS' photometry can be contaminated by the accretion variability, variable extinction, and disk-scattered light \citep[e.g.,][]{parks2014,rebull2020,rampalli2021}. An alternative way to measure the star's \prot{} is to trace the apparent effective temperature (\teff) variation caused by changes in the area of cool spot coverage on the surface of a rotating star \citep[e.g.,][]{catalano2002}. This provides a powerful, photometry-independent tool for the measurement of stellar \prot{}.

Line-depth ratios (LDRs) have been an established technique for estimating the \teff{} of a stellar object since \citet[][]{gray1989}. By estimating the LDR of a pair of isolated lines (e.g., V\,\textsc{i} at 6251.83 \AA{} and Fe\,\textsc{i} at 6252.57 \AA), this technique is immune to spectral broadening such as that caused by stellar rotation and instrumental resolution. With improvements over time, the precision of LDR-based relative \teff{} measurements for main-sequence and giant stars has reached $\leq$10 K \citep[e.g.,][]{gray1991,gray2001}. This method has also facilitated the measurement of rotationally-modulated temperature variations for mature main-sequence stars such as $\epsilon$ Eri \citep[$\sim$15 K,][]{gray1995} and $\sigma$ Dra \citep[$\sim$5 K,][]{gray1992}, and more active stars like VY Ari \citep[$\sim$177 K,][]{catalano2002}.

As TTSs have typical spectral types of late K to M, observing them in the near-IR is more efficient given that the spectral energy distributions of late-type stars peak in this wavelength region. Studies of the LDR--\teff{} relationship in the near-IR began about two decades after work in the optical region. For example, \citet{fukue2015} used absorption lines in the $H$ band to determine the temperature for G- and K-type giants and supergiants, \citet{taniguchi2018} used lines in $Y$ and $J$ bands to establish the LDR-\teff{} for G- and M-type giants, \citet{kovtyukh2023} used near-IR lines for obtaining Cepheids' \teff{}, and \citet{afsar2023a} used atomic lines in $H$ and $K$ bands to establish LDR--\teff{} for late-type stars with a range of gravity (\logg{}) and metallicity ([M/H]). The effect of metallicity and \logg{} on the near-IR LDR--\teff{} relationship has also been studied \citep{jian2019,jian2020}.

The \teff{} sensitive OH lines at $\sim$1.56310 \micron{}\footnote{All wavelengths used in this study are in vacuum.} and 1.56317 \micron{} were first used by \citet{oneal1997} and \citet{oneal2001} to study starspots on active stars. They reported a nearly linear relationship between the OH total equivalent width and the \teff{}, shown an increasing trend from 5000\,K to 3000\,K. These OH lines, along with the nearby Fe\,\textsc{i} lines at $\sim$1.56259 \micron{} and 1.56362 \micron{} were later shown by \citet{prato2002} to display inverse depth growth with spectral type from G0 to M9, i.e., the line depths increase for OH lines from G0 to M9, but the line depths of Fe lines get shallower from G0 to M9. \citet{lopez-valdivia2019} went on to use LDs (as opposed to LDRs) of the aforementioned atomic and molecular lines to measure \teff{} for 162 K- and M-type dwarfs with the Immersion GRating INfrared Spectrometer \citep[IGRINS,][]{yuk2010,park2014,mace2016}. These authors determined the temperature of each target, which they call $T_{\rm spec}$, by matching the targets' LDs to those from the BT-Settl spectral model grid \citep{allard2011,allard2012}. They then calibrated systematic offsets between the models and observations using the color–temperature relation from \citet{mann2015}, ultimately establishing a linear $T_{\rm spec}$--LDR relationship from about 3300 to 3900\,K. This relationship is more broadly applicable than that between $T_{\rm spec}$ and LDs, because LDRs are less affected by spectral broadening effects such as those induced by the instrumental profile. However, \citet{lopez-valdivia2019} did not discuss how the $T_{\rm spec}$--LDR relationship might possibly change with stellar parameters such as \logg{}, the line-of-sight projected equatorial rotation velocity (\vsini{}), metallicity, and average surface magnetic field strength ($\bar B$)\, all of which have a larger range of values among TTSs than among main-sequence dwarfs \citep{sokal2020,lopez-valdivia2021}.

In this work, we explore the use of the OH/Fe LDR and equivalent width ratio (EWR) methods for assessing the relative apparent \teff{} variability of TTSs. We further apply these techniques to estimate the \prot{} of selected TTSs through their periodic LDR and EWR variations. The following sections lay out the structure of our study: Section~\ref{sec:sample} first describes our target samples, the TTS and the \teff{} calibration sources, following with the IGRINS observations and data reduction. Section~\ref{sec:spec_analysis} introduces the relevant spectral lines and analysis of the LD and EW ratios, taking into account the impact of spectral broadening. Section~\ref{sub:emprical} establishes empirical \teff{}--EWR relationships from IGRINS data, which we show are more robust to use compared to the \teff{}--LDR relationship. Section~\ref{sec:model} compares the observation results to spectral models. In Section~\ref{sec:discussion} we interpret the \teff{} variability observed in the TTS sample, present our \prot{} estimates, including a potential 1:1 \prot{}:\porb{} commensurability in the CI\,Tau system, and explore the dependence of \teff{} variations on other known properties of these stars. We conclude with a summary in Section~\ref{sec:summary}.

%-----------------------------------
\section{Target Samples, Observation, and Data Reduction}\label{sec:sample}

This study uses two target samples: the science target TTSs and the \teff{} calibration targets. The \teff{} calibration targets are main-sequence stars less active than TTSs, meaning almost no \teff{} variation. The \teff{} calibration targets also have known physical parameters, such as \teff{}, allowing us to establish an empirical \teff{}--EWR relationship after measuring the EWR from their spectra. This relationship is then used to estimate the \teff{} of TTSs after measuring their EWR from their spectra. In the following sections, we describe the two target samples and then discuss the IGRINS observation and data reduction.

%%----------------------------------
\subsection{T Tauri Star Sample} \label{sub:TTSsample}

The TTS sample was selected from targets in the Young Exoplanets Spectroscopic Survey (YESS) \citep{prato2008,crockett2012,johns-krull2016,tang2023}, a long-term program to search for stellar \citep[e.g.,][]{tang2023} and substellar \citep[e.g.,][]{johns-krull2016} companions to young active stars using optical and near-IR RVs. Of the $\sim$120 TTSs in the YESS target list, we focussed on the 13 with at least six IGRINS observations taken during a single season. This cutoff maximizes our sample size while also ensuring sufficient data in each season for a substantive analysis of \teff{} variability. 

The 13 targets can be broadly divided into two categories: those with two or more seasons of at least six observations in each season, and those with only one such season. The former category includes AA\,Tau, CI\,Tau, DK\,Tau, Hubble\,4, V827\,Tau, V830\,Tau, and V1075\,Tau; these targets' data not only allow the study of short-term (i.e., within a rotation period) \teff{} variations but also track long-term (seasonal) changes. The latter category includes DH\,Tau, DM\,Tau, DS\,Tau, GI\,Tau, IQ\,Tau, and LkCa\,15. 

\begin{figure}[tb!]
\centering
\includegraphics[width=1.\columnwidth]{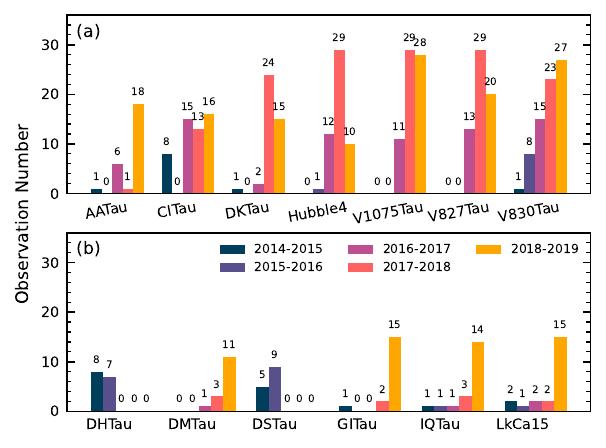}
\caption{
    The number of IGRINS observations of each TTS target over the five observing seasons between 2014--2019. 
    (a) Targets with two or more seasons having at least six observations; (b) Targets with only one season of at least six observations. 
    }
\label{fig:obs_count}
\end{figure}

The observations spanned five observational seasons, from 2014 to 2019, with a typical cadence of one observation per night for one or more weeks during each season. Figure~\ref{fig:obs_count} displays the number of IGRINS observations for each target during these periods, while basic information on the targets can be found in Table~\ref{tab:tts_basic}. In Table~\ref{tab:tts_basic}, columns (2) to (11) contain values from existing literature, whereas column (12) present measurements from this study. The results of the analysis from the time series data for these TTSs are detailed in Table~\ref{tab:tts}, where column (1) gives the target name, columns (2) and (3) list the observation UT dates and Julian Dates (JDs), respectively. The median signal-to-noise ratio (S/N) of the spectra from the reduction pipeline \citep[plp\,v2.2.0;][for further details, see Section~\ref{sec:data}]{jae_joon_lee_2017_845059} is provided in column (4). For our TTS spectra the typical median S/N is around 200. Finally, measurements from this study are documented in columns (5) through (9).

% =========== FIGURE % FIGURE =========== %

% =========== TABLE % TABLE =========== %
\begin{deluxetable*}{r C c RRRRRR c CCC}
\tablecaption{Basic Information for the TTS Sample\label{tab:tts_basic}
             }
\tabletypesize{\scriptsize}
\tablehead{
    \colhead{Name}      & \colhead{Hmag$^a$}             & \colhead{SpT$^b$} &
    \colhead{\teff{}}   & \colhead{$\sigma$\teff{}}  &
    \colhead{\logg{}}   & \colhead{$\sigma$\logg{}}  &
    \colhead{$\bar B$}  & \colhead{$\sigma \bar B$}  &&
    \colhead{\vsini{}}  & \colhead{$\sigma$\vsini{}} & \colhead{avg. RV} \vspace{-.2cm} \\ 
	\colhead{}          & \colhead{(mag)}   & \colhead{}   &
	\multicolumn{2}{c}{(K)}     &
    \colhead{}          & \colhead{}        &
    \multicolumn{2}{c}{(kG)}  & \colhead{}  &
    \multicolumn{3}{c}{(\kms)} \\
    \cline{4-5} \cline{8-9} \cline{11-13} 
	\colhead{(1)} & \colhead{(2)} & \colhead{(3)} & 
	\colhead{(4)} & \colhead{(5)} & 
    \colhead{(6)} & \colhead{(7)} &
    \colhead{(8)} & \colhead{(9)} & \colhead{} &
    \colhead{(10)} & \colhead{(11)} & \colhead{(12)}
     }
\startdata 
% \multicolumn{12}{c}{CTTS}\\
% AA\,Tau & 8.5 & M0.6 & 3751 & 171 & 3.9 & 0.3 & 2.16 & 0.41 && 17.7 & 12.7 \\
% CI\,Tau & 8.4 & K5.5 & 3951 & 94 & 3.8 & 0.2 & 1.95 & 0.31 && 17.1 & 11.2 \\
% DH\,Tau & 8.8 & M2.3 & 3477 & 125 & 3.9 & 0.2 & 2.21 & 0.32 && 15.9 & 7.1 \\
% DK\,Tau & 7.8 & K8.5 & 3809 & 183 & 4.0 & 0.3 & 2.55 & 0.64 && 15.9 & 14.3 \\
% DM\,Tau & 9.8 & M3.0 & 3449 & 104 & 4.1 & 0.2 & 1.63 & 0.29 && 18.5 & 4.7 \\
% DS\,Tau & 8.6 & M0.4 & 3879 & 138 & 3.9 & 0.2 & 2.09 & 0.41 && 15.9 & 12.4 \\
% GI\,Tau & 8.4 & M0.4 & 3689 & 186 & 3.8 & 0.3 & 1.92 & 0.44 && 17.1 & 10.6 \\
% IQ\,Tau & 8.4 & M1.1 & 3612 & 177 & 3.7 & 0.3 & 1.81 & 0.53 && 15.5 & 13.1 \\
% LkCa\,15 & 8.6 & K5.5 & 4156 & 123 & 4.1 & 0.2 & 1.83 & 9.99 && 17.9 & 13.6 \\
% \hline
% \multicolumn{12}{c}{WTTS} \\
% HubbleI\,4 & 7.6 & K8.0 & 3806 & 82 & 3.7 & 0.1 & 2.63 & 0.27 && 17.0 & 16.2 \\
% V1075\,Tau & 9.1 & K6.0 & 4122 & 97 & 4.2 & 0.2 & 2.62 & 9.99 && 18.2 & 31.5 \\
% V827\,Tau & 8.5 & M2.0 & 3610 & 92 & 3.9 & 0.1 & 2.42 & 9.99 && 18.0 & 20.3 \\
% V830\,Tau & 8.6 & K7.5 & 3878 & 80 & 3.9 & 0.1 & 2.4 & 9.99 && 17.0 & 31.6 \\
\multicolumn{13}{c}{CTTS}\\
AATau & 8.5 & M0.6 & 3751 & 171 & 3.9 & 0.3 & 2.16 & 0.41 && 12.5 & 2.3 & 17.7 \\
CITau & 8.4 & K5.5 & 3951 & 94 & 3.8 & 0.2 & 1.95 & 0.31 && 12.5 & 1.9 & 17.1 \\
DHTau & 8.8 & M2.3 & 3477 & 125 & 3.9 & 0.2 & 2.21 & 0.32 && 8.4 & 2.1 & 15.9 \\
DKTau & 7.8 & K8.5 & 3809 & 183 & 4.0 & 0.3 & 2.55 & 0.64 && 17.6 & 3.3 & 15.9 \\
DMTau & 9.8 & M3.0 & 3449 & 104 & 4.1 & 0.2 & 1.63 & 0.29 && 5.7 & 2.0 & 18.5 \\
DSTau & 8.6 & M0.4 & 3879 & 138 & 3.9 & 0.2 & 2.09 & 0.41 && 13.4 & 2.3 & 15.9 \\
GITau & 8.4 & M0.4 & 3689 & 186 & 3.8 & 0.3 & 1.92 & 0.44 && 12.1 & 2.6 & 17.1 \\
IQTau & 8.4 & M1.1 & 3612 & 177 & 3.7 & 0.3 & 1.81 & 0.53 && 14.3 & 2.5 & 15.5 \\
LkCa15 & 8.6 & K5.5 & 4156 & 123 & 4.1 & 0.2 & 1.83 & 9.99 && 15.4 & 2.3 & 17.9 \\
\multicolumn{13}{c}{WTTS} \\
HubbleI4 & 7.6 & K8.0 & 3806 & 82 & 3.7 & 0.1 & 2.63 & 0.27 && 16.8 & 1.7 & 17.0 \\
V1075Tau & 9.1 & K6.0 & 4122 & 97 & 4.2 & 0.2 & 2.62 & 9.99 && 30.9 & 2.1 & 18.2 \\
V827Tau & 8.5 & M2.0 & 3610 & 92 & 3.9 & 0.1 & 2.42 & 9.99 && 20.8 & 1.8 & 18.0 \\
V830Tau & 8.6 & K7.5 & 3878 & 80 & 3.9 & 0.1 & 2.4 & 9.99 && 31.6 & 1.8 & 17.0 \\
\enddata
\tablecomments{
    $^a$ 2MASS H magnitude \citep{skrutskie2006}. 
    $^b$ Spectra type from \citet{luhman2017}. 
    Information in columns (4)--(11) is from \citet{lopez-valdivia2021}.
    Column (12) shows results from this study using \texttt{IGRINS RV v.1.5.1} \citep{stahl2021,tang2021a,tang2023}. The average radial velocity (avg. RV) values are calculated using data in Table~\ref{tab:tts}.
    }
\end{deluxetable*}

%%----------------------------------
\subsection{Effective Temperature Calibration Sample}\label{sub:Calsample}

In order to derive an empirical \teff{}--EWR relationship, we first identified targets for the calibration of effective temperature. These sources are nearby main-sequence late K and M dwarfs with IGRINS observations that have also been studied by \citet[][hereafter B12]{boyajian2012}, \citet[][hereafter M15]{mann2015}, and are included in the Gaia FGK benchmark star sample \citep[GBS,][]{jofre2014,heiter2015}. In the following, we first describe how \teff{} and \logg{} were calculated for each source and then introduce the [Fe/H] values adopted or derived in these studies. 

%%%---------------------------------
\subsubsection{Effective Temperature \& Surface Gravity}\label{subsub:teff_cal_funphy}

The \teff{} values from \citetalias{boyajian2012} and GBS \citep{heiter2015} were estimated from the bolometric flux and the angular diameter of each stellar object using the Stefan–Boltzmann law \citepalias[e.g., Section~3.1 in][]{boyajian2012}. Stellar angular diameter can be measured directly using long-baseline interferometry (LBI), and bolometric flux can be calculated from combined photometry and/or spectra. This method therefore provides by far the most accurate \teff{} estimates \citep[e.g.,][]{vonbraun2011,boyajian2013,baines2021}. A physical stellar radius can be calculated with parallax and the angular diameter of the star. The stellar mass can be estimated from the evolutionary tracks (GBS), or via the mass–luminosity relationship \citepalias{boyajian2012}. In this study, we use radii and masses from GBS and \citetalias{boyajian2012} to derive \logg{} of the target from the fundamental relation $g=GM/R^2$, where $M$ is the stellar mass, $R$ the stellar radius, and $G$ the gravitational constant. These \logg{} estimates are in agreement with literature values, when available. \citet{passegger2022} reports \logg{} values for 10 of the targets in our final \teff{} calibration sample (Section~\ref{subsub:final_teff_cal}) and in all cases the \logg{} adopted in this study is within the reported uncertainties around the literature median value.

\begin{figure}[tb!]
\centering
\includegraphics[width=.99\columnwidth]{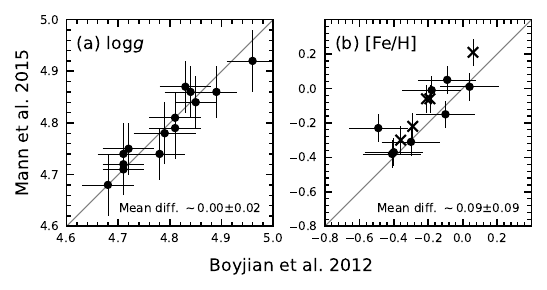}
\caption{
    Comparison of (a) \logg{} and (b) [Fe/H] for the \teff{} calibration sample targets common to both \citet{mann2015} and \citet{boyajian2012}. The mean difference between the two studies for each parameter is shown; X symbols denote targets without a reported [Fe/H] uncertainty in \citet{boyajian2012}. %\citep{boyajian2012,neves2012,anderson2011,rojas-ayala2012}.
    }
\label{fig:ma_bo_comp}
\end{figure}

The \teff{} values from \citetalias{mann2015} were estimated by fitting observed spectra to a model grid based on the method described in \citet{mann2013}. As this approach was tuned to match the results of LBI measurements to spectral fits with careful selection of the atmospheric models and wavelength regions used, the \teff{} measurements of \citetalias{mann2015} are comparable to those of GBS and \citetalias{boyajian2012}. The mean difference in \teff{} between \citetalias{mann2015} and literature LBI measurements is only $\sim$20$ \pm$11 K. Using the Stefan–Boltzmann law with parallax and bolometric flux as described above, \citetalias{mann2015} also estimated the angular diameter and physical radius for all their targets. The mean difference in angular diameter between their measurements and those derived from LBI is $\sim$1.4\% $\pm$0.7\%. Like \citetalias{boyajian2012}, \citetalias{mann2015} also adopted a mass–luminosity relationship to estimate the masses and thus the \logg{} for each target. Figure~\ref{fig:ma_bo_comp}a shows good agreement in \logg{} for the 13 targets common to both \citetalias{mann2015} and \citetalias{boyajian2012}.

The accuracy of \teff{} measurements via interferometry critically hinges on the the measurement of angular diameters and on the calculation of bolometric fluxes, with the primary uncertainties in the latter stemming from the establishment of photometric zero points, model atmosphere selection, and extinction estimation. Interferometric calibration further introduces discrepancies, especially given systematic differences at between different facilities and beam combiners \citep{casagrande2014, white2018}. A comprehensive analysis by \citet{tayar2022} identified systematic uncertainty floors of 2.4\% $\pm$ 0.6\% for bolometric fluxes and 4\% $\pm$ 1\% for angular diameters, contributing to an overarching \teff{} measurement uncertainty minimum of 2.0\% $\pm$ 0.5\%. For a typical \teff{} of TTS, 3800\,K, the minimum uncertainty is about 75\,K. 

For targets with angular diameters $\lesssim$ 1 mas, \citet{casagrande2014} identified a systematic offset where interferometric \teff{} measurements were found to be about 100 K higher compared to those derived using the infrared flux method (IRFM). To assess the impact of such discrepancies on our study, we performed IRFM \teff{} calculations for targets in \citetalias{boyajian2012} and \citetalias{mann2015} with available bolometric flux data. Utilizing the BT-Settl spectral library \citep{allard2011} for model flux and the IRTF Spectral Library \citep{cushing2005,rayner2009} for estimating monochromatic flux on Earth, we narrowed our analysis to four targets with \citetalias{mann2015}'s bolometric fluxes (GJ\,411, GJ\,581, GJ\,806, and GJ\,846) and two targets with data from \citetalias{boyajian2012} (GJ\,411 and GJ\,581). A range of discrepancies in \teff{} are observed based on the bolometric flux source: $-81$ K to 87 K for \citetalias{mann2015}'s fluxes and $-119$ K to 132 K for \citetalias{boyajian2012}'s, with an average absolute discrepancy of 7\,K across all measurements. Our analysis of this limited sample does not reveal a significant \teff{} offset. Instead, the differences seem largely tied to the bolometric flux calculation method.

% =========== TABLE % TABLE =========== %
\begin{deluxetable*}{ccCR RRRRR}
\tablecaption{Measurements of the TTS Samples\label{tab:tts}
             }
\tabletypesize{\scriptsize}
\tablehead{
    \colhead{Name}  & \colhead{UT}          & \colhead{JD$-$2450000} &
    \colhead{S/N}   &
    \colhead{RV}    & \colhead{$\sigma$RV}  &
    \colhead{EWR}   & \colhead{$\sigma$EWR} &
    \colhead{\teff{}} \vspace{-.2cm} \\ 
	\colhead{}      & \colhead{(yyyy-mm-dd)}& \colhead{(days)}   &
    \colhead{}      &
	\multicolumn{2}{c}{(\kms{})}            &
    \colhead{}      & \colhead{}            &
    \colhead{(K)}   \\
    \cline{5-6} 
	\colhead{(1)} & \colhead{(2)} & \colhead{(3)} & 
	\colhead{(4)} & \colhead{(5)} & 
    \colhead{(6)} & \colhead{(7)} &
    \colhead{(8)} & \colhead{(9)} 
     }
\startdata 
AATau & 2015-01-05 & 7028.647 & 155 & 17.37 & 0.07 & 0.680 & 0.017 & 3839 \\
AATau & 2016-11-14 & 7707.980 & 105 & 17.63 & 0.08 & 0.695 & 0.020 & 3827 \\
AATau & 2016-11-17 & 7710.916 & 66 & 17.87 & 0.16 & 0.793 & 0.028 & 3756 \\
AATau & 2016-11-22 & 7715.891 & 68 & 18.77 & 0.08 & 0.756 & 0.027 & 3782 \\
AATau & 2016-11-23 & 7716.824 & 117 & 17.95 & 0.09 & 0.748 & 0.019 & 3788 \\
\multicolumn{9}{c}{$\vdots$}
\enddata
\tablecomments{
    (This table is available in its entirety in machine-readable form.)
    }
\end{deluxetable*} 

% =========== FIGURE % FIGURE =========== %

%%%---------------------------------
\subsubsection{Metallicity}\label{subsub:teff_cal_feh}

The metallicity calculation for the GBS \citep{jofre2014} was based on high-resolution and high S/N ratio optical spectra (i.e., HARPS, NARVAL, and UVES). The reported final [Fe/H] values were computed based on the abundances of selected Fe$\textsc{i}$ lines with an absolute solar Fe abundance from \citet{grevesse2007}. During the fitting process, the \teff{} and \logg{} were fixed to values determined from fundamental relations \citep[][as described in Section~\ref{subsub:teff_cal_funphy}]{heiter2015}. In \citetalias{mann2015}, [Fe/H] values were calculated using the relationship between the near-IR atomic lines' EWs and [Fe/H], established in \citet{mann2013a,mann2014}, which made use of 156 binary systems with both a solar-type star of known [Fe/H] and a late K- or M-dwarf companion \citep[metallicities for the solar-type stars were measured by][]{valenti2005}. Lastly, \citetalias{boyajian2012} reported metallicities derived from a variety of sources: \citet{anderson2011,neves2012,rojas-ayala2012}. Figure~\ref{fig:ma_bo_comp}b compares [Fe/H] between the 13 targets common to both \citetalias{mann2015} and \citetalias{boyajian2012}. Unlike the \logg{} values shown in Figure~\ref{fig:ma_bo_comp}a, the [Fe/H] display a systematic offset and scatter of $\sim$0.1, but overall the metallicities from \citetalias{boyajian2012} are consistent with those of \citetalias{mann2015} within the uncertainties. Also, not all [Fe/H] values collected by \citetalias{boyajian2012} have reported uncertainties.

% =========== TABLE % TABLE =========== %
\begin{deluxetable*}{r CCCCCCCCc RR cCC CC}
\tablecaption{Calibration Sources\label{tab:cals}
             }
\tabletypesize{\scriptsize}
\tablehead{
    \colhead{Name}  &
    \colhead{\teff{}} & \colhead{$\sigma$\teff{}}   &
    \colhead{Mass} & \colhead{$\sigma$Mass}   &
    \colhead{R} & \colhead{$\sigma$R}   &
    \colhead{\logg{}} & \colhead{$\sigma$\logg{}}   & 
    \colhead{Ref.1$^a$}  &
    \colhead{[Fe/H]}  & \colhead{$\sigma${\rm[Fe/H]}} & 
    \colhead{Ref.2$^b$}  & \colhead{S/N$^c$} & \colhead{\vsini}   &
    \colhead{EWR}   & \colhead{$\sigma${\rm EWR}} \vspace{-.2cm} \\
%    \cline{2-3} \cline{8-9} \cline{11-12}  
	\colhead{(1)} & \colhead{(2)} & \colhead{(3)} & 
	\colhead{(4)} & \colhead{(5)} &\colhead{(6)} & \colhead{(7)} &
    \colhead{(8)} & \colhead{(9)} & \colhead{(10)} & \colhead{(11)} &
    \colhead{(12)} & \colhead{(13)} & \colhead{(14)} & \colhead{(15)} & 
    \colhead{(16)} & \colhead{(17)} 
     }
\startdata 
GJ 581 & 3418 & 80 & 0.29 & 0.04 & 0.30 & 0.02 & 4.94 & 0.08 & 1,2 & -0.15 & 0.08 & 1 & 535 & 0.00 & 1.419 & 0.021 \\
GJ 725A & 3424 & 61 & 0.33 & 0.05 & 0.35 & 0.01 & 4.85 & 0.07 & 1,2 & -0.23 & 0.08 & 1 & 545 & 1.00 & 1.566 & 0.021 \\
GJ 687 & 3426 & 66 & 0.41 & 0.06 & 0.42 & 0.02 & 4.81 & 0.07 & 1,2 & 0.05 & 0.08 & 1 & 597 & 0.00 & 1.470 & 0.019 \\
GJ 436 & 3447 & 80 & 0.46 & 0.06 & 0.45 & 0.03 & 4.79 & 0.08 & 1,2 & 0.01 & 0.08 & 1 & 767 & 0.00 & 1.258 & 0.015 \\
GJ 615.2C & 3454 & 63 & 0.42 & 0.04 & 0.44 & 0.02 & 4.78 & 0.06 & 1 & -0.06 & 0.03 & 1 & 236 & \nodata & 1.256 & 0.024 \\
GJ 70 & 3458 & 60 & 0.40 & 0.04 & 0.41 & 0.02 & 4.80 & 0.06 & 1 & -0.13 & 0.08 & 1 & 265 & \nodata & 1.404 & 0.025 \\
GJ 625 & 3475 & 60 & 0.32 & 0.03 & 0.33 & 0.01 & 4.89 & 0.05 & 1 & -0.35 & 0.08 & 1 & 374 & 0.00 & 1.653 & 0.034 \\
GJ 745B & 3494 & 62 & 0.30 & 0.03 & 0.32 & 0.01 & 4.90 & 0.06 & 1 & -0.35 & 0.08 & 1 & 327 & 0.00 & 1.593 & 0.033 \\
GJ 745A & 3500 & 60 & 0.30 & 0.03 & 0.31 & 0.01 & 4.92 & 0.06 & 1 & -0.33 & 0.08 & 1 & 345 & 0.00 & 1.584 & 0.030 \\
GJ 3195 & 3500 & 61 & 0.41 & 0.04 & 0.42 & 0.02 & 4.80 & 0.06 & 1 & -0.12 & 0.08 & 1 & 235 & \nodata & 1.502 & 0.029 \\
PM I18007+2933 & 3509 & 61 & 0.46 & 0.05 & 0.46 & 0.02 & 4.76 & 0.06 & 1 & -0.06 & 0.03 & 1 & 218 & \nodata & 1.289 & 0.026 \\
GJ 411 & 3513 & 61 & 0.39 & 0.06 & 0.39 & 0.01 & 4.84 & 0.06 & 1,2 & -0.38 & 0.08 & 1 & 978 & 0.00 & 1.640 & 0.018 \\
GJ 806 & 3542 & 61 & 0.43 & 0.04 & 0.44 & 0.02 & 4.78 & 0.06 & 1 & -0.15 & 0.08 & 1 & 552 & 0.00 & 1.279 & 0.018 \\
GJ 393 & 3548 & 60 & 0.43 & 0.04 & 0.42 & 0.02 & 4.82 & 0.06 & 1 & -0.18 & 0.08 & 1 & 466 & 0.00 & 1.335 & 0.020 \\
GJ 412A & 3558 & 71 & 0.40 & 0.06 & 0.39 & 0.02 & 4.85 & 0.07 & 1,2 & -0.37 & 0.08 & 1 & 589 & 0.00 & 1.540 & 0.022 \\
GJ 752A & 3558 & 60 & 0.47 & 0.05 & 0.47 & 0.02 & 4.76 & 0.05 & 1 & 0.10 & 0.08 & 1 & 551 & 0.20 & 1.099 & 0.015 \\
GJ 15A & 3585 & 61 & 0.41 & 0.06 & 0.39 & 0.01 & 4.88 & 0.06 & 1,2 & -0.30 & 0.08 & 1 & 810 & \nodata & 1.446 & 0.019 \\
GJ 382 & 3623 & 60 & 0.53 & 0.05 & 0.52 & 0.02 & 4.72 & 0.05 & 1 & 0.13 & 0.08 & 1 & 414 & 1.50 & 0.897 & 0.015 \\
GJ 526 & 3633 & 67 & 0.49 & 0.07 & 0.48 & 0.02 & 4.76 & 0.07 & 1,2 & -0.31 & 0.08 & 1 & 514 & 0.00 & 1.489 & 0.022 \\
GJ 87 & 3638 & 62 & 0.44 & 0.04 & 0.44 & 0.02 & 4.79 & 0.06 & 1 & -0.36 & 0.08 & 1 & 434 & \nodata & 1.648 & 0.028 \\
GJ 908 & 3646 & 60 & 0.41 & 0.04 & 0.41 & 0.01 & 4.83 & 0.05 & 1 & -0.45 & 0.08 & 1 & 435 & 0.00 & 1.913 & 0.035 \\
GJ 3408B & 3656 & 62 & 0.43 & 0.04 & 0.42 & 0.02 & 4.83 & 0.06 & 1 & -0.26 & 0.03 & 1 & 374 & \nodata & 1.369 & 0.025 \\
GJ 686 & 3657 & 60 & 0.44 & 0.04 & 0.42 & 0.01 & 4.83 & 0.05 & 1 & -0.25 & 0.08 & 1 & 539 & 0.00 & 1.343 & 0.020 \\
GJ 176 & 3680 & 60 & 0.49 & 0.05 & 0.45 & 0.02 & 4.82 & 0.06 & 1 & 0.14 & 0.08 & 1 & 227 & 0.00 & 1.048 & 0.019 \\
GJ 887 & 3682 & 92 & 0.51 & 0.07 & 0.47 & 0.02 & 4.80 & 0.08 & 1,2 & -0.06 & 0.08 & 1 & 348 & \nodata & 1.294 & 0.021 \\
PM I02441+4913W & 3685 & 60 & 0.52 & 0.05 & 0.50 & 0.02 & 4.76 & 0.05 & 1 & 0.06 & 0.03 & 1 & 353 & \nodata & 0.958 & 0.016 \\
GJ 134 & 3700 & 61 & 0.64 & 0.06 & 0.63 & 0.03 & 4.65 & 0.06 & 1 & 0.53 & 0.08 & 1 & 442 & 1.40 & 0.603 & 0.013 \\
GJ 649 & 3700 & 60 & 0.53 & 0.05 & 0.51 & 0.02 & 4.75 & 0.05 & 1 & 0.03 & 0.08 & 1 & 329 & 0.30 & 0.905 & 0.016 \\
GJ 505B & 3709 & 60 & 0.54 & 0.05 & 0.54 & 0.02 & 4.71 & 0.05 & 1 & -0.12 & 0.03 & 1 & 396 & \nodata & 0.938 & 0.015 \\
GJ 880 & 3716 & 61 & 0.57 & 0.08 & 0.55 & 0.02 & 4.71 & 0.06 & 1,2 & 0.21 & 0.08 & 1 & 993 & 1.30 & 0.751 & 0.012 \\
GJ 514 & 3727 & 61 & 0.53 & 0.05 & 0.48 & 0.02 & 4.79 & 0.05 & 1 & -0.09 & 0.08 & 1 & 468 & 0.00 & 0.998 & 0.015 \\
GJ 809 & 3741 & 63 & 0.58 & 0.08 & 0.54 & 0.02 & 4.73 & 0.07 & 1,2 & -0.06 & 0.08 & 1 & 804 & 0.00 & 0.785 & 0.013 \\
GJ 212 & 3765 & 60 & 0.59 & 0.06 & 0.57 & 0.02 & 4.70 & 0.06 & 1 & 0.19 & 0.03 & 1 & 327 & 1.90 & 0.695 & 0.014 \\
GJ 281 & 3771 & 60 & 0.63 & 0.06 & 0.63 & 0.03 & 4.64 & 0.06 & 1 & 0.12 & 0.08 & 1 & 468 & 1.70 & 0.523 & 0.013 \\
GJ 96 & 3785 & 62 & 0.61 & 0.06 & 0.60 & 0.02 & 4.67 & 0.05 & 1 & 0.14 & 0.08 & 1 & 920 & \nodata & 0.606 & 0.012 \\
GJ 205 & 3801 & 60 & 0.62 & 0.09 & 0.58 & 0.02 & 4.71 & 0.06 & 1,2 & 0.49 & 0.08 & 1 & 612 & 2.10 & 0.571 & 0.013 \\
GJ 685 & 3846 & 61 & 0.59 & 0.06 & 0.54 & 0.02 & 4.74 & 0.05 & 1 & 0.10 & 0.08 & 1 & 225 & 0.90 & 0.742 & 0.015 \\
GJ 846 & 3848 & 60 & 0.59 & 0.06 & 0.55 & 0.02 & 4.73 & 0.05 & 1 & 0.02 & 0.08 & 1 & 396 & 1.10 & 0.676 & 0.013 \\
GJ 338B & 3867 & 37 & 0.60 & 0.06 & 0.57 & 0.01 & 4.71 & 0.05 & 2 & -0.15 & 0.17 & 2,4 & 549 & 0.50 & 0.579 & 0.013 \\
GJ 338A & 3913 & 69 & 0.61 & 0.09 & 0.56 & 0.03 & 4.72 & 0.08 & 1,2 & -0.01 & 0.08 & 1 & 617 & 1.00 & 0.559 & 0.013 \\
GJ 208 & 3966 & 60 & 0.65 & 0.07 & 0.60 & 0.02 & 4.69 & 0.05 & 1 & 0.05 & 0.08 & 1 & 337 & 2.00 & 0.519 & 0.013 \\
GJ 820B & 4044 & 32 & 0.64 & 0.06 & 0.60 & 0.02 & 4.67 & 0.04 & 1,2,3 & -0.22 & 0.08 & 1 & 552 & \nodata & 0.542 & 0.013 \\
GJ 169 & 4124 & 62 & 0.74 & 0.07 & 0.69 & 0.02 & 4.63 & 0.05 & 1 & 0.39 & 0.08 & 1 & 1085 & 1.30 & 0.352 & 0.012 \\
GJ 673 & 4124 & 60 & 0.71 & 0.07 & 0.65 & 0.02 & 4.66 & 0.05 & 1 & 0.19 & 0.08 & 1 & 550 & \nodata  & 0.374 & 0.013 \\
GJ 820A & 4374 & 22 & 0.68 & 0.07 & 0.66 & 0.00 & 4.63 & 0.04 & 2,3 & -0.33 & 0.38 & 3 & 779 & \nodata & 0.328 & 0.012 \\
GJ 702B & 4393 & 149 & 0.70 & 0.07 & 0.67 & 0.01 & 4.63 & 0.05 & 2 & 0.06 & 0.04 & 2,5,6 & 624 & \nodata & 0.237 & 0.013 \\
GJ 570A & 4507 & 58 & 0.74 & 0.07 & 0.74 & 0.02 & 4.57 & 0.05 & 2 & 0.05 & 0.04 & 2,6,7 & 698 & \nodata & 0.172 & 0.013 \\
GJ 892 & 4699 & 16 & 0.76 & 0.08 & 0.78 & 0.01 & 4.54 & 0.04 & 2 & 0.01 & 0.04 & 2,5,6 & 913 & \nodata & 0.134 & 0.013 \\
eps Vir & 4983 & 61 & \nodata & \nodata & \nodata & \nodata & 2.77 & 0.02 & 3 & 0.15 & 0.16 & 3 & 418 & \nodata & 0.183 & 0.014 \\
\enddata
\tablecomments{
    $^a$ Reference for \teff{} and \logg{} measurements:
        (1) \citet{mann2015}, (2) \citet{boyajian2012}, and 
        (3) Gaia FGK benchmark stars \citep[GBS,][]{jofre2014,heiter2015}.
    $^b$ Reference for [Fe/H] measurement: (1)--(3) the same as reference 
    for \teff{} and \logg{}. (4) \citet{rojas-ayala2012}, (5) \citet{ramirez2012},
    (6) \citet{luck2006}, and (7) \citet{ghezzi2010}.
    $^c$ S/N refers to the median S/N values between 1.561--1.565 \micron{} from the 
    IGRINS pipeline package version 2.2.0 \citep[plp\,v2.2.0;][]{jae_joon_lee_2017_845059}.
    Column (15) give \vsini{} measurements from \citet{reiners2022}, and columns
    (16) and (17) give equivalent width ratio (EWR) measurements and their uncertainties 
    from this study.
    (This table is available in machine-readable form.)
    }
\end{deluxetable*} 

%%%---------------------------------
\subsubsection{Final Samples}\label{subsub:final_teff_cal}

Targets with \teff{} between 3200 and 5000\,K in \citetalias{boyajian2012}, \citetalias{mann2015}, and GBS were cross-matched with The Raw \& Reduced IGRINS Spectral Archive \citep[RRISA,][]{sawczynec2022,sawczynec2023}\footnote{\url{https://igrinscontact.github.io}}. With a S/N cut of $>$100, we are left with 49 final \teff{} calibration targets with high-quality IGRINS spectra, of which 19 have measurements from \citetalias{boyajian2012}, 43 have measurements from \citetalias{mann2015}, and three have measurements from GBS. 

For targets with multiple measurements in \teff{} and \logg{}, we took the mean and adopted an uncertainty based on the quadratic sum of all uncertainties. For [Fe/H], we adopted measurements from \citetalias{mann2015} and GBS, if available. If not, we used the values in \citetalias{boyajian2012}, except for the three targets that lack [Fe/H] uncertainties in \citetalias{boyajian2012}. For these targets (GJ\,702B, GJ\,570A, and GJ\,892), we draw [Fe/H] values and uncertainties from \citet{luck2006}, \citet{ramirez2012}, and \citet{ghezzi2010}. 

\begin{figure}[tbh]
\centering
\includegraphics[width=1.\columnwidth]{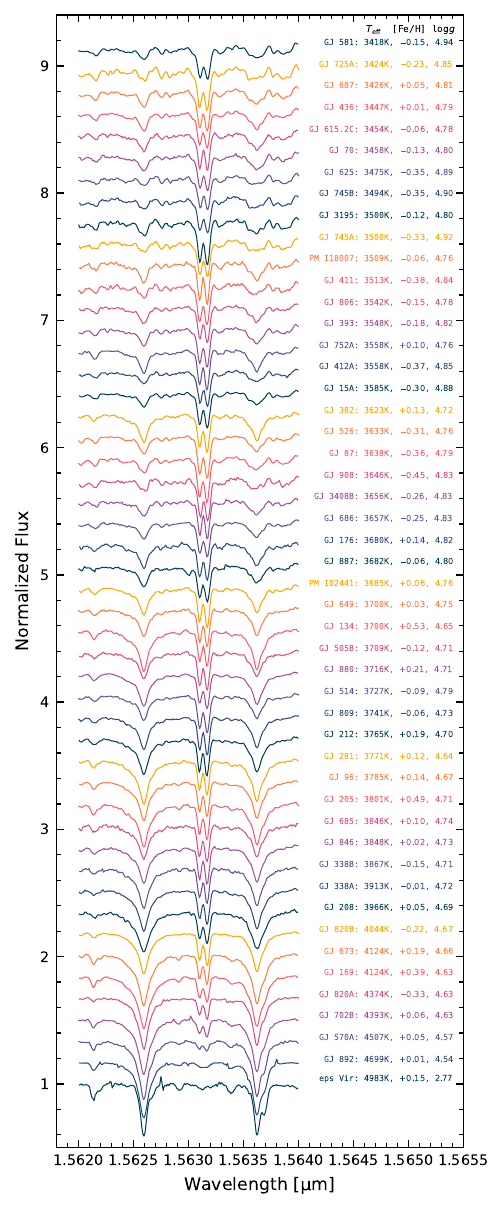}
\caption{
    IGRINS spectra for the \teff{} calibration sample. Spectra are sorted by \teff{}, with high \teff{} on the top and low \teff{} at bottom. The \teff{}, [Fe/H], and \logg{} values are from Table~\ref{tab:cals}.
    }
\label{fig:teffSTD_spec}
\end{figure}

Table~\ref{tab:cals} presents the final parameters for the \teff{} calibration targets. Column (1) shows the target name, and (2) and (3) provides the \teff{} values and their uncertainties, respectively. Columns (4) to (9) give stellar mass, stellar radii ($R$), \logg{}, and their respective uncertainties. The sources for literature values of mass and radii are cited in column (10). Columns (11) and (12) provide the [Fe/H] values and uncertainties, with their references listed in column (13). The median S/N values from the plp reduction pipeline are included in column (14). Column (15) shows \vsini{} values from \citet{reiners2022}. Measurements conducted in this study are in columns (16) and (17). Figure~\ref{fig:teffSTD_spec} displays the IGRINS $H$ band spectra centered on the Fe and OH lines, arranged from top to bottom by increasing \teff{}. The values of \teff{}, [Fe/H], and \logg{} for the targets, as listed in Table~\ref{tab:cals}, are also illustrated in the figure.

% =========== FIGURE % FIGURE =========== %

%-----------------------------------
\subsection{Observations and Data Reduction}\label{sec:data}

All spectra used in this study were taken with IGRINS, a cross-dispersed echelle spectrograph which can simultaneously cover the entire $H$ (1.49--1.80 \micron{}) and $K$ bands (1.96--2.46 \micron{}) while delivering a resolution of $R\sim$45,000 with a fixed width slit (0\farcs8). With no moving parts, IGRINS has been installed at a number of different sites, including the McDonald Observatory's 2.7 m Harlan J. Smith Telescope, the 4.3 m Lowell Discovery Telescope, and the 8-m Gemini South telescope. Our target spectra were observed at these three sites from July 2014 to April 2019. Observations were all taken with one or more A-B nodding sequence(s), and the reduction was carried out using the IGRINS pipeline package version 2.2.0 \citep[plp\,v2.2.0\footnote{\url{https://github.com/igrins/plp}};][]{jae_joon_lee_2017_845059}. 

We downloaded the reduced spectra with telluric line correction from RRISA. The telluric line correction was performed by dividing the science spectrum by an A0V spectrum observed shortly before or after the science target. (typically within an air-mass difference of 0.1). As RRISA also provides a cross-match (XMatch) catalog with SIMBAD that has been checked by hand to ensure the XMatch accuracy, we used the RVs given in the XMatch catalog for the \teff{} calibration sample. For the TTS sample, we calculated RVs for each epoch using \texttt{IGRINS RV v.1.5.1}\footnote{\url{https://github.com/shihyuntang/igrins_rv}} \citep{stahl2021,tang2021a,tang2023} because RV variations in these young active stars can be as high as several \kms{} \citep[e.g.,][]{crockett2012}. The average RV for each target TTS, determined from \texttt{IGRINS RV}, is given in column (12) of Table~\ref{tab:tts_basic}.

% =========== FIGURE % FIGURE =========== %
\begin{figure*}[tbh!]
\centering
\includegraphics[width=1.0\textwidth]{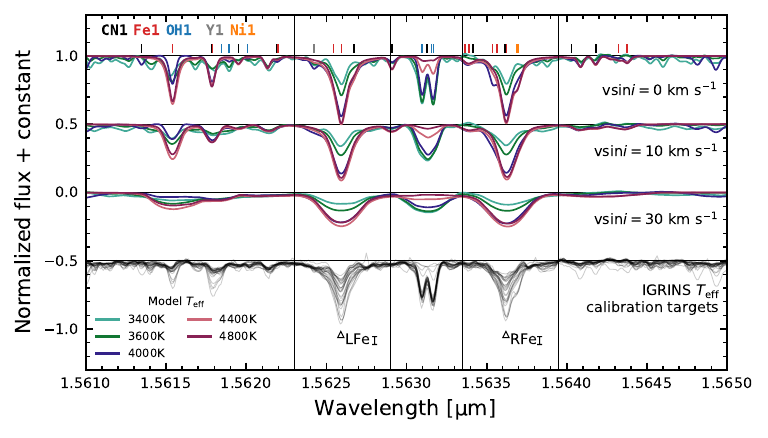}
\caption{
    Spectra showing the $H$ band region centered around the Fe and OH lines, all normalized as per Section~\ref{sub:cal}. 
    The three sets of colored lines on the top are model spectra (see Section~\ref{sec:model}) of various \teff{} with \logg{} = 4.0, and $R$ = 45,000; black lines at the bottom are the IGRINS spectra of all \teff{} calibration sources.
    The locations of several strong lines (e.g., CN\,\textsc{i}, Fe\,\textsc{i}, OH\,\textsc{i}, Y\,\textsc{i}, and Ni\,\textsc{i}) are marked and a legend is shown for these in the upper left corner. Relatively strong, unlabelled lines that appear in sources with \teff{} $\lesssim$ 3600 K are mostly water lines. Vertical black lines mark the wavelength regions for the EW calculation. The two main Fe\,\textsc{i} lines discussed in Section~\ref{sec:spec_analysis}, LFe\,\textsc{i} at $\sim$1.56259 \micron{} and RFe\,\textsc{i} at $\sim$1.56362 \micron{} are highlighted. Line information is taken from the VALD3 line list \citep{ryabchikova2015a}.
    }
\label{fig:normal_spectra}
\end{figure*}
% ======================================= %

%-----------------------------------
\section{Spectral Analysis} \label{sec:spec_analysis}
%%----------------------------------
\subsection{The OH and Fe Region} \label{sub:OHFe}

The OH lines at $\sim$1.56310 \micron{} and $\sim$1.56317 \micron{} and the Fe\,\textsc{i} lines at $\sim$1.56259 \micron{} (the left-hand side Fe\,\textsc{i}, LFe\,\textsc{i}) and $\sim$1.56362 \micron{} (the right-hand side Fe\,\textsc{i}, RFe\,\textsc{i}, Figure~\ref{fig:normal_spectra}) in the near-IR $H$ band are advantageous for LD and EW analysis for a number of reasons. The lines are close in wavelength, so they will be minimally affected by imperfect spectral normalization, and, for CTTSs, the wavelength dependent veiling. Additionally, the OH and Fe\,\textsc{i} lines have contrasting responses to changes in \teff{}: LD increases with \teff{} for the Fe\,\textsc{i} lines from $\sim$3200\,K to $\sim$5000\,K, but for the OH lines, LD decreases with increasing \teff{} from $\sim$3400\,K to $\sim$5000\,K (Fig.~\ref{fig:teffSTD_spec}).

The telluric contamination in this wavelength region is also minimal. The main source of telluric absorption is water vapor, which is only present at substantial levels during winter and at lower altitude observing sites (in this case, the McDonald Observatory and the Lowell Discovery Telescope). Telluric water absorption only affects the LFe\,\textsc{i} line and the shorter wavelength of the OH lines ($\sim$1.56310 \micron{}). The typical flux difference in this wavelength region, when choosing an A0V standard star with an air-mass difference of 0.1, is less than 0.3\%, which is about a change of 0.005 in EWR (see Section~\ref{sub:cal}).

By contrast, the effects of line blending are both significant and unavoidable given the large \vsini{} values of our TTS sample (up to about $30$ \kms{}, Table~\ref{tab:tts_basic}). To show the level of significance of blending, we display model spectra with various values of \teff{} and \vsini{} in colored lines in Figure~\ref{fig:normal_spectra}. The line blending effect can be seen, for example, in the LFe\,\textsc{i} line. It is blended with another Fe\,\textsc{i} line and a CN\,\textsc{i} line. Another example is the blending of the RFe\,\textsc{i} with two other Fe\,\textsc{i} lines, a CN\,\textsc{i} line, and a Ni\,\textsc{i} line. Some lines can be distinguished in the low \vsini{} spectra but start to blend in with each other at a higher \vsini{}; for example, the two separate Fe\,\textsc{i} lines within 1.5635--1.5636 \micron{} are distinguishable when \vsini{}=0 \kms but blend together at \vsini{}=10 \kms.

%%----------------------------------
\subsection{Equivalent Width and Line Depth Calculation} \label{sub:cal}

To ensure LD and EW measurements are consistent between synthetic models and IGRINS spectra, as well as between stars with different \vsini{}, we develop a standardized normalization approach: all spectra are normalized to the 95th percentile value of the flux measured between 1.561--1.565 \micron{} (Fig.~\ref{fig:normal_spectra}). This wavelength region was chosen to ensure that spectra maintain a similar continuum level at $\sim$1.5612 and $\sim$1.5646 \micron{} regardless of the rotational broadening exhibit by our targets, even with \vsini{} as high as 30 \kms{}.

To streamline the measurement process and minimize the blending effect discussed in Section~\ref{sub:OHFe}, we calculated LD and EW using wavelength regions which include lines that might be blended at higher \vsini{}. The EW is measured from 1.56230--1.56290 \micron{} for the LFe\,\textsc{i} line, between 1.56290--1.56335 \micron{} for the OH lines, and from 1.56335--1.56395 \micron{} for the RFe\,\textsc{i} line (these regions are demarcated by vertical black lines in Figure~\ref{fig:normal_spectra}). The final LD and EW for Fe is the mean of the values measured in the two Fe\,\textsc{i} regions. The OH lines form a close doublet that is effectively unresolved at the IGRINS resolution of 45,000.

\begin{figure}[tb!]
\centering
\includegraphics[width=1.\columnwidth]{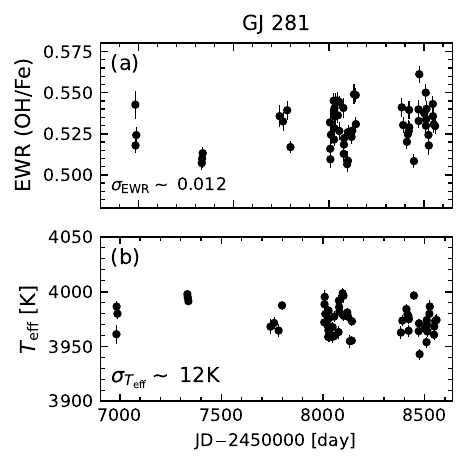}
\caption{
    The (a) EWR and (b) \teff{} time series for GJ\,281 spanning about 4 years. The \teff{} values in (b) are calculated from the EWRs in (a) using the ``solar'' relationship given in Figure~\ref{fig:empirical}. The small scatter in EWR, $\sigma_{\rm EWR}\sim$ 0.012, and in the \teff{}, $\sigma_{T_{\rm eff}}\sim$ 12 K, demonstrate the precision of our technique at $\sim$3950 K (EWR $\sim$ 0.53)}.
\label{fig:gj281}
\end{figure}

We used the \texttt{Python} package \texttt{specutils v1.7.0} \citep{earl2022} to measure the EW of lines in the three wavelength regions. Uncertainties were calculated using the Monte Carlo (MC) method with 5000 spectra generated based on the measured S/N. We tested the robustness of this procedure by measuring the EWR in spectra of a known quiescent star, GJ\,281, which has been used as a standard because of its low RV variability \citep[$\sigma_{\rm RV}<$30 \ms{},][]{endl2003,stahl2021}. Our EWR measurements of GJ281 show a scatter of $\sigma_{\rm EWR}\sim$ 0.012 over 61 nights of observation spanning a period of 4 years (Fig.~\ref{fig:gj281}~a). This scatter is about 2.8 times larger than the median uncertainty of individual EWR values calculated based on the S/N from the plp pipeline. To check for systematic errors in the EWR, we assumed GJ\,281 has negligible intrinsic variation between observations and we used the average scatter in the residual between adjacent spectra, sorted by S/N, to infer a more realistic S/N. We find these residuals to be systematically larger than implied by the plp S/N estimate, and this yields scaling factors for the plp spectral uncertainty: 0.75 for plp S/N below 225, and 0.58 for those above. The revised S/N values lead to a median uncertainty of 0.0051, which is still significantly less (by a factor of 2.4) than the observed $\sigma_{\rm EWR}$ of GJ\,281. Consequently, we define an additional systematic uncertainty, $\sigma_{\rm sys} = \sqrt{0.012^2 - 0.0051^2} \sim 0.011$, which is incorporated quadratically into all final EWR measurement uncertainties. Possible sources of this systematic uncertainty include inaccurate continuum normalization and the influence of spot coverage \citep[field M dwarfs are known to have spots, e.g,][]{claytor2024}. The EWR uncertainty introduced by adopting different A0V standards during the telluric correction (Section~\ref{sub:OHFe}), could be one of these sources. Using the \teff{}--EWR relationship established for solar metallicity targets (Section~\ref{sub:emprical}, Figure~\ref{fig:empirical}), we convert GJ\,281's EWRs to \teff{}, as shown in Figure~\ref{fig:gj281}b. The resultant \teff{} scatter, $\sigma_{T_{\rm eff}}\sim$ 12 K, indicates our measurement precision for \teff{} around 3950 K (an EWR of approximately 0.53). This analysis employs results from Section~\ref{sub:emprical}, discussed later, to maintain the logical flow of our findings and discussions.

% . Therefore, we do not incorporate it in the final uncertainties as we consider the effect minimal compared to 0.011

% =========== FIGURE % FIGURE =========== %
% =========== FIGURE % FIGURE =========== %

There are three common techniques to measure the LD: fit a Gaussian function to the main profile of the line \citep[e.g.,][]{afsar2023a}, fit a parabolic curve to pixels around the lowest flux \citep[e.g.,][]{gray1994,catalano2002,jian2020}, and calculate the difference between the average flux near the minimum flux point and the continuum \citep[e.g.,][]{lopez-valdivia2019}. We tested all three methods in the three wavelength regions used for EW measurement (Fig.~\ref{fig:normal_spectra}) by (a) fitting a Gaussian function within each region, (b) fitting a second-order polynomial to the 7 pixels ($\sim$2 resolution elements) around the minimum flux in each region, and (c) calculating the mean flux of the 3 pixels around the minimum flux in each region. 

Because the spectra in our TTS sample exhibit generally lower S/N than the \teff{} calibration sample, we smoothed the spectra using a Gaussian (standard normal) kernel with $\sigma =$ 1 pixel before performing the LD measurements. This smoothing process helps minimize the effect of outlier flux measurements in the lower S/N spectra. 

%%----------------------------------
\subsection{Spectral Broadening Effect}\label{sub:vsini}

\begin{figure}[t!]
\centering
\includegraphics[width=1.0\columnwidth]{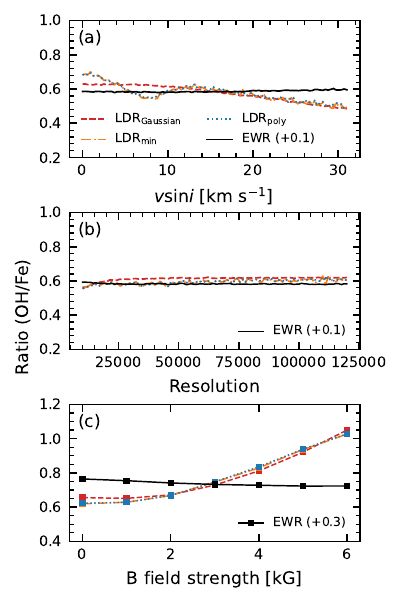}
\caption{
    Simulated EWR and LDR as a function of (a) \vsini{}, (b) spectral resolution ($R$), and (c) magnetic field strength.
    Model spectra are generated with \teff{} = 4000 K, \logg{} = 4.0, [Fe/H] = 0.0, and S/N = 150. A \vsini{} of 10 \kms{} is adopted for model spectra used in (b) and (c), and an $R$ of 45,000 is adopted for model spectra used in (a) and (c).
    Red lines show LDR calculated by fitting a Gaussian, orange lines show LDR calculated by adopting the mean flux of 3 pixels around the minimum flux, blue lines are LDR calculated by fitting a second-order polynomial to the 7 pixels around the absorption lines' minimum flux, and the black lines show the EWR (see Section~\ref{sub:cal}).
    In all panels, the EWR lines are shifted by the amount given in the legend for a better comparison. Also, in all panels, the orange and the blue lines are on top of each others. The EWR method is less affected by all three spectral broadening effects.}
\label{fig:vsinieffect}
\end{figure}

In order to be robust and comprehensive, an empirical \teff{}--EWR (and/or LDR) relationship must not significantly depend on other factors such as \vsini{}, spectral resolution ($R$), or the surface average magnetic field strength ($\bar B$). We study the effect of these parameters on LDRs and EWRs using synthetic spectra with \teff{} = 4000\,K, \logg{} = 4.0, and S/N = 150. Figure~\ref{fig:vsinieffect} illustrates the impact on EWR and LDR of (a) \vsini{}, (b) $R$, and (c) magnetic field strength. Each LDR measurement method (Gaussian fit, polynomial fit, minimum) is displayed in a different color; note the high resolution of IGRINS spectra make the minimum (yellow) and polynomial (blue) results often indistinguishable.

% =========== FIGURE % FIGURE =========== %
% =========== FIGURE % FIGURE =========== %

% =========== FIGURE % FIGURE =========== %
\begin{figure*}[tbh!]
\centering
\includegraphics[width=1.\textwidth]{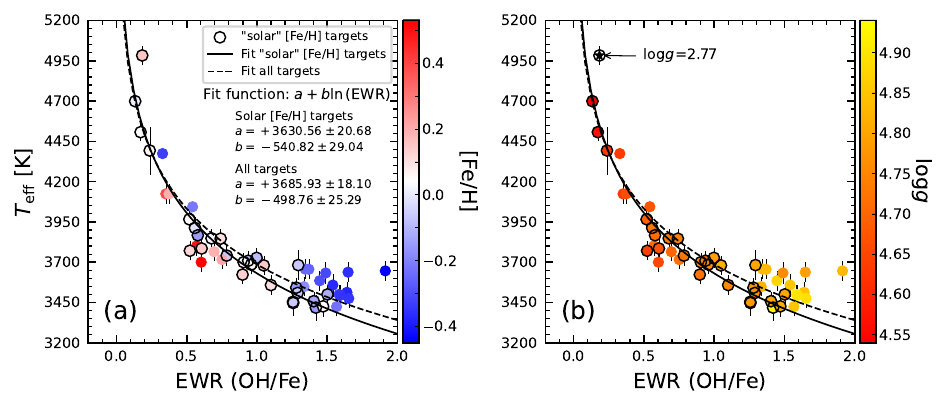}
\caption{
    The \teff{}--EWR plot for the \teff{} calibration sample.
    (a) sample color coded by [Fe/H], and (b) sample color coded by \logg{}. The empirical relationship of \teff{}--EWR is fitted using a natural log function. The solid black line shows the fitting result for solar-type ($-0.15<$[Fe/H]$<0.15$) stars, highlighted with plotted outlined with black circles, and the dashed line shows the fitting result of all stars in the sample. The best-fit results and their associated uncertainties are given in panel (a). In (b), an outlier with a low \logg{} of 2.77 is marked with a star symbol.
    }
\label{fig:empirical}
\end{figure*}
% =========== FIGURE % FIGURE =========== %

First, we observe that LDRs clearly vary with \vsini{} regardless of measurement method (Fig.~\ref{fig:vsinieffect}a) because of the blending of lines. For example, the dip in LDR around a \vsini{} of 4--10 \kms{} with the minimum and polynomial fit methods is primarily the result of blending of the OH doublet. The maximum variation in the LDR from \vsini{} of 0 to 30 \kms{} is about 0.21, which is equivalent to a temperature difference of about 230 K at 4000\,K if we adopt the solar \teff{}--EWR relationship derived in the next section (Section~\ref{sub:emprical}, and Figure~\ref{fig:empirical}). The EWR measurements, on the other hand, only yield a maximum variation of about 0.02, equivalent to a $\sim$20 K difference at 4000 K. 

To account for variations in the $R$ across the IGRINS detector \citep[as shown in Figure~4 of][]{stahl2021}, and to generalize this method for use with other spectrographs, Figure~\ref{fig:vsinieffect}b illustrates the relationship of LDR and EWR with $R$, ranging from 10,000 to 120,000. Notably, there is a marked decrease in LDR for $R$ values below approximately 30,000. This decline is attributed to the increasingly dominant influence of spectral resolution blending close lines, which becomes similarly significant to a \vsini{} of 10 \kms. In general, the impact of $R$ on the ratio is relatively minor (with a maximum LDR variation of 0.07) in comparison to the effect of \vsini{}.

The $B$ strength has the largest impact on both the LDR and EWR measurements. Given a typical range of $\bar B$ values from $\sim$1--3 kG for TTSs \citep[e.g.,][]{johns-krull2007,lopez-valdivia2021,sokal2020}, the effect on the ratio measurements are $\lesssim$ 0.04 for EWR and $\lesssim$ 0.13 for LDR. A difference of 0.04 in EWR corresponds to an approximate $\Delta$\teff{} of $\sim$15 K at an EWR of 1.5 (equivalent to $\sim$3400 K) and $\sim$72 K at an EWR of 0.3 (equivalent to $\sim$4300 K), according to the solar \teff{}--EWR relationship. Unlike \vsini{} and $R$, the response of different atomic lines to the magnetic field strength is unique to that particular line. Hence, the findings presented here are specific to the spectral region examined in this study. Ultimately, we adopt EWR, not LDR, for deriving the \teff{} relationship as it demonstrates reduced sensitivity to spectral broadening effects.

%%----------------------------------
\subsection{\teff{}--EWR Empirical Relationship}\label{sub:emprical}

Figure~\ref{fig:empirical} shows EWR versus \teff{} for our calibration sources. In Figure~\ref{fig:empirical}a, the sample is color-coded by [Fe/H], and in \ref{fig:empirical}b, it is color-coded by \logg{}. The tight \teff{}--EWR trend at high \teff{} is weakened at temperatures cooler than $\sim$3700 K because of the presence of sub-solar metallicities and higher \logg{} targets (we later show in Section~\ref{sub:tuning_results} and Figure~\ref{fig:grid} that the scatter in low temperature is mainly caused by metallicity differences). At EWR of $\sim$1.5, the \teff{} difference between a solar metallicity target and a sub-solar ([Fe/H]$\sim-0.4$) metallicity target can be as large as 200K. We therefore fit two empirical \teff{}--EWR relationships: one including the entire sample, and the other using only a ``solar'' metallicity ($-0.15<$ [Fe/H] $<0.15$) sub-sample. Both \teff{}--EWR relationships were fit with a natural log function (\teff{} $=a + b\ln{\rm EWR}$) using the orthogonal distance regression (\texttt{Python} package \texttt{scipy.odr}) to consider uncertainties in both \teff{} and EWR. Best-fit coefficients are given in Figure~\ref{fig:empirical}a. These empirical \teff{}--EWR relationships are best characterized for EWR from about 0.1 (\teff{} $\sim 5000$\,K), before it asymptotes, to about 1.5, where we start to lose constraints from the observed samples (\teff{} $\sim 3400$\,K).

\begin{figure*}[tb!]
\centering
\includegraphics[width=1.\textwidth]{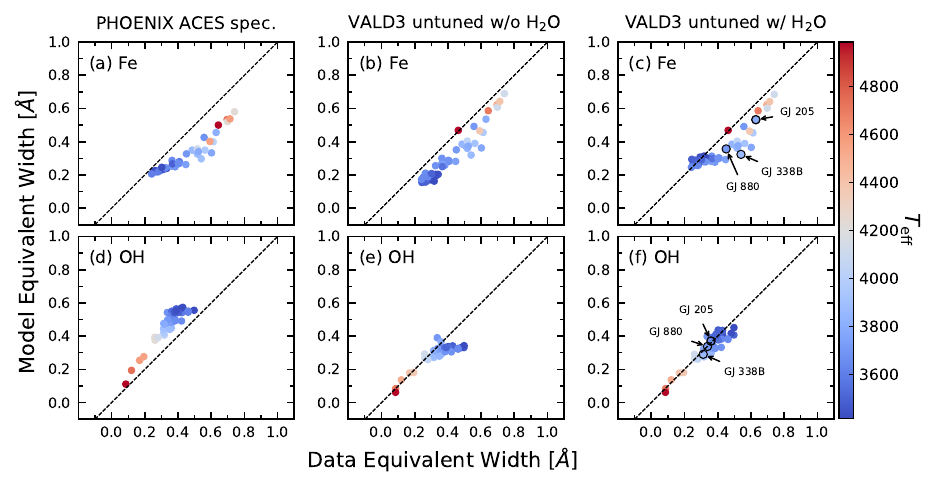}
    \caption{
    Comparison plots of Fe and OH equivalent widths (EW) from model spectra versus IGRINS observational data for the \teff{} calibration targets. The top row (panels a, b, and c) show the Fe EW outcomes, while the bottom row (panels d, e, and f) details the OH EW results. Column one illustrates the EW derived from the PHOENIX ACES spectral grid, column two from \textsc{synmast} using the VALD3 line list without water lines, and column three incorporates the same line list with additional BT2 water lines \citep{barber2006} in \textsc{synmast}. Data points are color-coded according to \teff{} as indicated by the color bar on the right. The dashed line across each panel represents the 1:1 correspondence line, serving as a reference for perfect agreement between modeled and observed EW values.
    }
\label{fig:model_data}
\end{figure*}

%-----------------------------------
\section{Model Comparison} \label{sec:model}

To obtain the most precise \teff{} estimates possible, it would be ideal to customize the \teff{}--EWR relationship for different \logg{}, [Fe/H], and magnetic field strengths using models calibrated with observations.
We compare our observational results against theoretical models to explore the possibility that spectral models might provide a useful extension of our observed sample in this way.

First, we compare our empirically-derived relationship with measurements based off of the publicly available pre-computed high-resolution model spectral library PHOENIX-ACES-AGSS-COND-2011 \citep[][hereafter PHOENIX-ACES spec]{husser2013}\footnote{\url{http://phoenix.astro.physik.uni-goettingen.de}}. We interpolate across the spectral grid to get model spectra for each of our \teff{} calibration stars with the adopted parameters from Table~\ref{tab:cals}. The adoption of the $\alpha$ element (O, Ne, Mg, Si, S, Ar, Ca, and Ti) enhanced spectra was based on the following rule:  
[$\alpha$/Fe] = 0.4 for $-5.0$ $\leq$ [M/H] $\leq-1.0$; 
[$\alpha$/Fe] = $-0.4 \times$ [M/H] for $-1.0 \leq$ [M/H] $\leq$ 0.0; and [$\alpha$/Fe] = 0.0 for [M/H] $\geq$ 0.0 \citep{gustafsson2008}. The same procedures introduced in Section~\ref{sec:spec_analysis} were then applied to get the model EW for PHOENIX-ACES spec. The first column of Figure~\ref{fig:model_data} compares the EW results derived from the PHOENIX-ACES spectra and IGRINS data for Fe (Fig.~\ref{fig:model_data}a) and OH (Fig.~\ref{fig:model_data}d). Deviation from the slope $=$ 1 line is evident, especially for Fe. This mismatch indicates that pre-computed spectral models are not a useful extension of observations when it comes to calibrating a \teff{}--EWR relationship. We therefore explore a number of means to improve on these spectral models, relying on spectral synthesis to generate our own model spectra for which we can specify the spectral line list and make other detailed changes.

We utilize the \textsc{synmast} spectral synthesis code \citep{piskunov1999a,kochukhov2007,kochukhov2010} because it allows us to incorporate the effects of $B$ fields. Given that TTSs can exhibit surface magnetic field strengths several kG higher than those found in M dwarfs, accounting for the influence of $B$ fields on the \teff{}--EWR relationship could be crucial. To this end, we employ a line list from the VALD3 database \citep{ryabchikova2015a}\footnote{\url{http://vald.astro.uu.se}}. Additionally, a comparison between the results of using different model atmosphere profiles, i.e., BT-NextGen \citep{allard2011,allard2012}, PHOENIX-ACES \citep[][PHOENIX-ACES atm.]{husser2013}, and MARCS \citep{gustafsson2008} is shown in Appendix~\ref{sec:atm_comp}. Because the difference is minimal, we opt only to show results using the BT-NextGen atmospheres with ASGG2009 \citep{asplund2009} solar abundance.

\begin{figure*}[tbh!]
\centering
\includegraphics[width=1\textwidth]{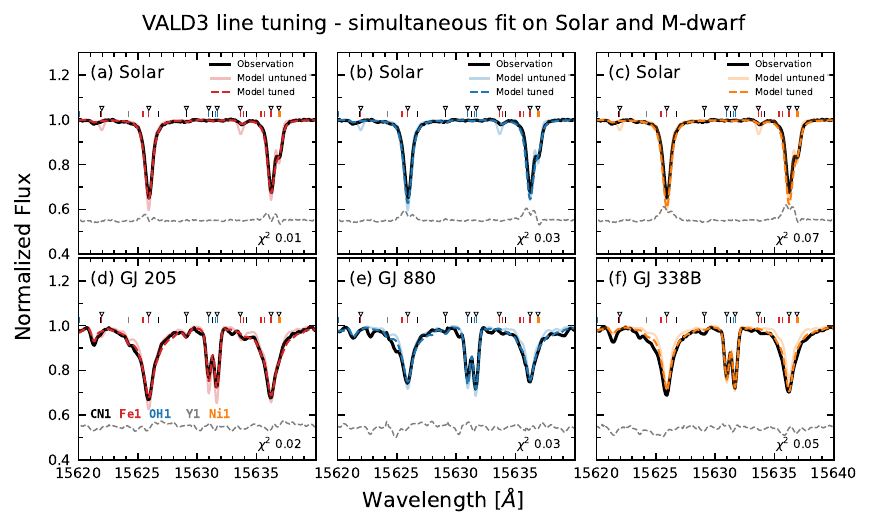}
\caption{
    Spectral tuning results using \textsc{synmast}, incorporating a tuned VALD3 line list, BT2 water lines, and adopted the BT-NextGen atmospheric profile. The tuning exercise involved a simultaneous fitting process applied to solar spectra (top row) and spectra from three M-dwarfs (bottom row): GJ\,205 (first column), GJ\,880 (second column), and GJ\,338B (third column). Each panel displays the observed spectrum as a solid black line with the model spectra in color. The model spectra generated with the un-tuned line list are depicted by faded color lines, and the model spectra generated with tuned line list are in dashed color lines. Residuals between the observations and models with tuned line list are indicated by grey dashed lines, offset 0.55 for visibility. The $\chi^2$ statistic of fit is also shown. Key spectral lines, including CN\,\textsc{i}, Fe\,\textsc{i}, OH\,\textsc{i}, Y\,\textsc{i}, and Ni\,\textsc{i}, are identified and labeled within each panel. The six lines that underwent specific tuning are highlighted by inverted triangles.
    }
\label{fig:tuned_spec}
\end{figure*}

\begin{figure*}[tbh!]
\centering
\includegraphics[width=1\textwidth]{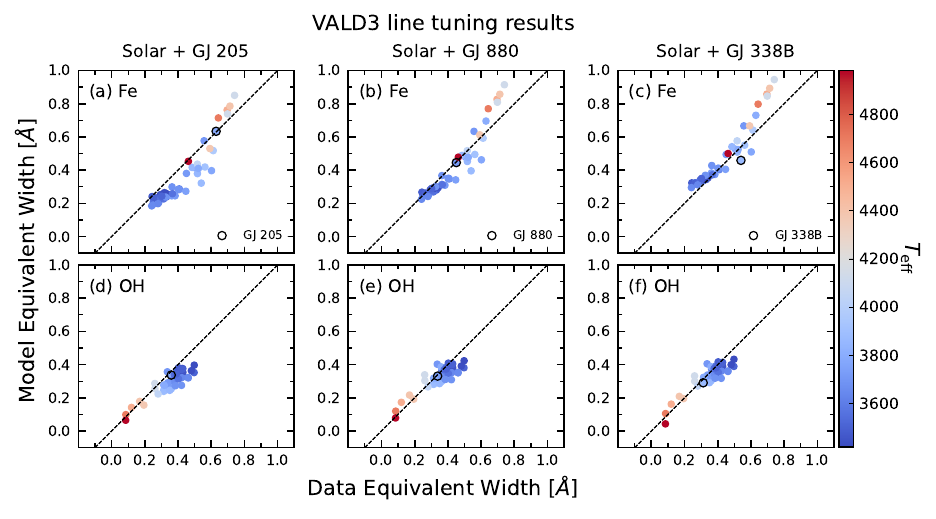}
\caption{
    Similar to Figure~\ref{fig:model_data} but showing \textsc{synmast} model results with VALD3 tuned line list, BT2 water lines, and adopted the BT-NextGen atmosphere profile. The simultaneous fit to solar and the three M-dwarfs are shown in the first column for GJ\,205, the second for GJ\,880, and the third for GJ\,338B. Each of the M-dwarfs used in the tuning process is highlighted in each panel. The top row, panels (a), (b), and (c), show Fe results, and the bottom row, panels (d), (e), and (f) show OH results. 
    }
\label{fig:tuned_EW}
\end{figure*}

The middle column of Figure~\ref{fig:model_data} displays these results versus our data-derived \teff{} relationships for Fe EW, Figure~\ref{fig:model_data}b, and OH EW, Figure~\ref{fig:model_data}e. The \textsc{synmast} spectra produce a better match than the PHOENIX-ACES spec for Fe EWs, displaying a strong correlation with the data-derived results albeit with a slight offset. The \textsc{synmast} spectra also show a marked improvement for OH EWs among stars warmer than $\sim$3800 K. The more divergent results from \textsc{synmast} spectra cooler than $\sim$3800 K may be the result of the lack of water lines in VALD3 at these temperatures, and/or the relative insensitivity of the OH line depth at lower temperatures in \textsc{synmast}.

To further improve on these results, we experimented with adding a water line list to our \textsc{synmast} model generation. We tested two different water ($^1{\rm H}_2$$^{16}{\rm O}$) line lists, the BT2 \citep{barber2006} and the POKAZATEL \citep{polyansky2018a} from the ExoMol database  \citep{tennyson2012}\footnote{\url{https://www.exomol.com}}. The difference between the resulting EWR by adding BT2 versus the POKAZATEL water lines is small, as shown in Appendix~\ref{sec:atm_comp} Figure~\ref{fig:atm_comp}; thus, here we only show results using the BT2 line list. The improvement in the model EWR after adding water lines is significant, as shown in the third column (panels c and f) of Figure~\ref{fig:model_data}. The extra opacity from water lines in cooler stars results in larger Fe and OH EWR values, making them move upward on the plot closer to the slope = 1 line. Nevertheless, a larger deviation from the unity line remains for Fe EWR in Figure~\ref{fig:model_data}c around targets with \teff{}$\sim$ 4000\,K. One reason for this deviation is that spectral lines from atomic and molecular databases are often not precise enough to match with observational data \citep[e.g.,][]{johns-krull2004,flores2019}. Below, we show our results on manually adjusting the log$gf$ and the van der Waals constant for six dominant lines. 

%%----------------------------------
\subsection{VALD3 Line Tuning} \label{sub:line_tuning}

It is well known that spectral line data collected in various databases are often not accurate enough to properly reproduce stellar spectra \citep[e.g.][]{valenti1996a}. Investigators often choose to tune line parameters such as the oscillator strength (in this case log$gf$, where $g$ is the statistical weight and $f$ is the oscillator strength) and the Van der Waals (VdW) damping parameter by fitting the spectrum of the quiet Sun. In our case, however, simply adjusting the log$gf$ values and VdW damping parameters of the two dominant Fe {\sc i} lines based on the NSO disk center solar atlas \citep{wallace1996} results in systematically weakened Fe {\sc i} lines for stars with \teff{} $\leq$ 4000 K. This suggests a degree of degeneracy between the log$gf$ values and VdW damping parameters that cannot be resolved by fitting the solar spectrum alone. A simultaneous fit incorporating a cooler star is necessary to disentangle these effects. We opt to calibrate our line parameters through a joint fit of both solar and cool (\teff{} $\sim$ 3800 K) star spectra, given our focus on stars substantially cooler than the Sun, which exhibit distinct features like OH lines absent in the solar spectrum (Figure~\ref{fig:tuned_spec}). The cool stellar spectra used were observed with IGRINS under similar S/N conditions, ensuring that variations in instrument, resolution, and S/N do not skew the tuning process.

We first created a line list using the VALD3 line database \citep{ryabchikova2015a} by performing two ``Extract Stellar" queries, one at \teff{} = 5775 K and one at \teff{} = 3800 K, and combining the two returned lists. Spectra generated using the original VALD3 line list are shown in Figure~\ref{fig:tuned_spec} as faded color lines. The line tuning process is done by simultaneously fitting the Solar spectrum (here we are using the spectrum of the asteroid Ceres from IGRINS) and a M-dwarf spectrum in the \teff{} calibration sample. When performing this tuning, it is key to use stars with accurately known stellar parameters given that parameters such as \teff{}, gravity, and metallicity affect line strength. While the Solar values are quite reliable, there is more uncertainty in these parameters for M stars. We therefore choose 3 stars from Table 2 with interferometrically determined \teff{} values and with metallicities from \citet{mann2015}, based on relations calibrated to FGK stars where metallicity determinations are more reliable. The log$g$ values are again based on the interferometric radii combined with a mass-luminosity relationship \citep{boyajian2012}. \citet{passegger2022} collected log$g$ values from the literature for all of these stars and they show generally excellent agreement, giving confidence in the values we adopt.  Finally, we also select stars with minimal contamination from water lines since the line data for water has significant uncertainties. Using the spectra of GJ\,205, GJ\,880, and GJ\,338B, we performed the line tuning three times using the Sun and each of these stars in turn. The locations of these M-dwarfs in EW--\teff{} space are highlighted in Figure~\ref{fig:model_data}c and Figure~\ref{fig:model_data}f. We used the parameters in Table~\ref{tab:cals} for the M stars when doing the tuning. For the Solar spectrum, we adopt \teff{} = 5775 K, \logg{} = 4.44, and [Fe/H] = 0.0. 

We performed the line tuning using a non-linear least squares fitting procedure based on the Marquardt method \citep{bevington1992} to simultaneously fit synthesized spectra to the observed spectrum of the Sun (Ceres) and the cooler star of choice. While the observed S/N for each spectrum varies somewhat, they are all above 100. We arbitrarily set the S/N for each observation to 100 in order to equally weight the fitting to each spectrum. The free parameters of the fit are the log$gf$ and VdW broadening terms of lines that showed significant mismatch between the initially synthesized spectra and the observations (Figure~\ref{fig:tuned_spec}). For weaker lines, only the log$gf$ value is fit. Model spectra were calculated with the synthesis code \textsc{synmast} without water lines included, and the BT-NextGen models were interpolated to the specific stellar parameters (\teff{}, \logg{}, [Fe/H]) of each star using the PySME code \citep{wehrhahn2023}.

%%----------------------------------
\subsection{Line tuning Results} \label{sub:tuning_results}

Figure~\ref{fig:tuned_spec} shows the line tuning results on the three M-dwarfs, and Figure~\ref{fig:tuned_EW} shows the resulting models EW compared to the observation for the entire sample. Line tuning using the sun and GJ\,205 seems to give the lowest residual between the model and observed spectra as shown in the first column of Figure~\ref{fig:tuned_spec} (panels a and d). However, this set of tuned lines gives the most significant deviation of Fe EW between the models and the observations of the whole \teff{} calibration target set (Fig.~\ref{fig:tuned_EW}a). GJ\,338B, on the other hand, shows an inverse result. While lines tuned with the sun and GJ\,338B give the largest residual between the model and the observed spectra for these stars (Fig.~\ref{fig:tuned_spec}c and Fig.~\ref{fig:tuned_spec}f), Fe EW comparison between the models and the observations have the tightest relationship for cool stars \teff{} $\lesssim$ 4400\,K (Fig.~\ref{fig:tuned_EW}c). The OH EWs (Fig.~\ref{fig:tuned_EW}f) show a relatively low-scatter correspondence. Nevertheless, this tuning systematically underestimates the EW of the model for the hotter stars. Of the three M-dwarfs, the result for GJ\,880 lies between GJ\,205's and GJ\,338B's results, thus this is the one we choose to use for the further analysis.

Line tuning effectively brings the model EW of selected targets closer to the equality line, as evidenced by the adjustments for GJ\,205 and GJ\,880 in Figure~\ref{fig:tuned_EW}a and \ref{fig:tuned_EW}b, respectively. Nonetheless, this method does not address deviations for the entire sample. Additionally, even targets selected for tuning, such as GJ\,338B, may still deviate from the equality line. This issue arises during simultaneous fitting with the Sun, which is presumed to have accurately determined physical parameters, limiting the enhancement of the VdW damping strength. The spectral fitting results for the Sun, when simultaneously fitting with GJ\,338B, illustrated in Figure~\ref{fig:tuned_EW}c, already shows excessively deep Fe cores, thus preventing the fitting routine from further increasing the strength of the VdW value.

As mentioned above, adopting an accurate set of stellar parameters is essential in the line tuning. However, this is challenging for the cool M-dwarfs, especially with respect to the [Fe/H] values. \citet{passegger2022} studied metallicities of 18 M-dwarfs determined with different techniques, and all three of the M-dwarfs we used in the tuning are included in their sample. \citet{passegger2022} collected literature [Fe/H] values for their entire sample and listed them in their Table~A.1, where they also provided literature median values and associated uncertainties. Although the [Fe/H] values adopted for the three targets in our tuning process are all within the uncertainties in the literature median from \citet{passegger2022}, the range in the literature [Fe/H] values can be as large as 0.69. For example, for GJ 205, the lowest [Fe/H] is $+0.0 \pm 0.09$ \citep{maldonado2015a} and the highest [Fe/H] is $+0.69 \pm 0.10$ \citep{terrien2015}. 

We evaluate our line tuning process on GJ\,205 with three different [Fe/H] values (0.20, 0.39, and 0.49) from the range of estimates in \citet{passegger2022}. Using 0.49 from \citetalias{mann2015} as our initial (default) value yielded the smallest discrepancy between the model and observed spectra when fitting the Sun and GJ\,205 simultaneously (Fig.~\ref{fig:tuned_spec}). Using the other two [Fe/H] values did not produce simultaneous fits as good as that when we assume [Fe/H] = 0.49. Line tuning critically relies on accurate stellar parameters when making adjustments to database-provided line parameters to enhance the observational fit. However, if a target's stellar parameters are poorly defined, biases can arise, limiting line tuning to merely compensating for these uncertainties. As a result, while line tuning may align the EW of a specifically tuned target with the line of equality (Fig.~\ref{fig:tuned_EW}), it cannot rectify overall sample deviations. Our experiments highlight the need for precisely defined physical parameters in calibration targets for \teff{} and improvements in spectral modeling for cooler M-dwarfs.

\begin{figure}[tbh!]
\centering
\includegraphics[width=1.\columnwidth]{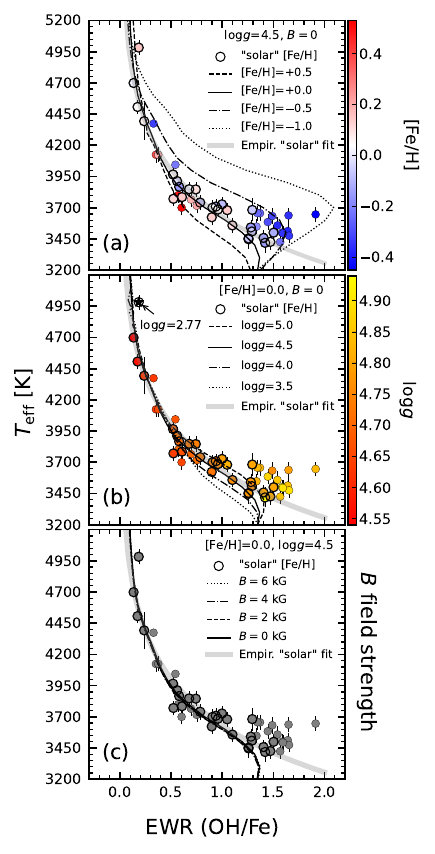}
\caption{
    The effective temperature (\teff{}) versus equivalent width ratio (EWR) for model grids and IGRINS observational data. The model grids were produced using \textsc{synmast} with a VALD3 line list calibrated on solar and GJ\,880 spectra, incorporating BT2 water lines within the BT-NextGen atmospheric profile. Panel (a) shows the effect of metallicity [Fe/H], (b) the influence of surface gravity (\logg{}), and (c) magnetic field strength impact. Models are shown as thin lines, observational data as circles, and a grey line denotes the empirical fit for solar metallicity ($-0.15<$[Fe/H]$<0.15$), detailed in Section~\ref{sub:emprical} and Figure~\ref{fig:empirical}. Panels (a) and (b) assume 0 $B$ field strength, with (a) set at \logg{}=4.5 and (b) using solar metallicity. Panel (c) is also at \logg{}=4.5 and solar metallicity.
    }
\label{fig:grid}
\end{figure}

%%----------------------------------
\subsection{EWR Comparison of Model and Observation} \label{sub:modelVSobs_EWR}

Figure~\ref{fig:grid} is similar to Figure~\ref{fig:empirical} showing \teff{} versus EWR, but now overplotted with model results generated using atomic lines tuned on GJ\,880, the BT2 water list, and BT-NextGen atmospheres. By employing models, we are able to investigate the EW across a continuous range of \teff{}, offering deeper understanding of the \teff{}--EWR relationship and its dependency on parameters such as \logg{}, [Fe/H], and the $B$ field. This approach overcomes the constraints of our limited sample.

In Figure~\ref{fig:grid}a, we see that [Fe/H] has a more significant effect on the OH/Fe EWR toward cooler \teff{} compared to the impact of \logg{} in Figure~\ref{fig:grid}b and $B$ field in Figure~\ref{fig:grid}c. This larger difference is expected as our EWR includes Fe lines. The more intriguing result occurs at sub-solar metallicities below $\sim$3700\,K where the models converge toward a single EWR value. The backtracking in sub-solar metallicity models below $\sim$3700\,K occurs as Fe lines effectively vanish, leaving the models to reflect predominantly water opacities. For \logg{}, the effect is indistinguishable between \logg{} of 3.5 to 5.0 above $\sim$3900\,K. Models only start to be slightly separated below $\sim$3900\,K to $\sim$3200\,K. The small separation between different \logg{} models implies the \teff{} estimated using the \teff{}--EWR relationship is less susceptible to the accuracy of \logg{} values. Lastly, we can see that there is almost no difference in models as a function of B field strength (Fig.~\ref{fig:grid}c).

In Figure~\ref{fig:grid}a and \ref{fig:grid}b we can also see that models with \logg{} = 4.5 and solar metallicity (solid black thin lines) match well with the empirical \teff{}--EWR fit for solar metallicity targets (data points with open black circles, also see Section~\ref{sub:emprical}). Still, below $\sim$3400\,K, the model deviates from the empirical fit, turning almost straight down to trace what would be a non-unique \teff{}--EWR relationship. Taken altogether, these results indicate a robust empirical \teff{}--EWR relationship for cool stars and TTSs in particular, so long as \teff{} is not below 3400\,K, in agreement with our observations in Section~\ref{sub:emprical}.

%%-----------------------------------
\section{Discussion} \label{sec:discussion}
%-----------------------------------
\subsection{Effective Temperature Variability in T Tauri Stars}\label{sec:teffvar}

To study the \teff{} variability of the 13 TTSs in our sample, we first measure their time series EWR using the procedures described in Section~\ref{sub:cal}. We then use the empirical \teff{}--EWR relationship in section~\ref{sub:emprical} to obtain their time series \teff{}. As TTSs mostly have solar metallicities, e.g., \citet{dorazi2011}, we use the ``solar'' empirical \teff{}--EWR relationship. With time series \teff{} measurements, we study the overall \teff{} variability, likely caused by the variation in spot(s) coverage, of the TTSs (nine CTTSs and four WTTSs), then explore the seasonal \teff{} variations of the seven TTSs (three CTTSs and four WTTSs) that were most densely observed.

%%----------------------------------
\subsubsection{Overall Variability} \label{sub:teffvar}

\begin{figure*}[tb!]
\centering
\includegraphics[width=0.95\textwidth]{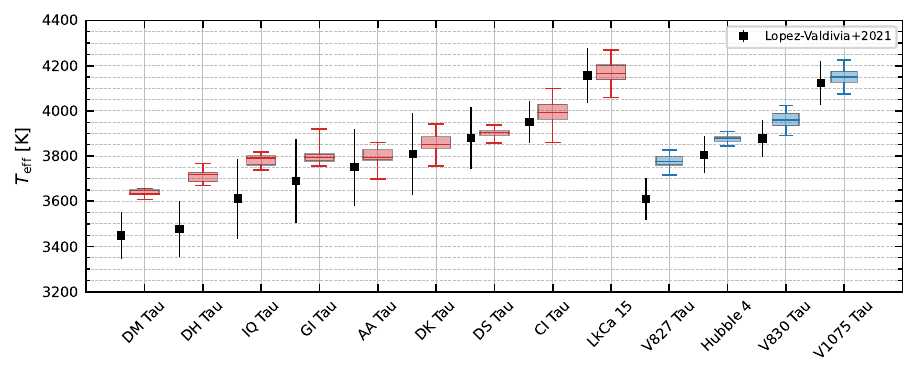}
\caption{
    Box plot showing the \teff{} ``variations'' for the TTS sample, with CTTSs in red and WTTSs in blue. The boxes span the range between the first quartile (Q1, 25th percentile) and the third quartile (Q3, 75th percentile); the median is indicated with a dark colored horizontal line. The whiskers (upper and lower bars) show the maximum and minimum values. Black squares show \teff{} values estimated by \citet{lopez-valdivia2021} using MCMC fitting to IGRINS spectra, and the errorbars show the ``uncertainties'' in the measurements.}
\label{fig:teffvar}
\end{figure*}

The box plots in Figure~\ref{fig:teffvar} display the \teff{} distribution of each of our TTS targets, with the extent of temperature variability illustrated by the size of the boxes, from the first quartile (Q1, 25th percentile) to the third quartile (Q3, 75th percentile). The two whiskers (vertical bars) correspond to the minimum and maximum values of the \teff{}, such that the range the whiskers span gives the total amplitude of the \teff{} variation. Targets exhibiting significant \teff{} variability (amplitude $\gtrsim$150~K), include V1075\,Tau ($\sim$150~K), DK\,Tau ($\sim$180~K), CI\,Tau ($\sim$240~K), and LkCa\,15 ($\sim$210~K). These variations are all more than 3.5 times the measured \teff{} sensitivity ($\Delta$\teff{}) of each target listed in Table~\ref{tab:tts_precision}. Targets with smaller \teff{} amplitudes ($\lesssim$60~K) include Hubble\,4 ($\sim$65~K), DS\,Tau ($\sim$80~K), and DM\,Tau ($\sim$47~K). For these sources, the \teff{} variation is at least 2.5 times greater than the respective $\Delta$\teff{}. 

% =========== TABLE % TABLE =========== %
\begin{deluxetable}{c CCCCC}
\tablecaption{TTS EWR Temperature Sensitivity  \label{tab:tts_precision}
             }
\tabletypesize{\scriptsize}
\tablehead{
    \colhead{Name}  & \colhead{avg. EWR} & \colhead{\teff{}} &
    \colhead{avg. $\sigma$EWR}  & \colhead{$\Delta$\teff{}}  \vspace{-.1cm} \\ 
	\colhead{(1)} & \colhead{(2)} & \colhead{(3)} & 
	\colhead{(4)} & \colhead{(5)} 
     }
\startdata 
AA\,Tau & 0.728 & 3804 & 0.020 & 30 \\
CI\,Tau & 0.512 & 3994 & 0.016 & 34 \\
DH\,Tau & 0.859 & 3714 & 0.018 & 23 \\
DK\,Tau & 0.659 & 3858 & 0.019 & 31 \\
DM\,Tau & 0.988 & 3637 & 0.016 & 18 \\
DS\,Tau & 0.604 & 3903 & 0.017 & 30 \\
GI\,Tau & 0.732 & 3801 & 0.019 & 28 \\
Hubble\,4 & 0.637 & 3875 & 0.014 & 24 \\
IQ\,Tau & 0.755 & 3783 & 0.018 & 26 \\
LkCa\,15 & 0.370 & 4170 & 0.016 & 47 \\
V1075\,Tau & 0.383 & 4151 & 0.014 & 40 \\
V827\,Tau & 0.763 & 3777 & 0.014 & 20 \\
V830\,Tau & 0.544 & 3960 & 0.014 & 28 \\
\enddata
\tablecomments{
    Column 2 displays each TTS's average EWR from all observations, while column 3 lists their corresponding \teff{} (K), estimated via the solar \teff{}--EWR relationship shown in Figure~\ref{fig:empirical}. Column 4 lists the average uncertainties in EWR. By applying the average EWR from column 2 and its uncertainties from column 4 to the \teff{}--EWR relationship, we calculate the systematic scatter in \teff{}, the $\Delta$\teff{} (K), presented in column 5. This scatter is interpreted as the sensitivity of the \teff{} measurement for each TTS.}
\end{deluxetable} 

No apparent trend links \teff{} variability to whether a star is classified as a CTTS or a WTTS. This may suggest similar spot distributions, sizes, and spot-to-photosphere temperature contrasts between both classes of stars. Moreover, as shown in Figure~\ref{fig:teffvar}, where targets are ordered by their increasing median \teff{} for CTTSs and WTTSs separately, a positive correlation between \teff{} and the activity level, i.e., the \teff{} variation amplitude, is observed. To quantify this result, we calculate the Pearson correlation coefficient and Spearman’s rank correlation coefficient. We use the inter-quartile range (IQR, Q3$-$Q1) instead of the \teff{} variation amplitude to mitigate the effects of outliers. Both tests yield a correlation of 0.6 with p-values of 0.02, indicating a significant linear relationship. This result might suggest higher surface activity in TTSs with higher average \teff{}. However, further observations and model testing are needed.

\citet{lopez-valdivia2021} analyzed the same TTSs as shown in Figure~\ref{fig:teffvar}, using the identical IGRINS dataset. However, they employed a forward modeling approach to determine \teff{} for each TTS, averaging between 1 and 10 observational epochs per target, with uncertainties assessed via MCMC fitting. This sets the stage for an insightful comparison between \teff{} ``uncertainty'' derived from forward modeling and the ``variability'' we identify through EWR measurements. In Figure~\ref{fig:teffvar}, we show \citet{lopez-valdivia2021}'s model fitting results (represented as black squares with error bars) against our findings. Notably, the \teff{} ``uncertainty'' reported by \citet{lopez-valdivia2021} often matches or surpasses the ``variability'' detected in our study. Additionally, \citet{lopez-valdivia2021}'s average \teff{} values are consistently lower than our median values derived through the EWR method. This systematic lower \teff{} is underscored by a significant negative correlation, indicated by a Pearson coefficient of $r=-0.75$ and a p-value of 0.003, suggesting a larger discrepancy at lower \teff{} values. This trend could be explained by the fact that \citet{lopez-valdivia2021}'s \teff{} estimates are based off of measurements in the $K$ band instead of the $H$ band. At these longer wavelengths, the \teff{} estimates would be more sensitive to cool stellar spots \citep{gully-santiago2017}: as the temperature of the photosphere decreases, the contrast between the spot and the photosphere increases more in the $K$ band than in the $H$ band. This would produce systematically lower temperature results, potentially explaining the  negative correlation observed.

\begin{figure*}[tbh!]
\centering
\includegraphics[width=0.95\textwidth]{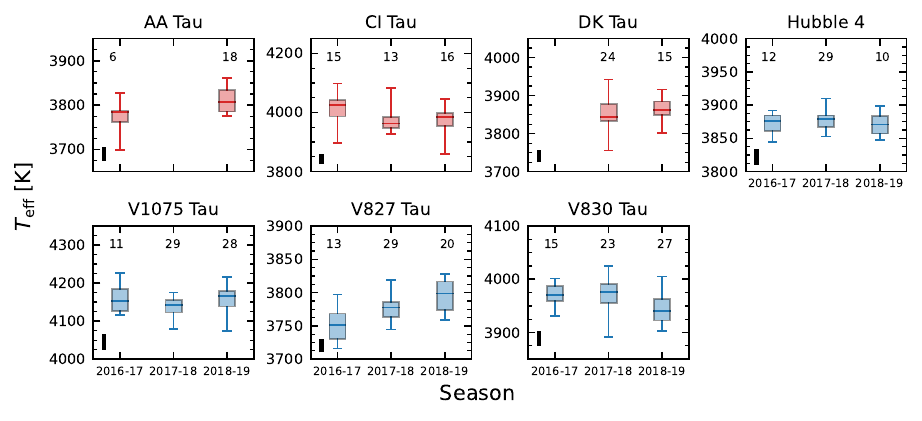}
\caption{
    Box plots showing the seasonal \teff{} ``variation'' of seven TTSs which have at least six observations in two or more seasons. Three observing seasons from 2016--2019 are shown. Red boxes are CTTSs, and blue boxes are WTTSs. The black bar in the lower left of each panel shows the \teff{} sensitivity from Table~\ref{tab:tts_precision}, column (5). The smaller the black bar's length compared to the variability, the more robustly the changes are characterized. The numbers at the top of each box show the total observations for each season.} 
\label{fig:SeasonalTeffvar}
\end{figure*}

%%----------------------------------
\subsubsection{Seasonal Variability} \label{sub:season}

In this section, we delve deeper into the seasonal variability of seven TTSs that feature two or more seasons with at least six observations each. Figure~\ref{fig:SeasonalTeffvar} illustrates the \teff{} distributions of these TTSs across three observation seasons (2016--2019). By focusing on the median values and the size of the IQR to minimize the impact of outliers, seasonal \teff{} variations become apparent for targets like AA\,Tau, CI\,Tau, V827\,Tau, and V830\,Tau. Interpreting these fluctuations is challenging without model simulations, as both spot size and temperature contrast with the stellar photosphere likely influence the apparent \teff{}. However, assuming minor temperature contrast changes between seasons, an increase in overall \teff{}, such as observed for V827\,Tau, may suggest a reduction in spot sizes. Additionally, the IQR's ``size'' serves as an indicator of changes in a star's activity level across seasons. An increase in either the number or size of spots, along with greater temperature contrast with the photosphere, suggests heightened stellar activity, reflected by a larger IQR. Consequently, an expanding IQR, as seen with AA\,Tau, V830\,Tau, and the last two seasons in V827\,Tau, could signal rising stellar activity.

\begin{figure*}[tbh!]
\centering
\includegraphics[width=0.87\textwidth]{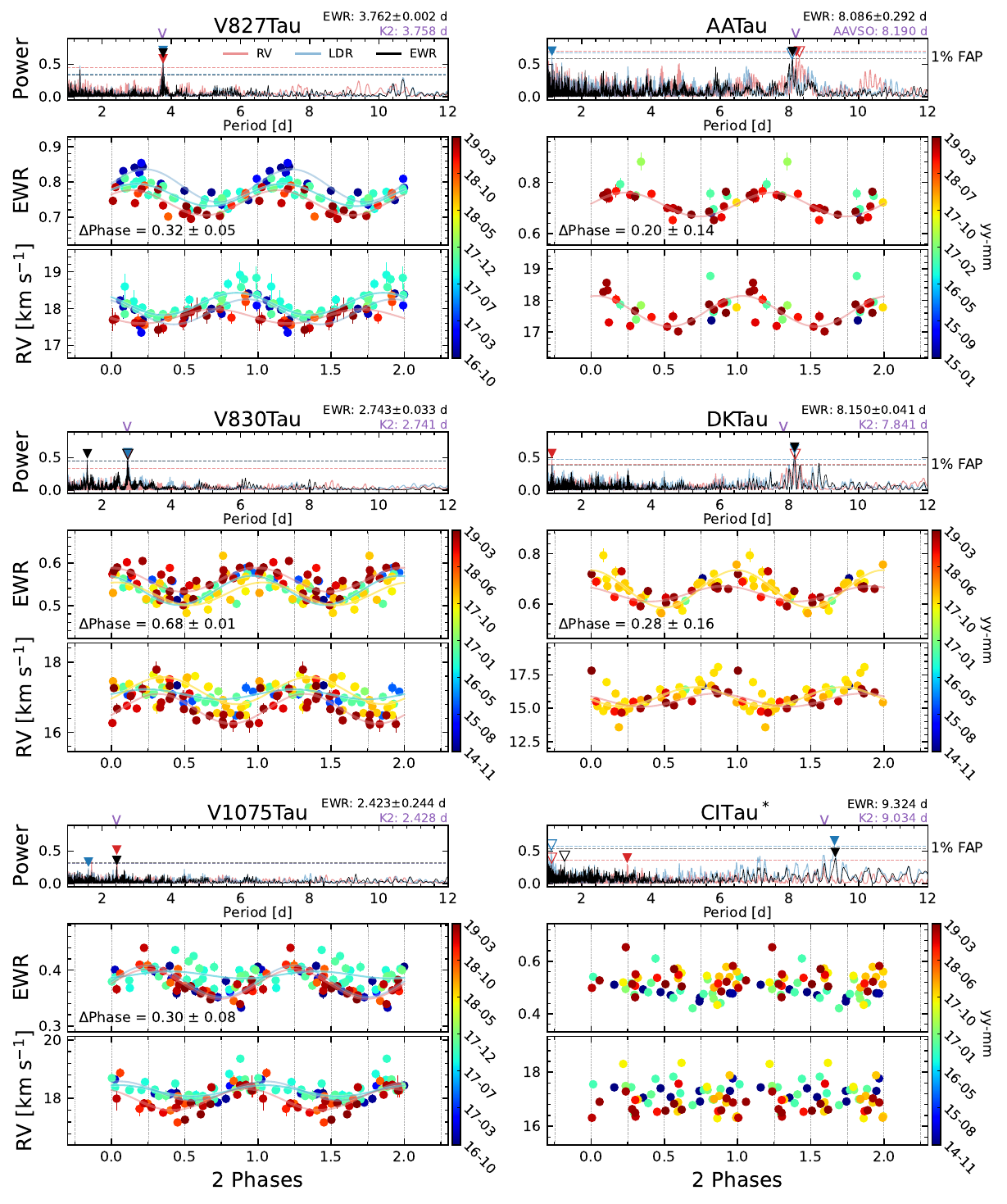}
\caption{
    Periodograms, as well as EWRs and RVs folded at the rotation period, for our TTS targets with at least six observations in two or more seasons (Hubble\,4 is not shown given its complex nature, see Section~\ref{sub:season}). Each data point is plotted twice as we are showing two phases. In the left column are the WTTSs (V827\,Tau, V830\,Tau, and V1075\,Tau) and in the right are the CTTSs (AA\,Tau, DK\,Tau, and CI\,Tau). Periodograms are calculated from RVs (red), LDRs (blue), and EWRs (black). The strongest periodogram peak for all three datasets is marked with a solid triangle, and the second strongest peak is marked with an open triangle if significant. The adopted rotation period from the EWR data is displayed on the upper right of each panel. For comparison, rotation periods derived from lightcurve analysis, e.g., K2 \citep{rebull2020} and AAVSO \citep{watson2006}, are printed as text and also indicated with purple carets above the periodogram. The 1\% false alarm probabilities (FAP) estimated using the bootstrap method are indicated with horizontal dotted lines. For each target, the middle and bottom panels correspond to the folded EWR and RV curves (to the EWR period), respectively; data points are color-coded by date (scale bars on right). The $\Delta$ phase values indicate the phase delay between the EWR and RV curves, as detailed in Section~\ref{sub:delay}. The sine waves used to calculate $\Delta$phases are shown in lines with matching color to the adopted season (Section~\ref{sub:delay}). $^*$ Final results for CI\,Tau are presented in Figure~\ref{fig:citau}.
    }
\label{fig:Pdiagram}
\end{figure*}

In addition to illuminating the overall \teff{} variability amplitudes, examining trends in EWR when folded to the target's \prot{} can provide a more insightful view as now we can examined the shape of the EWR curve. Figure~\ref{fig:Pdiagram} displays, for each target, a Lomb-Scargle periodogram \citep[LSP,][]{lomb1976,scargle1982,vanderplas2018} analysis of RVs (red), LDRs (blue), and EWRs (black) in an upper panel. The lower panels show the EWRs and RVs phase folded at \prot{}. We leave the discussion of the periodograms to the next section (Section~\ref{sec:est_prot}), and only focus on the shapes of the folded EWR and RV curves here. 

Figure 14 shows that a strong periodic signal is present for all targets \footnote{Hubble\,4 is excluded from Figure~\ref{fig:Pdiagram} because of the complexities arising from its binary system nature \citep{rizzuto2020a}. Despite detecting a periodic signal when phase-folding the LDR and EWR to the reported \prot{} of 1.5459 days \citep{carvalho2021}, a significant phase shift was observed, consistent with findings from \citet{carvalho2021}'s optical RV data. Additionally, our observations were conducted during the periastron phase of the binary orbit. Without well-defined orbital parameters, detrending the long-term RV variations attributable to the binary system is challenging. V827\,Tau is also reported as a binary system, but we include it in our analysis because only one prominent signal is present in its EWR and RV curves, which we expect to be solely from the primary star. This expectation is supported by a primary-to-secondary flux ratio of about 1.7 in the $H$ band, indicating that we mainly observe the primary star \citep{kraus2011}.}. Moreover, color-coding each observation by time often reveals changes in the EWR and RV variations between seasons, i.e., on an annual time scale. For example, the EWRs of V827\,Tau peak more sharply during the winter of 2016 (blue), around phase 1.25, than in the winters of 2017 (cyan) and 2018 (red). Its RV data, meanwhile, appears to trace a higher amplitude trend during the 2017 (cyan) season than in either 2016 (blue) or 2018 (red). The characteristics of the EWR and RV curves could offer insights into the size and configurations of stellar spots through model simulations. However, accurately interpreting these features is challenging without such modeling. It is plausible that spots appearing as patches could cause significant scatter in the EWR curve, whereas a larger, more uniform spot might result in a more concentrated EWR curve, as observed in V1075\,Tau. Specifically, V1075\,Tau exhibited a wide scatter in EWRs during the winter of 2017 (depicted in cyan), in contrast to a more defined pattern in the winter of 2018 (shown in red). Continued investigation with simulations will be crucial for a deeper understanding of these dynamics but is beyond the scope of this paper. 

The potential of using the EWR and RV curves to study spot lifetimes is promising, as significant changes in the size of the IQR are observed from season to season in our TTS sample. Additionally, the shape of the EWR and RV folded curves varies dramatically from season to season in several of our TTSs. By observing over a longer period, we should gain a clearer understanding of the lifetimes of spots on these young active stars.

%%----------------------------------
\subsection{A New Way to Estimate Stellar Rotation Period?}\label{sec:est_prot}

In Figure~\ref{fig:Pdiagram}, for the WTTSs, all datasets (EWRs, LDRs, and RVs) display the strongest periodicity at the same period determined by lightcurve analysis to within about 0.02 d. This suggests that, even with our limited samples, power spectrum analysis of EWRs and LDRs have similar sensitivity to rotation periods as photometry for WTTSs \citep[e.g.,][]{rebull2020}. We show the adopted \prot{} from the EWR data on the upper right of each panel. The quoted period uncertainties are calculated using MC sampling of the data, with a simulated scatter equivalent to the standard deviation of the data around a sinusoidal fit. This effectively takes into account the influence of astrophysical variability on the rotation period determination process. Moreover, to avoid signals that are too different from the adopted \prot{}, we only search for periods within $\pm$1 day of the adopted \prot{} for each sample in the MC process.

The periodograms for the CTTSs are much noisier than those of the WTTSs. Greater discrepancies between the periodicities can be seen in EWR, LDR, RV, and the \prot{} determined from photometric studies. Strong peaks around 1 day are discarded as they often resemble the nightly observation cadence. To fortify the results on CTTSs, we perform another period-searching algorithm that is designed to process non-sinusoidal and irregularly spaced time-series data, the phase dispersion minimization (PDM) technique \citep{stellingwerf1978}, on the EWR data. Using the same MC method to estimate the uncertainties, we find the following \prot{} values: for AA\,Tau, the LSP method yields 8.086 $\pm$ 0.245 d, while the PDM method yields 8.085 $\pm$ 0.292 d; for DK\,Tau, the LSP method gives 8.150 $\pm$ 0.025 d, and the PDM method gives 8.151 $\pm$ 0.041 d; for CI\,Tau, the LSP method results in 9.324 d, and the PDM method results in 9.017 d. For CI\,Tau, the strongest signal (9.324 d) from the LSP does not reveal any periodicity as shown in Figure~\ref{fig:Pdiagram}. Thus, the uncertainties of its EWR \prot{} are not reported.

For both AA\,Tau and DK\,Tau, the \prot{} values determined from the LSP and PDM methods are nearly identical, with only a 0.001 d difference. However, both methods provide slightly different uncertainties. We choose to adopt the higher values to be conservative.

In Figure~\ref{fig:Pdiagram}, DK\,Tau's EWR, LDR, and RV data are all periodic at $\sim$8.15 $\pm$ 0.15 d, about 0.31 days longer than the period from the K2 data analysis \citep{rebull2020}. However, our EWR \prot{} of 8.15 $\pm$ 0.041 days agrees well with AAVSO's result of \prot{} $\sim$ 8.18 days \citep{watson2006}. For AA\,Tau, the EWRs are periodic at $\sim$ 8.086 d, the LDRs at $\sim$ 8.077 d, and the RVs at $\sim$ 8.288 d. These periods are, on average, $\sim$0.1 days different from the \prot{} determined from AAVSO photometric data \citep[8.19 d,][]{watson2006}, which is within the uncertainty of EWR's \prot{}, 0.292 d.

\begin{figure}[tb!]
\centering
\includegraphics[width=1.\columnwidth]{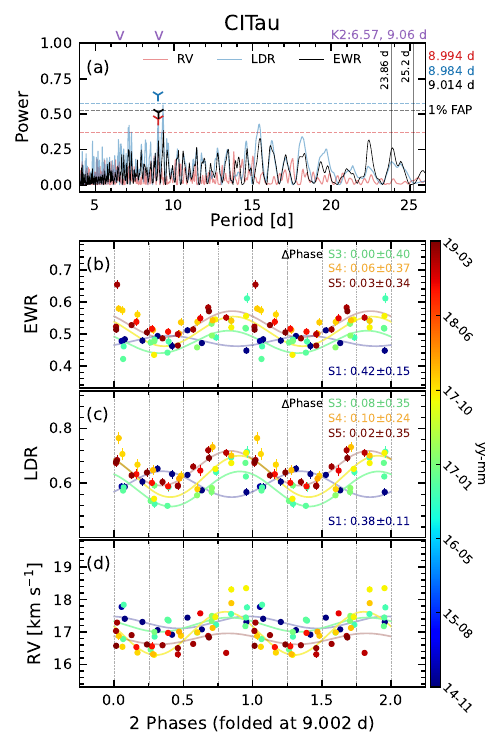}
\caption{
    A re-analysis of CI\,Tau structured similarly to Figure~\ref{fig:Pdiagram}. The \textsc{y} symbols in  the periodogram (panel a) identify strong peaks around 9.3 d that produce more coherent signals in the folded EWR, LDR, and RV data. The color-coded periods of the selected peaks are given just outside the upper right of panel (a). The two strong periods identified by \citet{biddle2018} using K2 lightcurve analysis are shown in purple carets above the periodogram. Panels (b), (c), and (d) show EWR, LDR, and RV data folded to 9.002 d and color-coded by date (scale bar on right). A coherent sinewave-like modulation is clearer for CI\,Tau when folded to 9.002 d, as opposed to the 9.324 d period used in Figure~\ref{fig:Pdiagram}. The magnitude of the phase delay ($\Delta$Phase) between the RV and EWR curves for four seasons (S1: 2014--2015, S3: 2016--2017, S4: 2017--2018, and S5: 2018--2019) are shown in panels (b) and (c) and the fitted sine curves for each season are plotted, color-coded to match the $\Delta$Phase text. 
    }
\label{fig:citau}
\end{figure}

In contrast to the LSP of other targets shown in Figure~\ref{fig:Pdiagram}, the periodogram of CI\,Tau shows less agreement between the signals dominating the RV, LDR, and EWR data. The strong periodicity of LDR and EWR near 9.324 $\pm$ 0.03 d is about 0.3 d different then the period determined by K2 data analysis $\sim$9.034 d \citep{rebull2020}. Folding the EWR and RV data at 9.324 d in the bottom two panels does not show any coherent signal. In search for an alternative signal, we pick next strongest peaks around 9.3 days for EWR, LDR, and RV, marked with a \textsc{y} symbol in Figure~\ref{fig:citau}a, and list their corresponding periods in the upper right corner of the plot (RV at 8.994 days, LDR at 8.984 days, and EWR at 9.014 days). Interestingly, these periods are all very close to the \prot{} determined from the PDM above, 9.017 d. Figures~\ref{fig:citau}b, \ref{fig:citau}c, and \ref{fig:citau}d illustrate the EWR, LDR, and RV data, respectively, folded to the average period of these four signals, 9.002 days. This new period caused the folded curves in Figures~\ref{fig:citau}b--d to exhibit a more pronounced sine wave-like pattern, indicating a more uniform trend across different seasons compared to CI\,Tau's results in Figure~\ref{fig:Pdiagram}. The reason that the best period that produces a coherent signal in all the data does not appear as the strongest peak in the LSP may be attributed to the scatter observed around the best-fit sine curve for each season in Figures~\ref{fig:citau}b--d. Additionally, the mean values of EWR, LDR, and RV are shifted from season to season. Together, these factors contribute to the LSP's decreased sensitivity in accurately detecting the correct period, and indicate the preferability of the PDM approach under these conditions.

%%----------------------------------
\subsection{Phase Delay Between EWR and RV Curves} \label{sub:delay}

If RV modulation originates from spot(s), we would expect to see a phase delay between the RV and the LDR/EWR curve. This is because the RV extrema occur when spot coverage on the stellar surface is at the most asymmetrical phases (when spot(s) are closest to each of the limbs). As for the LDR and EWR, in the case of OH/Fe, the signal's maximum occurs during the greater cool spot(s) coverage phase, and the minimum occurs during the lower cool spot(s) coverage phase. Here, we define a positive delay when the RV curve leads the EWR curve. Depending on the exact spot configuration, the phase delay between the RV and LDR/EWR curves can vary, but is expected to be a quarter-phase for a single dominant cool spot. As long as spots dominate the RV signal, the phase delay should not be near zero.

We fit a sine function to the EWR and RV curves to measure this phase offset (Fig.~\ref{fig:Pdiagram}). To obtain the best fitting results for each target, we only fit data from the same observation season for which we have dense sampling. For WTTSs, we fit data from 2016--2019; for DK\,Tau, we fit from 2017--2019, and AA\,Tau we only fit for the last season of data, 2018--2019. The phase differences between the EWR and RV curves are calculated with uncertainties estimated via the MC method. The weighted results among seasons are given in Figure~\ref{fig:Pdiagram} as $\Delta$Phase. The $\Delta$Phase, excluding V830\,Tau, ranges from 0.20 to 0.32 with a typical uncertainty of 0.1. The average of these targets' $\Delta$Phase is $\sim$0.28, close to a quarter-phase delay. Although V830\,Tau also shows a significant phase delay between the RV and the EWR curves, its $\Delta$Phase, about 0.68, is almost half a phase larger than the targets mentioned before. One possibility for this result is for V830\,Tau to be dominated by cool regions and have one significant relatively hot spot. Such a configuration would produce RV and EWR curves with the phase delay as defined here equal to $\sim 0.75$. Future work on modeling can help confirm this hypothesis and can also help explore more potential scenarios.

% to have a cool spot(s) that covers almost the entire photosphere and a few small hot spot(s) at 1/4 phase behind the center line of the cool spot. 

\subsection{Possible Commensurability of \prot:\porb{} = 1:1 for CI Tau} \label{sub:citau}

CI\,Tau's $\Delta$Phase significantly deviates from the expected quarter-phase delay, showing almost no delay during Season 3 (S3, 2016--2017) and Season 5 (S5, 2018--2019), as shown in Figure~\ref{fig:citau}. Season 4 (S4, 2017--2018) exhibits a marginally larger $\Delta$Phase in LDR, approximately 0.1, yet it still lacks significance relative to a quarter-phase delay. Notably, for both LDR and EWR, Season 1 (S1, 2014--2015) has the greatest $\Delta$Phase, around 0.4. Although the modulation of EWR's S1 data might not be convincing because of the small variation magnitude, the LDR data surprisingly follows the best-fit sine curve. This result in S1 may be attributed to the significantly lower number of data points, only half as many as in S3 and S5, but if it is real, this change in the value of $\Delta$Phase can indicate a migrating of spot(s). 

The large uncertainties in the $\Delta$Phase value mainly reflect the RV curves' small amplitude. With each season's data points tightly following the best-fit sine curve for both the EWR and LDR data, we consider the near-zero phase delay in CI\,Tau to be genuine. The weighted average $\Delta$Phase is 0.29 $\pm$ 0.07, and by excluding S1, we get 0.06 $\pm$ 0.13. Therefore, CI\,Tau stands out from other targets shown in Figure~\ref{fig:Pdiagram} by demonstrating a near zero, non-quarter phase delay.

% This marginally larger $\Delta$Phase could merely reflect the curve's non-sinusoidal nature. 

% EWR and LDR avg results
% S1 Avg delay: 0.36 +- 0.10
% S3 Avg delay: 0.10 +- 0.08
% S4 Avg delay: 0.10 +- 0.08
% S5 Avg delay: 0.19 +- 0.15

This lack of a quarter-phase delay between the EWR and the RV curve might indicate that CI\,Tau's RV signal is not solely caused by spot(s). A candidate hot Jupiter (HJ) companion to CI\,Tau with an \porb{} of $\sim$9 d, was identified by the YESS team based on optical and near-IR RVs \citep{johns-krull2016}. The $\sim$11.6 $M_{\rm Jup}$ HJ candidate was later confirmed by \citet{flagg2019} through a direct detection of CO in the planet's atmosphere, modulated by a period of $\sim$8.9891 d. Moreover, this 9 d signal was also present in the K2 lightcurve, and \citet{biddle2018} associates this signal with the planet-disk interaction. However, the optical spectropolarimetric analysis of \citet{donati2020} led the authors to suggest that the 9 d RV modulation was instead associated with stellar activity. 

A recent study by the GRAVITY team \citep{gravitycollaboration2023} using the Very Large Telescope Interferometer (VLTI) provided an updated look at the CI\,Tau system on inner sub-au scales. They concluded that a massive inner planet orbiting at $\sim$0.08 au can explain the extension of the inner disk edge compared to the dust sublimation radius. This orbital radius of 0.08 au is approximately the stellar corotation radius of CI\,Tau, when estimated with a stellar rotation period of 9 d and a stellar mass of 0.9 $M_{\odot}$ \citep[][]{donati2020}. Additionally, \citet{kozdon2023}'s spectroscopic analysis of data taken at the NASA IRTF revealed a nine-day period in the accretion variability measured with the hydrogen Pf$\beta$ line, hinting at companion-driven accretion dynamics. The discrepancy between the inner disk's inclination ($\sim$70 degrees) and the outer disk's ($\sim$49 degrees) further supports the existence of a close-in massive planet \citep{gravitycollaboration2023, clarke2018}. As the measurement of EWR is not sensitive
to accretion, our independent determination of CI\,Tau's rotation period at approximately 9.002 days using EWR (Section~\ref{sec:est_prot}),  suggests that CI\,Tau and CI\,Tau b might be in a \prot:\porb{} = 1:1 commensurability, a result of the in situ planet formation and/or stalling at the corotation radius during disk migration \citep{dawson2018}.

\citet{manick2024} recently identified a potential planet of 3.6 $\pm$ 0.3 Jupiter masses orbiting CI\,Tau with a \porb{} of about 25.2 d, a finding consistently observed across all their photometric and spectroscopic data. Concurrently, \citet{donati2024} observed a similar periodic signal, but determined it to be 23.86 d, noting it was distinct in CO lines but absent in atomic lines. Consequently, \citet{donati2024} suggested this signal likely corresponds to a non-axisymmetric structure within CI\,Tau's inner disk, rather than a planetary signal. Our periodogram analysis of CI\,Tau, shown in Figure~\ref{fig:citau}a, aligns with these findings; while we do not detect a signal around 25.2 d, we do identify a peak around 23.86 d in the LDR and EWR data, although the false alarm probability is 76\%. Interestingly, \citet{biddle2021} had previously noted a signal around 24 d, specifically at 24.44 d, suggesting an earlier observation of this phenomenon. The absence of an RV signal at 23.86 d in our data merits further investigation, especially considering that \texttt{IGRINS RV} measurements heavily depend on CO lines in the K band. However, exploring this result falls outside of the scope of this paper.

%%-----------------------------------
\subsection{Dependence of the Effective Temperature Variation} \label{sub:dep_teff_par}

\begin{figure}[tb!]
\centering
\includegraphics[width=1.0\columnwidth]{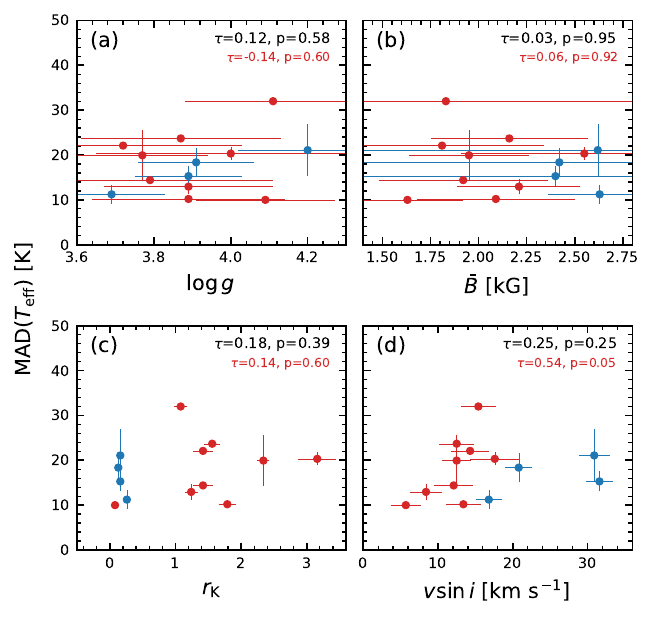}
\caption{
    The median absolute deviation (MAD) of \teff{} for the TTS sample as a function of (a) \logg{}, (b) average surface magnetic field strength ($\bar B$), (c) $K$ band veiling (r$_{\rm K}$), and (d) \vsini{}. 
    Red corresponds to CTTSs, blue to WTTSs.
    The Kendall rank correlation coefficient ($\tau$) and associated p-values 
    % (a hypothesis test giving the probability of a random chance we see a correlation as strong as $\tau$ calculated) 
    are given as determined for all objects (black) and for CTTSs alone (red). Measurement uncertainties are shown for \logg{}, $\bar B$, r$_{\rm K}$, and \vsini{}. The y-axis errorbars show the standard deviations of the \teff{} variation amplitudes from different seasons, so not all targets have y-axis errorbars.
    }
\label{fig:mad}
\end{figure}

To explore whether the magnitude of the \teff{} variations correlates with parameters such as \logg{}, average surface magnetic field strength ($\bar B$), $K$ band veiling (r$_{\rm K}$), and \vsini{}, we plot each target's median absolute deviation (MAD) of their EWR-derived \teff{} against these four properties (Fig.~\ref{fig:mad}). Values of \logg{}, $\bar B$, r$_{\rm K}$, and \vsini{} are from \citet{lopez-valdivia2021} (Table~\ref{tab:tts_basic}). The MAD of the \teff{} is meant to represent the level of the average stellar activity. Therefore, for targets with multi-season observation, the \teff{} MAD shown is the average of each season's values. 

We use a Kendall rank correlation coefficient to test the dependence between parameters, as we have no assumptions about their underlying relationships. We calculate the $\tau$ and p-values for the full sample (13 TTSs) as well as for just the CTTSs. For a sample size of 13 (full sample), the critical $\tau_c$ is 0.359 at $\alpha=0.05$, and for a sample size of 9 (CTTSs only) $\tau_c$ is 0.5 at $\alpha=0.05$. As shown in Figure~\ref{fig:mad}, only \vsini{} for the CTTS sample has a $\tau$ value, 0.54, larger than the critical value of 0.5, i.e., that shows a significant correlation at the 95\% confidence level; this is also evident from the corresponding p-value of 0.05. However, given that \vsini{} is a function of both the stellar inclination angle and the stellar rotation velocity, interpreting this correlation becomes challenging. This complexity further motivates our investigation into the correlation between the MAD of \teff{} and the inclination angle, as detailed below.

\begin{figure}[tb!]
\centering
\includegraphics[width=1.0\columnwidth]{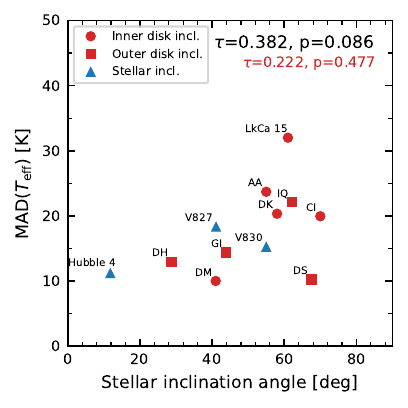}
\caption{
    The median absolute deviation (MAD) of \teff{} for all our TTS samples as a function of stellar inclination angle ($i$). For CTTSs (red symbols), the $i$ values are estimated either from the outer circumstellar disk's inclination angle (circles) or the inner disk's inclination angle (squares). Values for $i$ are estimated from \vsini{}, stellar radius, and stellar rotation period (triangles, see Section~\ref{sub:dep_teff_par}). The Kendall rank correlation coefficient ($\tau$) and the p-values are given for the full sample (black), and only for the CTTSs alone (red).
    }
\label{fig:incl}
\end{figure}

Figure~\ref{fig:incl} shows EWR \teff{}'s MAD against the stellar inclination angle ($i$). For WTTSs, we obtain $i$ from the literature, if possible, such as for V830\,Tau from \citet{donati2015} and for Hubble\,4 from \citet{carvalho2021}. Both studies estimate the $i$ from \vsini{} using estimates of stellar radius ($R_\ast$) and measurements of stellar rotation period ($P_{\rm rot}$). For V827\,Tau, we calculated an $i \sim$ 41.1 deg from \vsini{} = 20.8 \kms{} (Table~\ref{tab:tts_basic}), $P_{\rm rot} = 3.758$ d \citep{rebull2020}, and $R_\ast = 2.35$ $R_\odot$ \citep{gangi2022}. We leave V1075\,Tau out of Figure~\ref{fig:incl} as we cannot find a dedicated study which determines a reliable $R_\ast$ for this star. 

For CTTSs, we assume that the circumstellar disk is perpendicular to the stellar rotation axis and adopt the disk inclination angle as the stellar inclination angle. Inner disk $i$ is preferred, if available, as it is more representative of the stellar rotation $i$ than outer disk $i$ \citep{nelissen2023}. Disk inclination angles collected from the literature are listed in Table~\ref{tab:incl}, and we take averages for targets with multiple measurements. 

The $\tau$ value for the full sample of 12 targets between the \teff{}'s MAD to the stellar $i$ is 0.351 as shown in Figure~\ref{fig:incl}. This value indicates a correlation between the two parameters at a 90\% confidence level as $\tau_c = 0.303$ at $\alpha=0.1$ with a sample size of 12. Although the sample size is small, this relationship, if real, can be explained by TTSs hosting high latitude spot(s) \citep[e.g.,][]{carroll2012,donati2014,donati2017,donati2020,yu2019}. With spot(s) at high latitude, the changes in the spot coverage area is minimized when the star is nearly pole on (small $i$), such that less temperature variability is expected to be observed. For larger $i$, we are closer to facing the equator of the TTS, and so spot coverage area will vary more over the visible hemisphere to produce a larger observed temperature variability.

\begin{deluxetable}{c L l}
\tablecaption{TTS Sample's Inclination Angles\label{tab:incl}
             }
\tabletypesize{\scriptsize}
\tablehead{
    \colhead{Target}        & \colhead{Incl.}        &
    \colhead{Reference}      \\
	\colhead{}              & \colhead{(deg)}      & 
	\colhead{}              
     }
\decimalcolnumbers
\startdata 
\multicolumn{3}{c}{Outer Disk Incl.}\\
AA\,Tau     &   59                      & \citet{francis2020}  \\ 
            &   59.1 \pm 0.3            & \citet{loomis2017}   \\
CI\,Tau     &   50.0 \pm 0.3             & \citet{long2019}     \\ 
            &   49.24                   & \citet{clarke2018}     \\ 
DH\,Tau     &   16.9 ^{+2.0}_{-2.2}     & \citet{long2019}     \\ 
            &   48.4 ^{+1.4}_{-1.5}     & \citet{sheehan2019}     \\
            &   21.4 ^{+4.2}_{-3.0}     & \citet{rota2022}     \\
DK\,Tau     &   12.8 ^{+2.5}_{-2.8}     & \citet{long2019}     \\ 
            &   21.1 ^{+3.3}_{-2.7}     & \citet{rota2022}     \\
DM\,Tau     &   35                      & \citet{francis2020}  \\
            &   35.2 \pm 0.7            & \citet{kudo2018}  \\
            &   36.0 ^{+0.12}_{-0.09}   & \citet{flaherty2020}  \\
DS\,Tau     &   65.2 \pm 0.3            & \citet{long2019}     \\
            &   70 \pm 3                & \citet{long2019}     \\
GI\,Tau     &   43.8 \pm 1.1            & \citet{long2019}     \\ 
IQ\,Tau     &   62.1 \pm 0.5            & \citet{long2019}     \\ 
LkCa\,15    &   55                      & \citet{francis2020}   \\
            &   50.16 \pm 0.03          & \citet{facchini2020}   \\
            &   43.95 ^{+2.39}_{-2.06}  & \citet{bohn2022}   \\
\hline
\multicolumn{3}{c}{Inner Disk Incl.}    \\
AA\,Tau     &   55 \pm 25           & \citet{francis2020}    \\
CI\,Tau     &   71 \pm  1           & \citet{gravitycollaboration2023}   \\
DK\,Tau     &   58 ^{+18}_{-11}     & \citet{nelissen2023}   \\
DM\,Tau     &   41 \pm  4           & \citet{francis2020}   \\
LkCa\,15    &   61.02 ^{+18.59}_{-20.80}  & \citet{bohn2022}   \\
\hline
\multicolumn{3}{c}{Stellar Incl.}\\
Hubble\,4   &   11.8                & \citet{carvalho2021}   \\
V827\,Tau   &   39.9                & This study   \\
V830\,Tau   &   55 \pm 10           & \citet{donati2015}   \\
\enddata
% \tablecomments{
%     }
\end{deluxetable}

%-----------------------------------
%-----------------------------------
\section{Summary} \label{sec:summary}

As part of the YESS project, we used the atomic Fe\,\textsc{i} and molecular OH lines in the near-infrared $H$ band to investigate the spot variability of T Tauri Stars (TTSs). We demonstrate a positive correlation between the line depth of the OH lines at $\sim$1.56310 \micron{} and $\sim$1.56317 \micron{} as the effective temperature (\teff{}) drops from $\sim$5000\,K to $\sim$3400\,K, and a negative correlation between the Fe\,\textsc{i} lines at $\sim$1.56259 \micron{} and $\sim$1.56362 \micron{} as \teff{} decreases. These temperature sensitivities make the lines ideal for probing the \teff{} variability in TTSs within the 3400--5000\,K range. Using data from 49 calibration targets observed with the Immersion GRating INfrared Spectrometer (IGRINS), we establish an empirical \teff{}--equivalent width ratio (EWR) relationship. Key findings and contributions of this study include:

\begin{itemize}
    \item We observe that line depth ratios (LDRs), heavily influenced by blending in our spectral analysis region, are susceptible to various broadening effects, including those resulting from rotational velocity (\vsini{}), spectral resolution, and magnetic field strength. In contrast, these influences have little to no impact on our EWR measurements. 

    \item We verify the reliability of our EWR methodology through observations of the quiescent M-dwarf GJ\,281. Across 61 nights spanning four years, the star exhibits a scatter in EWR of $\sigma_{\rm EWR} \approx 0.012$, equivalent to $\sigma_{T_{\rm eff}} \approx$ 12\,K.

    \item A comparison between our empirical \teff{}--EWR relationship and our model grid, particularly for solar metallicity targets in the $\sim$3400\,K to 5000\,K range, shows general agreement. However, there are noticeable discrepancies in equivalent width (EW) measurements between observed and modeled spectra. In order to extend the applicability of the EWR technique, refined physical parameters, such as metallicity, for the \teff{} calibration targets are needed, as well as improvements in spectral modeling and molecular line lists for cooler M-dwarfs.

    \item Our analysis of \teff{} variability in a sample of 13 TTSs reveals no noticeable trend distinguishing Weak-line TTSs and Classical TTSs, with both groups exhibiting substantial \teff{} variations ($\gtrsim$150\,K) within a two-year observational period. This might indicate a common spot structure between these two types of TTSs. Furthermore, we discover a strong positive correlation between the median \teff{} of the TTSs to their \teff{} variation magnitudes, hinting a higher surface activity with higher average \teff{}.

    \item We observe a quarter-phase delay between most of our targets' radial velocity (RV) and EWR modulations when folding them to their rotation periods. V830\,Tau and CI\,Tau stand out for having non-quarter-phase delay results. The clean and stable $0.68$ phase delay seen in V830\,Tau, over a period of 4 years, suggests the potential involvement of hot spot(s). With a phase delay of 0.06 $\pm$ 0.13, the CTTS CI\,Tau suggesting additional influences on its RVs and/or EWRs, possibly from a planetary companion, hinting at a 1:1 commensurability between CI\,Tau's rotation period and the orbital period of its Hot Jupiter companion.

    \item We report a positive correlation between \teff{} variability amplitude and stellar inclination angle with 90\% confidence, in agreement with previous studies that have identified high-latitude spots on TTSs.

\end{itemize}
%-----------------------------------
%-----------------------------------
% \clearpage

\begin{acknowledgments}

We are thankful for the comments and suggestions from an anonymous referee to help improve the quality of this paper. We also thank the technical and logistical staff at McDonald and Lowell Observatories for their excellent support of the Immersion Grating Infrared Spectrograph (IGRINS) installations, software, and observation program described here. In particular, D. Doss, C. Gibson, J. Kuehne, K. Meyer, B. Hardesty, F. Cornelius, M. Sweaton, J. Gehring, S. Zoonematkermani, E. Dunham, S. Levine, H. Roe, W. DeGroff, G. Jacoby, T. Pugh, A. Hayslip, and H. Larson. Partial support for this work was provided by NASA Exoplanet Research Program grant 80-NSSC19K-0289 (to LP). CMJ acknowledges partial support for this work through grants to Rice University provided by NASA (award 80-NSSC18K-0828) and the NSF (awards AST-2009197 and AST-1461918).

We are grateful for the generous donations of John and Ginger Giovale, the BF Foundation, and others which made the IGRINS-LDT program possible. Additional funding for IGRINS at the LDT was provided by the Mt. Cuba Astronomical Foundation and the Orr Family Foundation. IGRINS was developed under a collaboration between the University of Texas at Austin and the Korea Astronomy and Space Science Institute (KASI) with the financial support of the US National Science Foundation under grant AST-1229522 and AST-1702267, of the University of Texas at Austin, and of the Korean GMT Project of KASI.

This work benefited from the language editing assistance of GPT-4, a language model developed by OpenAI in San Francisco, CA, USA \url{http://openai.com}. This work made use of the VALD database, operated at Uppsala University, the Institute of Astronomy RAS in Moscow, and the University of Vienna. This study also made use of the SIMBAD database and the VizieR catalogue access tool, both operated at CDS, Strasbourg, France. Some observations were obtained at the Lowell Discovery Telescope (LDT) at Lowell Observatory. Lowell is a private, non-profit institution dedicated to astrophysical research and public appreciation of astronomy and operates the LDT in partnership with Boston University, the University of Maryland, the University of Toledo, Northern Arizona University and Yale University. We have used IGRINS archival data older than the 2 year proprietary period. 
\end{acknowledgments}

\clearpage
% \vspace{5mm}
\facilities{LDT (IGRINS), Smith (IGRINS)}

\software{
        \texttt{astropy} \citep{astropycollaboration2013,astropycollaboration2018,astropycollaboration2022},  
        \texttt{matplotlib} \citep{hunter2007},
        \texttt{NumPy}, \citep{harris2020}
        \texttt{IGRINS RV} \citep{stahl2021,tang2021a}, 
        \texttt{pysynphot} \citep{stscidevelopmentteam2013}
        }

%-----------------------------------
%-----------------------------------
% \clearpage

\appendix{} 
\counterwithin{figure}{section}
\counterwithin{table}{section}

%%----------------------------------
\section{Atmosphere Profile Comparison} \label{sec:atm_comp}

This appendix section describes the effects of employing different atmospheric profiles to determine equivalent widths (EWs) of the atomic Fe\textsc{i} and molecular OH line region using the \textsc{synmast} code. The results presented here use the modified VALD3 line data tuned on GJ\,205 (see Section~\ref{sub:line_tuning}) and supplemented with BT2 water lines. We investigate three distinct atmospheric profiles: BT-NextGen \citep{allard2011,allard2012}, PHOENIX-ACES\footnote{Atmospheric profiles retrieved from \url{ftp://phoenix.astro.physik.uni-goettingen.de/AtmosFITS/} under the v2.0 directory.}, and MARCS \citep{gustafsson2008}\footnote{Accessed via \url{https://marcs.astro.uu.se}}. The elemental abundances for the BT-NextGen and PHOENIX-ACES profiles are based on the ASGG2009 standard \citep{asplund2009}, whereas the MARCS profiles conform to GAS2007 \citep{grevesse2007}. For all considered models, the scaling of $\alpha$-element abundances is consistent with the methodologies described in Section~\ref{sec:model}.

\begin{figure*}[tbh!]
\centering
\includegraphics[width=0.9\textwidth]{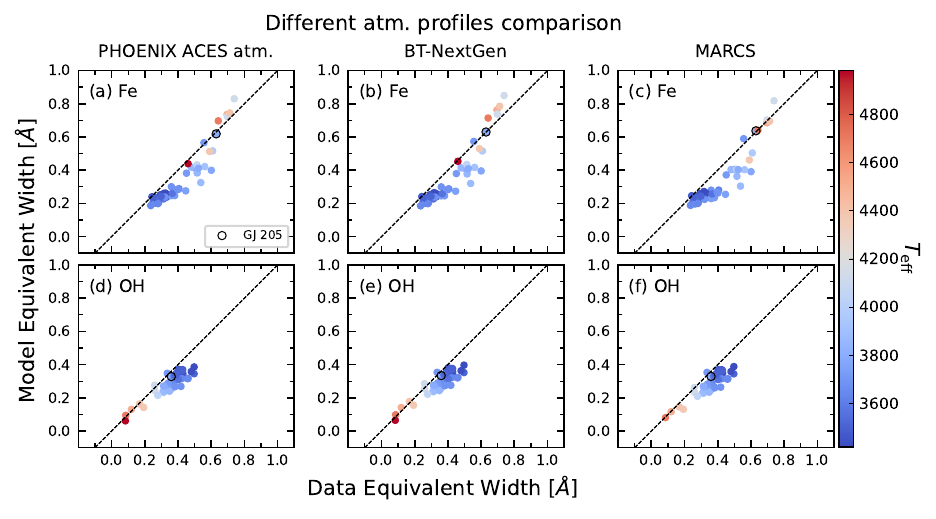}
\caption{
    Comparison of Fe and OH equivalent widths from the \textsc{synmast} model spectra using different atmospheric profiles and IGRINS data for the \teff{} calibration targets. Results derived from the PHOENIX ACES atmospheric profile are presented in the first column, while those obtained from the BT-NextGen and MARCS profiles are displayed in the second and third columns, respectively. Panels (a), (b), and (c) in the top row illustrate the results for Fe EW, and panels (d), (e), and (f) in the bottom row detail the OH EW findings. The color coding of data points corresponds to the effective temperature (\teff{}) scale, as indicated by the color bar on the right. The dashed line across each panel denotes the line of unity.
    }
\label{fig:atm_comp}
\end{figure*}

As depicted in Figure~\ref{fig:atm_comp}, the comparison of EW measurements across different atmospheric profiles reveals minimal variance for both the Fe and OH line regions. The choice of atmospheric model exhibits negligible influence on the precision of derived physical parameters for the targets. This consistency underscores the robustness of the employed spectroscopic techniques in yielding reliable stellar parameters irrespective of the chosen atmospheric profile.

%%----------------------------------
\section{Water line list Comparison} \label{sec:water_comp}

\begin{figure}[tbh!]
\centering
\includegraphics[width=0.4\textwidth]{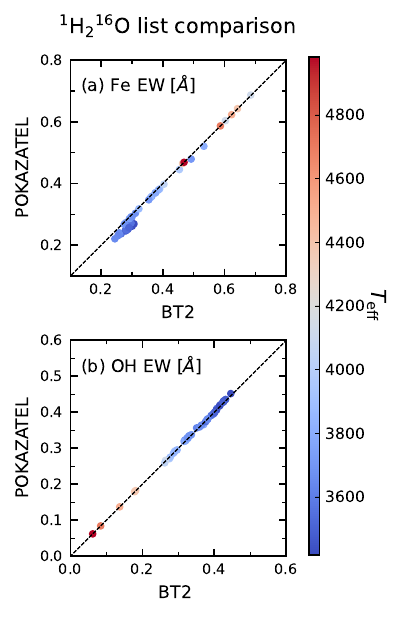}
\caption{
    POKAZATEL \citep{polyansky2018a} and BT2 \citep{barber2006} water line list comparisons for Fe EW in (a) and OH EW in (b). All models are generated using \textsc{synmast} with an unmodified VALD3 line list and adopting the BT-NextGen atmosphere profile. Circles show results generated based on the \teff{} calibration targets' parameters, color-coded by their \teff{}.  
    }
\label{fig:water_comp}
\end{figure}

Figure~\ref{fig:water_comp} illustrates the negligible impact on the equivalent width (EW) measurements of Fe (a) and OH (b) regions from employing different water ($^1{\rm H}_2$$^{16}{\rm O}$) line lists\,---BT2 \citep{barber2006} and POKAZATEL \citep{polyansky2018a}---\,within the \textsc{synmast} spectral synthesis code. The differences, as shown, are minimal, with only a marginal shift in the Fe EW at an overall 6\% difference.

%-----------------------------------
\clearpage
%-----------------------------------
%-----------------------------------
\bibliography{main}

\begin{thebibliography}{}
\expandafter\ifx\csname natexlab\endcsname\relax\def\natexlab#1{#1}\fi
\providecommand{\url}[1]{\href{#1}{#1}}
\providecommand{\dodoi}[1]{doi:~\href{http://doi.org/#1}{\nolinkurl{#1}}}
\providecommand{\doeprint}[1]{\href{http://ascl.net/#1}{\nolinkurl{http://ascl.net/#1}}}
\providecommand{\doarXiv}[1]{\href{https://arxiv.org/abs/#1}{\nolinkurl{https://arxiv.org/abs/#1}}}

\bibitem[{Af{\c s}ar {et~al.}(2023)Af{\c s}ar, Bozkurt, Topcu, {\"O}zdemir,
  Sneden, Mace, Jaffe, \& {L{\'o}pez-Valdivia}}]{afsar2023a}
Af{\c s}ar, M., Bozkurt, Z., Topcu, G.~B., {et~al.} 2023, The Astrophysical
  Journal, 949, 86, \dodoi{10.3847/1538-4357/acc946}

\bibitem[{Allard {et~al.}(2011)Allard, Homeier, \& Freytag}]{allard2011}
Allard, F., Homeier, D., \& Freytag, B. 2011, 448, 91.
\newblock \url{https://ui.adsabs.harvard.edu/abs/2011ASPC..448...91A}

\bibitem[{Allard {et~al.}(2012)Allard, Homeier, Freytag, \& Sharp}]{allard2012}
Allard, F., Homeier, D., Freytag, B., \& Sharp, C.~M. 2012, 57, 3,
  \dodoi{10.1051/eas/1257001}

\bibitem[{Anderson \& Francis(2011)}]{anderson2011}
Anderson, E., \& Francis, C. 2011, VizieR Online Data Catalog, V/137.
\newblock \url{https://ui.adsabs.harvard.edu/abs/2011yCat.5137....0A}

\bibitem[{Asplund {et~al.}(2009)Asplund, Grevesse, Sauval, \&
  Scott}]{asplund2009}
Asplund, M., Grevesse, N., Sauval, A.~J., \& Scott, P. 2009, Annual Review of
  Astronomy and Astrophysics, 47, 481,
  \dodoi{10.1146/annurev.astro.46.060407.145222}

\bibitem[{{Astropy Collaboration} {et~al.}(2013){Astropy Collaboration},
  Robitaille, Tollerud, Greenfield, Droettboom, Bray, Aldcroft, Davis,
  Ginsburg, {Price-Whelan}, Kerzendorf, Conley, Crighton, Barbary, Muna,
  Ferguson, Grollier, Parikh, Nair, Unther, Deil, Woillez, Conseil, Kramer,
  Turner, Singer, Fox, Weaver, Zabalza, Edwards, Azalee~Bostroem, Burke, Casey,
  Crawford, Dencheva, Ely, Jenness, Labrie, Lim, Pierfederici, Pontzen, Ptak,
  Refsdal, Servillat, \& Streicher}]{astropycollaboration2013}
{Astropy Collaboration}, Robitaille, T.~P., Tollerud, E.~J., {et~al.} 2013,
  Astronomy and Astrophysics, 558, A33, \dodoi{10.1051/0004-6361/201322068}

\bibitem[{{Astropy Collaboration} {et~al.}(2018){Astropy Collaboration},
  {Price-Whelan}, Sip{\H o}cz, G{\"u}nther, Lim, Crawford, Conseil, Shupe,
  Craig, Dencheva, Ginsburg, VanderPlas, Bradley, {P{\'e}rez-Su{\'a}rez}, {de
  Val-Borro}, Aldcroft, Cruz, Robitaille, Tollerud, Ardelean, Babej, Bach,
  Bachetti, Bakanov, Bamford, Barentsen, Barmby, Baumbach, Berry, Biscani,
  Boquien, Bostroem, Bouma, Brammer, Bray, Breytenbach, Buddelmeijer, Burke,
  Calderone, Cano~Rodr{\'i}guez, Cara, Cardoso, Cheedella, Copin, Corrales,
  Crichton, D'Avella, Deil, Depagne, Dietrich, Donath, Droettboom, Earl, Erben,
  Fabbro, Ferreira, Finethy, Fox, Garrison, Gibbons, Goldstein, Gommers, Greco,
  Greenfield, Groener, Grollier, Hagen, Hirst, Homeier, Horton, Hosseinzadeh,
  Hu, Hunkeler, Ivezi{\'c}, Jain, Jenness, Kanarek, Kendrew, Kern, Kerzendorf,
  Khvalko, King, Kirkby, Kulkarni, Kumar, Lee, Lenz, Littlefair, Ma, Macleod,
  Mastropietro, McCully, Montagnac, Morris, Mueller, Mumford, Muna, Murphy,
  Nelson, Nguyen, Ninan, N{\"o}the, Ogaz, Oh, Parejko, Parley, Pascual, Patil,
  Patil, Plunkett, Prochaska, Rastogi, Reddy~Janga, Sabater, Sakurikar,
  Seifert, Sherbert, {Sherwood-Taylor}, Shih, Sick, Silbiger, Singanamalla,
  Singer, Sladen, Sooley, Sornarajah, Streicher, Teuben, Thomas, Tremblay,
  Turner, Terr{\'o}n, {van Kerkwijk}, {de la Vega}, Watkins, Weaver, Whitmore,
  Woillez, Zabalza, \& {Astropy Contributors}}]{astropycollaboration2018}
{Astropy Collaboration}, {Price-Whelan}, A.~M., Sip{\H o}cz, B.~M., {et~al.}
  2018, The Astronomical Journal, 156, 123, \dodoi{10.3847/1538-3881/aabc4f}

\bibitem[{{Astropy Collaboration} {et~al.}(2022){Astropy Collaboration},
  {Price-Whelan}, Lim, Earl, Starkman, Bradley, Shupe, Patil, Corrales,
  Brasseur, N{\"o}the, Donath, Tollerud, Morris, Ginsburg, Vaher, Weaver,
  Tocknell, Jamieson, {van Kerkwijk}, Robitaille, Merry, Bachetti, G{\"u}nther,
  Aldcroft, {Alvarado-Montes}, Archibald, B{\'o}di, Bapat, Barentsen,
  Baz{\'a}n, Biswas, Boquien, Burke, Cara, Cara, Conroy, Conseil, Craig, Cross,
  Cruz, D'Eugenio, Dencheva, Devillepoix, Dietrich, Eigenbrot, Erben, Ferreira,
  {Foreman-Mackey}, Fox, Freij, Garg, Geda, Glattly, Gondhalekar, Gordon,
  Grant, Greenfield, Groener, Guest, Gurovich, Handberg, Hart,
  {Hatfield-Dodds}, Homeier, Hosseinzadeh, Jenness, Jones, Joseph, Kalmbach,
  Karamehmetoglu, Ka{\l}uszy{\'n}ski, Kelley, Kern, Kerzendorf, Koch, Kulumani,
  Lee, Ly, Ma, MacBride, Maljaars, Muna, Murphy, Norman, O'Steen, Oman,
  Pacifici, Pascual, {Pascual-Granado}, Patil, Perren, Pickering, Rastogi,
  Roulston, Ryan, Rykoff, Sabater, Sakurikar, Salgado, Sanghi, Saunders,
  Savchenko, Schwardt, {Seifert-Eckert}, Shih, Jain, Shukla, Sick, Simpson,
  Singanamalla, Singer, Singhal, Sinha, Sip{\H o}cz, Spitler, Stansby,
  Streicher, {\v S}umak, Swinbank, Taranu, Tewary, Tremblay, {de Val-Borro},
  Van~Kooten, Vasovi{\'c}, Verma, {de Miranda Cardoso}, Williams, Wilson,
  Winkel, {Wood-Vasey}, Xue, Yoachim, Zhang, Zonca, \& {Astropy Project
  Contributors}}]{astropycollaboration2022}
{Astropy Collaboration}, {Price-Whelan}, A.~M., Lim, P.~L., {et~al.} 2022, The
  Astrophysical Journal, 935, 167, \dodoi{10.3847/1538-4357/ac7c74}

\bibitem[{Baines {et~al.}(2021)Baines, Thomas~Armstrong, Clark, Gorney, Hutter,
  Jorgensen, Kyte, Mozurkewich, Nisley, Sanborn, Schmitt, \& {van
  Belle}}]{baines2021}
Baines, E.~K., Thomas~Armstrong, J., Clark, J.~H., {et~al.} 2021, The
  Astronomical Journal, 162, 198, \dodoi{10.3847/1538-3881/ac2431}

\bibitem[{Baliunas {et~al.}(1997)Baliunas, Henry, Donahue, Fekel, \&
  Soon}]{baliunas1997}
Baliunas, S.~L., Henry, G.~W., Donahue, R.~A., Fekel, F.~C., \& Soon, W.~H.
  1997, The Astrophysical Journal, 474, L119, \dodoi{10.1086/310442}

\bibitem[{Barber {et~al.}(2006)Barber, Tennyson, Harris, \&
  Tolchenov}]{barber2006}
Barber, R.~J., Tennyson, J., Harris, G.~J., \& Tolchenov, R.~N. 2006, Monthly
  Notices of the Royal Astronomical Society, 368, 1087,
  \dodoi{10.1111/j.1365-2966.2006.10184.x}

\bibitem[{Benatti {et~al.}(2020)Benatti, Damasso, Desidera, Marzari, Biazzo,
  Claudi, Di~Mauro, Lanza, Pinamonti, Barbato, Malavolta, Poretti, Sozzetti,
  Affer, Bignamini, Bonomo, Borsa, Brogi, Bruno, Carleo, Cosentino, Covino,
  Frustagli, Giacobbe, Gonzalez, Gratton, Harutyunyan, Knapic, Leto, Lodi,
  Maggio, Maldonado, Mancini, Martinez~Fiorenzano, Micela, Molinari, Molinaro,
  Nardiello, Nascimbeni, Pagano, Pedani, Piotto, Rainer, \&
  Scandariato}]{benatti2020}
Benatti, S., Damasso, M., Desidera, S., {et~al.} 2020, A\&A, 639, A50,
  \dodoi{10.1051/0004-6361/202037939}

\bibitem[{Bevington \& Robinson(1992)}]{bevington1992}
Bevington, P.~R., \& Robinson, D.~K. 1992, Data Reduction and Error Analysis
  for the Physical Sciences.
\newblock \url{https://ui.adsabs.harvard.edu/abs/1992drea.book.....B}

\bibitem[{Biddle {et~al.}(2018)Biddle, {Johns-Krull}, Llama, Prato, \&
  Skiff}]{biddle2018}
Biddle, L.~I., {Johns-Krull}, C.~M., Llama, J., Prato, L., \& Skiff, B.~A.
  2018, ApJ, 853, L34, \dodoi{10.3847/2041-8213/aaa897}

\bibitem[{Biddle {et~al.}(2021)Biddle, Llama, Cameron, Prato, Jardine, \&
  {Johns-Krull}}]{biddle2021}
Biddle, L.~I., Llama, J., Cameron, A., {et~al.} 2021, The Astrophysical
  Journal, 906, 113, \dodoi{10.3847/1538-4357/abc889}

\bibitem[{Bohn {et~al.}(2022)Bohn, Benisty, Perraut, {van der Marel},
  W{\"o}lfer, {van Dishoeck}, Facchini, Manara, Teague, Francis, Berger,
  {Garcia-Lopez}, Ginski, Henning, Kenworthy, Kraus, M{\'e}nard, M{\'e}rand, \&
  P{\'e}rez}]{bohn2022}
Bohn, A.~J., Benisty, M., Perraut, K., {et~al.} 2022, Astronomy and
  Astrophysics, 658, A183, \dodoi{10.1051/0004-6361/202142070}

\bibitem[{Bonomo {et~al.}(2017)Bonomo, Desidera, Benatti, Borsa, Crespi,
  Damasso, Lanza, Sozzetti, Lodato, Marzari, Boccato, Claudi, Cosentino,
  Covino, Gratton, Maggio, Micela, Molinari, Pagano, Piotto, Poretti,
  Smareglia, Affer, Biazzo, Bignamini, Esposito, Giacobbe, H{\'e}brard,
  Malavolta, Maldonado, Mancini, Martinez~Fiorenzano, Masiero, Nascimbeni,
  Pedani, Rainer, \& Scandariato}]{bonomo2017}
Bonomo, A.~S., Desidera, S., Benatti, S., {et~al.} 2017, Astronomy and
  Astrophysics, 602, A107, \dodoi{10.1051/0004-6361/201629882}

\bibitem[{Borsa {et~al.}(2015)Borsa, Scandariato, Rainer, Bignamini, Maggio,
  Poretti, Lanza, Di~Mauro, Benatti, Biazzo, Bonomo, Damasso, Esposito,
  Gratton, Affer, Barbieri, Boccato, Claudi, Cosentino, Covino, Desidera,
  Fiorenzano, Gandolfi, Harutyunyan, Maldonado, Micela, Molaro, Molinari,
  Pagano, Pillitteri, Piotto, Shkolnik, Silvotti, Smareglia, Southworth,
  Sozzetti, \& Stelzer}]{borsa2015}
Borsa, F., Scandariato, G., Rainer, M., {et~al.} 2015, Astronomy and
  Astrophysics, 578, A64, \dodoi{10.1051/0004-6361/201525741}

\bibitem[{Boyajian {et~al.}(2012)Boyajian, {von Braun}, {van Belle}, McAlister,
  {ten Brummelaar}, Kane, Muirhead, Jones, White, Schaefer, Ciardi, Henry,
  {L{\'o}pez-Morales}, Ridgway, Gies, Jao, {Rojas-Ayala}, Parks, Sturmann,
  Sturmann, Turner, Farrington, Goldfinger, \& Berger}]{boyajian2012}
Boyajian, T.~S., {von Braun}, K., {van Belle}, G., {et~al.} 2012, The
  Astrophysical Journal, 757, 112, \dodoi{10.1088/0004-637X/757/2/112}

\bibitem[{Boyajian {et~al.}(2013)Boyajian, {von Braun}, {van Belle},
  Farrington, Schaefer, Jones, White, McAlister, {ten Brummelaar}, Ridgway,
  Gies, Sturmann, Sturmann, Turner, Goldfinger, \& Vargas}]{boyajian2013}
---. 2013, The Astrophysical Journal, 771, 40,
  \dodoi{10.1088/0004-637X/771/1/40}

\bibitem[{Brogi {et~al.}(2012)Brogi, Snellen, {de Kok}, Albrecht, Birkby, \&
  {de Mooij}}]{brogi2012}
Brogi, M., Snellen, I. A.~G., {de Kok}, R.~J., {et~al.} 2012, Nature, 486, 502,
  \dodoi{10.1038/nature11161}

\bibitem[{Cale {et~al.}(2021)Cale, Reefe, Plavchan, Tanner, Gaidos, Gagn{\'e},
  Gao, Kane, B{\'e}jar, Lodieu, {Anglada-Escud{\'e}}, Ribas, Pall{\'e},
  Quirrenbach, Amado, Reiners, Caballero, Rosa Zapatero~Osorio, Dreizler,
  Howard, Fulton, Xuesong~Wang, Collins, El~Mufti, Wittrock, Gilbert, Barclay,
  Klein, Martioli, Wittenmyer, Wright, Addison, Hirano, Tamura, Kotani, Narita,
  Vermilion, Lee, Geneser, Teske, Quinn, Latham, Esquerdo, Calkins, Berlind,
  Zohrabi, Stibbards, Kotnana, Jenkins, Twicken, Henze, Kidwell, Burke,
  Villase{\~n}or, \& Boyd}]{cale2021}
Cale, B.~L., Reefe, M., Plavchan, P., {et~al.} 2021, AJ, 162, 295,
  \dodoi{10.3847/1538-3881/ac2c80}

\bibitem[{Carroll {et~al.}(2012)Carroll, Strassmeier, Rice, \&
  K{\"u}nstler}]{carroll2012}
Carroll, T.~A., Strassmeier, K.~G., Rice, J.~B., \& K{\"u}nstler, A. 2012,
  Astronomy and Astrophysics, 548, A95, \dodoi{10.1051/0004-6361/201220215}

\bibitem[{Carvalho {et~al.}(2021)Carvalho, {Johns-Krull}, Prato, \&
  Anderson}]{carvalho2021}
Carvalho, A., {Johns-Krull}, C.~M., Prato, L., \& Anderson, J. 2021, ApJ, 910,
  33, \dodoi{10.3847/1538-4357/abe237}

\bibitem[{Casagrande {et~al.}(2014)Casagrande, Portinari, Glass, Laney,
  Silva~Aguirre, Datson, Andersen, Nordstr{\"o}m, Holmberg, Flynn, \&
  Asplund}]{casagrande2014}
Casagrande, L., Portinari, L., Glass, I.~S., {et~al.} 2014, Monthly Notices of
  the Royal Astronomical Society, 439, 2060, \dodoi{10.1093/mnras/stu089}

\bibitem[{Catalano {et~al.}(2002)Catalano, Biazzo, Frasca, \&
  Marilli}]{catalano2002}
Catalano, S., Biazzo, K., Frasca, A., \& Marilli, E. 2002, Astronomy and
  Astrophysics, 394, 1009,
  \dodoi{https://ui.adsabs.harvard.edu/abs/2002A/\%26A...394.1009C/abstract}

\bibitem[{Clarke {et~al.}(2018)Clarke, Tazzari, Juhasz, Rosotti, Booth,
  Facchini, Ilee, {Johns-Krull}, Kama, Meru, \& Prato}]{clarke2018}
Clarke, C.~J., Tazzari, M., Juhasz, A., {et~al.} 2018, ApJ, 866, L6,
  \dodoi{10.3847/2041-8213/aae36b}

\bibitem[{Claytor {et~al.}(2024)Claytor, {van Saders}, Cao, Pinsonneault,
  Teske, \& Beaton}]{claytor2024}
Claytor, Z.~R., {van Saders}, J.~L., Cao, L., {et~al.} 2024, The Astrophysical
  Journal, 962, 47, \dodoi{10.3847/1538-4357/ad159a}

\bibitem[{Crockett {et~al.}(2012)Crockett, Mahmud, Prato, {Johns-Krull}, Jaffe,
  Hartigan, \& Beichman}]{crockett2012}
Crockett, C.~J., Mahmud, N.~I., Prato, L., {et~al.} 2012, ApJ, 761, 164,
  \dodoi{10.1088/0004-637X/761/2/164}

\bibitem[{Cushing {et~al.}(2005)Cushing, Rayner, \& Vacca}]{cushing2005}
Cushing, M.~C., Rayner, J.~T., \& Vacca, W.~D. 2005, The Astrophysical Journal,
  623, 1115, \dodoi{10.1086/428040}

\bibitem[{Dawson \& Johnson(2018)}]{dawson2018}
Dawson, R.~I., \& Johnson, J.~A. 2018, Annu. Rev. Astron. Astrophys., 56, 175,
  \dodoi{10.1146/annurev-astro-081817-051853}

\bibitem[{Dittmann {et~al.}(2009)Dittmann, Close, Green, \&
  Fenwick}]{dittmann2009}
Dittmann, J.~A., Close, L.~M., Green, E.~M., \& Fenwick, M. 2009, The
  Astrophysical Journal, 701, 756, \dodoi{10.1088/0004-637X/701/1/756}

\bibitem[{Donati {et~al.}(2014)Donati, H{\'e}brard, Hussain, Moutou, Grankin,
  Boisse, Morin, Gregory, Vidotto, Bouvier, Alencar, Delfosse, Doyon, Takami,
  Jardine, Fares, Cameron, M{\'e}nard, Dougados, Herczeg, \& {Matysse
  Collaboration}}]{donati2014}
Donati, J.~F., H{\'e}brard, E., Hussain, G., {et~al.} 2014, Monthly Notices of
  the Royal Astronomical Society, 444, 3220, \dodoi{10.1093/mnras/stu1679}

\bibitem[{Donati {et~al.}(2015)Donati, H{\'e}brard, Hussain, Moutou, Malo,
  Grankin, Vidotto, Alencar, Gregory, Jardine, Herczeg, Morin, Fares,
  M{\'e}nard, Bouvier, Delfosse, Doyon, Takami, Figueira, Petit, Boisse, \&
  {the MaTYSSE Collaboration}}]{donati2015}
Donati, J.-F., H{\'e}brard, E., Hussain, G. A.~J., {et~al.} 2015, Mon. Not. R.
  Astron. Soc., 453, 3707, \dodoi{10.1093/mnras/stv1837}

\bibitem[{Donati {et~al.}(2017)Donati, Yu, Moutou, Cameron, Malo, Grankin,
  H{\'e}brard, Hussain, Vidotto, Alencar, Haywood, Bouvier, Petit, Takami,
  Herczeg, Gregory, Jardine, Morin, \& {the MaTYSSE
  collaboration}}]{donati2017}
Donati, J.-F., Yu, L., Moutou, C., {et~al.} 2017, Mon. Not. R. Astron. Soc.,
  465, 3343, \dodoi{10.1093/mnras/stw2904}

\bibitem[{Donati {et~al.}(2020)Donati, Bouvier, Alencar, Moutou, Malo, Takami,
  M{\'e}nard, Dougados, Hussain, \& {the MaTYSSE collaboration}}]{donati2020}
Donati, J.-F., Bouvier, J., Alencar, S.~H., {et~al.} 2020, MNRAS, 491, 5660,
  \dodoi{10.1093/mnras/stz3368}

\bibitem[{Donati {et~al.}(2024)Donati, Finociety, Cristofari, Alencar, Moutou,
  Delfosse, Fouqu{\'e}, Arnold, Baruteau, K{\'o}sp{\'a}l, M{\'e}nard, Carmona,
  Grankin, Takami, Artigau, Doyon, H{\'e}brard, \&
  {collaboration}}]{donati2024}
Donati, J.-F., Finociety, B., Cristofari, P.~I., {et~al.} 2024, The Classical
  {{T Tauri}} Star {{CI Tau}} Observed with {{SPIRou}}: Magnetospheric
  Accretion and Planetary Formation,  arXiv.
\newblock \doeprint{2403.02166}

\bibitem[{D'Orazi {et~al.}(2011)D'Orazi, Biazzo, \& Randich}]{dorazi2011}
D'Orazi, V., Biazzo, K., \& Randich, S. 2011, Astronomy and Astrophysics, 526,
  A103, \dodoi{10.1051/0004-6361/201015616}

\bibitem[{Earl {et~al.}(2022)Earl, Tollerud, Jones, O'Steen, Kerzendorf, Busko,
  {shaileshahuja}, D'Avella, Robitaille, Ginsburg, Homeier, Sip{\H o}cz,
  Averbukh, Tocknell, Cherinka, Ogaz, Geda, Lim, Davies, G{\"u}nther, Barbary,
  Foster, Conroy, Droettboom, Torres, Bray, Casey, Teuben, Crawford, \&
  Ferguson}]{earl2022}
Earl, N., Tollerud, E., Jones, C., {et~al.} 2022, Astropy/Specutils:
  {{V1}}.7.0, Zenodo, \dodoi{10.5281/zenodo.6207491}

\bibitem[{Endl {et~al.}(2003)Endl, Cochran, Tull, \& MacQueen}]{endl2003}
Endl, M., Cochran, W.~D., Tull, R.~G., \& MacQueen, P.~J. 2003, The
  Astronomical Journal, 126, 3099, \dodoi{10.1086/379137}

\bibitem[{Facchini {et~al.}(2020)Facchini, Benisty, Bae, Loomis, Perez,
  Ansdell, Mayama, Pinilla, Teague, Isella, \& Mann}]{facchini2020}
Facchini, S., Benisty, M., Bae, J., {et~al.} 2020, Astronomy and Astrophysics,
  639, A121, \dodoi{10.1051/0004-6361/202038027}

\bibitem[{Flagg {et~al.}(2019)Flagg, {Johns-Krull}, Nofi, Llama, Prato,
  Sullivan, Jaffe, \& Mace}]{flagg2019}
Flagg, L., {Johns-Krull}, C.~M., Nofi, L., {et~al.} 2019, ApJ, 878, L37,
  \dodoi{10.3847/2041-8213/ab276d}

\bibitem[{Flaherty {et~al.}(2020)Flaherty, Hughes, Simon, Qi, Bai, Bulatek,
  Andrews, Wilner, \& K{\'o}sp{\'a}l}]{flaherty2020}
Flaherty, K., Hughes, A.~M., Simon, J.~B., {et~al.} 2020, The Astrophysical
  Journal, 895, 109, \dodoi{10.3847/1538-4357/ab8cc5}

\bibitem[{Flores {et~al.}(2019)Flores, Connelley, Reipurth, \&
  Boogert}]{flores2019}
Flores, C., Connelley, M.~S., Reipurth, B., \& Boogert, A. 2019, The
  Astrophysical Journal, 882, 75, \dodoi{10.3847/1538-4357/ab35d4}

\bibitem[{Francis \& {van der Marel}(2020)}]{francis2020}
Francis, L., \& {van der Marel}, N. 2020, The Astrophysical Journal, 892, 111,
  \dodoi{10.3847/1538-4357/ab7b63}

\bibitem[{Fukue {et~al.}(2015)Fukue, Matsunaga, Yamamoto, Kondo, Kobayashi,
  Ikeda, Hamano, Yasui, Arasaki, Tsujimoto, Bono, \& Inno}]{fukue2015}
Fukue, K., Matsunaga, N., Yamamoto, R., {et~al.} 2015, ApJ, 812, 64,
  \dodoi{10.1088/0004-637X/812/1/64}

\bibitem[{Gangi {et~al.}(2022)Gangi, Antoniucci, Biazzo, Frasca, Nisini,
  Alcal{\'a}, Giannini, Manara, Giunta, Harutyunyan, Munari, \&
  Vitali}]{gangi2022}
Gangi, M., Antoniucci, S., Biazzo, K., {et~al.} 2022, Astronomy and
  Astrophysics, 667, A124, \dodoi{10.1051/0004-6361/202244042}

\bibitem[{Garnett(2023)}]{garnett2023}
Garnett, R. 2023, Bayesian {{Optimization}} (Cambridge University Press),
  \dodoi{10.1017/9781108348973}

\bibitem[{Ghezzi {et~al.}(2010)Ghezzi, Cunha, Smith, \& {de la
  Reza}}]{ghezzi2010}
Ghezzi, L., Cunha, K., Smith, V.~V., \& {de la Reza}, R. 2010, The
  Astrophysical Journal, 724, 154, \dodoi{10.1088/0004-637X/724/1/154}

\bibitem[{{Gravity Collaboration} {et~al.}(2023){Gravity Collaboration},
  Soulain, Perraut, Bouvier, Pantolmos, Caratti O~Garatti, Caselli, Garcia,
  Lopez, Aimar, Amorin, Benisty, Berger, Bourdarot, Brandner, Cl{\'e}net, {de
  Zeeuw}, Davies, Drescher, Eckart, Eisenhauer, Schreiber, Gendron, Genzuel,
  Gillessen, Hei{\ss}el, Henning, Hippler, Horrobin, Jocou, Kervella, Labadie,
  Lacour, Lapeyrere, Le~Bouquin, L{\'e}na, Lutz, Mang, Ott, Paumard, Perrin,
  Sanchez, Scheithauer, Shangguan, Shimizu, Straub, Straubmeier, Sturm,
  Tacconi, Vincent, {van Dishoeck}, Widmann, Wieprecht, Wiezorrek, \&
  Yazici}]{gravitycollaboration2023}
{Gravity Collaboration}, Soulain, A., Perraut, K., {et~al.} 2023, Astronomy and
  Astrophysics, 674, A203, \dodoi{10.1051/0004-6361/202346446}

\bibitem[{Gray(1989)}]{gray1989}
Gray, D.~F. 1989, The Astrophysical Journal, 347, 1021, \dodoi{10.1086/168192}

\bibitem[{Gray(1994)}]{gray1994}
---. 1994, Publications of the Astronomical Society of the Pacific, 106, 1248,
  \dodoi{10.1086/133502}

\bibitem[{Gray \& Baliunas(1995)}]{gray1995}
Gray, D.~F., \& Baliunas, S.~L. 1995, The Astrophysical Journal, 441, 436,
  \dodoi{10.1086/175368}

\bibitem[{Gray {et~al.}(1992)Gray, Baliunas, Lockwood, \& Skiff}]{gray1992}
Gray, D.~F., Baliunas, S.~L., Lockwood, G.~W., \& Skiff, B.~A. 1992, The
  Astrophysical Journal, 400, 681, \dodoi{10.1086/172030}

\bibitem[{Gray \& Brown(2001)}]{gray2001}
Gray, D.~F., \& Brown, K. 2001, Publications of the Astronomical Society of the
  Pacific, 113, 723, \dodoi{10.1086/320811}

\bibitem[{Gray \& Johanson(1991)}]{gray1991}
Gray, D.~F., \& Johanson, H.~L. 1991, PASP, 103, 439, \dodoi{10.1086/132839}

\bibitem[{Grevesse {et~al.}(2007)Grevesse, Asplund, \& Sauval}]{grevesse2007}
Grevesse, N., Asplund, M., \& Sauval, A.~J. 2007, Space Science Reviews, 130,
  105, \dodoi{10.1007/s11214-007-9173-7}

\bibitem[{{Gully-Santiago} {et~al.}(2017){Gully-Santiago}, Herczeg, Czekala,
  Somers, Grankin, Covey, Donati, Alencar, Hussain, Shappee, Mace, Lee,
  Holoien, Jose, \& Liu}]{gully-santiago2017}
{Gully-Santiago}, M.~A., Herczeg, G.~J., Czekala, I., {et~al.} 2017, ApJ, 836,
  200, \dodoi{10.3847/1538-4357/836/2/200}

\bibitem[{Gustafsson {et~al.}(2008)Gustafsson, Edvardsson, Eriksson,
  J{\o}rgensen, Nordlund, \& Plez}]{gustafsson2008}
Gustafsson, B., Edvardsson, B., Eriksson, K., {et~al.} 2008, Astronomy and
  Astrophysics, 486, 951, \dodoi{10.1051/0004-6361:200809724}

\bibitem[{Harris {et~al.}(2020)Harris, Millman, {van der Walt}, Gommers,
  Virtanen, Cournapeau, Wieser, Taylor, Berg, Smith, Kern, Picus, Hoyer, {van
  Kerkwijk}, Brett, Haldane, {del R{\'i}o}, Wiebe, Peterson,
  {G{\'e}rard-Marchant}, Sheppard, Reddy, Weckesser, Abbasi, Gohlke, \&
  Oliphant}]{harris2020}
Harris, C.~R., Millman, K.~J., {van der Walt}, S.~J., {et~al.} 2020, Nature,
  585, 357, \dodoi{10.1038/s41586-020-2649-2}

\bibitem[{Haywood {et~al.}(2014)Haywood, Collier~Cameron, Queloz, Barros,
  Deleuil, Fares, Gillon, Lanza, Lovis, Moutou, Pepe, Pollacco, Santerne,
  S{\'e}gransan, \& Unruh}]{haywood2014}
Haywood, R.~D., Collier~Cameron, A., Queloz, D., {et~al.} 2014, Monthly Notices
  of the Royal Astronomical Society, 443, 2517, \dodoi{10.1093/mnras/stu1320}

\bibitem[{Heiter {et~al.}(2015)Heiter, Jofr{\'e}, Gustafsson, Korn, Soubiran,
  \& Th{\'e}venin}]{heiter2015}
Heiter, U., Jofr{\'e}, P., Gustafsson, B., {et~al.} 2015, Astronomy and
  Astrophysics, 582, A49, \dodoi{10.1051/0004-6361/201526319}

\bibitem[{Hunter(2007)}]{hunter2007}
Hunter, J.~D. 2007, Comput. Sci. Eng., 9, 90, \dodoi{10.1109/MCSE.2007.55}

\bibitem[{Husser {et~al.}(2013)Husser, {Wende-von Berg}, Dreizler, Homeier,
  Reiners, Barman, \& Hauschildt}]{husser2013}
Husser, T.-O., {Wende-von Berg}, S., Dreizler, S., {et~al.} 2013, A\&A, 553,
  A6, \dodoi{10.1051/0004-6361/201219058}

\bibitem[{Jian {et~al.}(2019)Jian, Matsunaga, \& Fukue}]{jian2019}
Jian, M., Matsunaga, N., \& Fukue, K. 2019, Monthly Notices of the Royal
  Astronomical Society, 485, 1310, \dodoi{10.1093/mnras/stz237}

\bibitem[{Jian {et~al.}(2020)Jian, Taniguchi, Matsunaga, Kobayashi, Ikeda,
  Yasui, Kondo, Sameshima, Hamano, Fukue, Arai, Otsubo, \& Kawakita}]{jian2020}
Jian, M., Taniguchi, D., Matsunaga, N., {et~al.} 2020, Monthly Notices of the
  Royal Astronomical Society, 494, 1724, \dodoi{10.1093/mnras/staa834}

\bibitem[{Jofr{\'e} {et~al.}(2014)Jofr{\'e}, Heiter, Soubiran,
  {Blanco-Cuaresma}, Worley, Pancino, {Cantat-Gaudin}, Magrini, Bergemann,
  Gonz{\'a}lez~Hern{\'a}ndez, Hill, Lardo, {de Laverny}, Lind, Masseron,
  Montes, Mucciarelli, Nordlander, Recio~Blanco, Sobeck, Sordo, Sousa,
  Tabernero, Vallenari, \& Van~Eck}]{jofre2014}
Jofr{\'e}, P., Heiter, U., Soubiran, C., {et~al.} 2014, Astronomy and
  Astrophysics, 564, A133, \dodoi{10.1051/0004-6361/201322440}

\bibitem[{Johns-Krull(2007)}]{johns-krull2007}
Johns-Krull, C.~M. 2007, ApJ, 664, 975, \dodoi{10.1086/519017}

\bibitem[{Johns-Krull {et~al.}(2004)Johns-Krull, Valenti, \&
  Saar}]{johns-krull2004}
Johns-Krull, C.~M., Valenti, J.~A., \& Saar, S.~H. 2004, ApJ, 617, 1204,
  \dodoi{10.1086/425652}

\bibitem[{{Johns-Krull} {et~al.}(2016){Johns-Krull}, McLane, Prato, Crockett,
  Jaffe, Hartigan, Beichman, Mahmud, Chen, Skiff, Cauley, Jones, \&
  Mace}]{johns-krull2016}
{Johns-Krull}, C.~M., McLane, J.~N., Prato, L., {et~al.} 2016, ApJ, 826, 206,
  \dodoi{10.3847/0004-637X/826/2/206}

\bibitem[{Kochukhov(2007)}]{kochukhov2007}
Kochukhov, O. 2007, arXiv:astro-ph/0701084.
\newblock \doeprint{astro-ph/0701084}

\bibitem[{Kochukhov {et~al.}(2010)Kochukhov, Makaganiuk, \&
  Piskunov}]{kochukhov2010}
Kochukhov, O., Makaganiuk, V., \& Piskunov, N. 2010, Astronomy and
  Astrophysics, 524, A5, \dodoi{10.1051/0004-6361/201015429}

\bibitem[{Kovtyukh {et~al.}(2023)Kovtyukh, Lemasle, Nardetto, Bono, {da Silva},
  Matsunaga, Yushchenko, Fukue, \& Grebel}]{kovtyukh2023}
Kovtyukh, V., Lemasle, B., Nardetto, N., {et~al.} 2023, Monthly Notices of the
  Royal Astronomical Society, 523, 5047, \dodoi{10.1093/mnras/stad1708}

\bibitem[{Kozdon {et~al.}(2023)Kozdon, Brittain, Fung, Kern, Jensen, Carr,
  Najita, \& Banzatti}]{kozdon2023}
Kozdon, J., Brittain, S.~D., Fung, J., {et~al.} 2023, The Astronomical Journal,
  166, 119, \dodoi{10.3847/1538-3881/ace903}

\bibitem[{Kraus {et~al.}(2011)Kraus, Ireland, Martinache, \&
  Hillenbrand}]{kraus2011}
Kraus, A.~L., Ireland, M.~J., Martinache, F., \& Hillenbrand, L.~A. 2011, ApJ,
  731, 8, \dodoi{10.1088/0004-637X/731/1/8}

\bibitem[{Kudo {et~al.}(2018)Kudo, Hashimoto, Muto, Liu, Dong, Hasegawa,
  Tsukagoshi, \& Konishi}]{kudo2018}
Kudo, T., Hashimoto, J., Muto, T., {et~al.} 2018, The Astrophysical Journal,
  868, L5, \dodoi{10.3847/2041-8213/aaeb1c}

\bibitem[{Lanza(2022)}]{lanza2022}
Lanza, A.~F. 2022, Astronomy and Astrophysics, 658, A195,
  \dodoi{10.1051/0004-6361/202142566}

\bibitem[{Lanza {et~al.}(2009)Lanza, Aigrain, Messina, Leto, Pagano, Auvergne,
  Baglin, Barge, Bonomo, Collier~Cameron, Cutispoto, Deleuil, De~Medeiros,
  Foing, \& Moutou}]{lanza2009}
Lanza, A.~F., Aigrain, S., Messina, S., {et~al.} 2009, A\&A, 506, 255,
  \dodoi{10.1051/0004-6361/200811487}

\bibitem[{Lee {et~al.}(2017)Lee, Gullikson, \&
  Kaplan}]{jae_joon_lee_2017_845059}
Lee, J.-J., Gullikson, K., \& Kaplan, K. 2017, Igrins/Plp 2.2.0, Zenodo,
  \dodoi{10.5281/zenodo.845059}

\bibitem[{Lomb(1976)}]{lomb1976}
Lomb, N.~R. 1976, Astrophysics and Space Science, 39, 447,
  \dodoi{10.1007/BF00648343}

\bibitem[{Long {et~al.}(2019)Long, Herczeg, Harsono, Pinilla, Tazzari, Manara,
  Pascucci, Cabrit, Nisini, Johnstone, Edwards, Salyk, Menard, Lodato, Boehler,
  Mace, Liu, Mulders, Hendler, Ragusa, Fischer, Banzatti, Rigliaco, {van de
  Plas}, Dipierro, {Gully-Santiago}, \& {Lopez-Valdivia}}]{long2019}
Long, F., Herczeg, G.~J., Harsono, D., {et~al.} 2019, The Astrophysical
  Journal, 882, 49, \dodoi{10.3847/1538-4357/ab2d2d}

\bibitem[{Loomis {et~al.}(2017)Loomis, {\"O}berg, Andrews, \&
  MacGregor}]{loomis2017}
Loomis, R.~A., {\"O}berg, K.~I., Andrews, S.~M., \& MacGregor, M.~A. 2017, The
  Astrophysical Journal, 840, 23, \dodoi{10.3847/1538-4357/aa6c63}

\bibitem[{{L{\'o}pez-Valdivia} {et~al.}(2019){L{\'o}pez-Valdivia}, Mace, Sokal,
  Hussaini, Kidder, Mann, Gosnell, Oh, Kesseli, Muirhead, {Johns-Krull}, \&
  Jaffe}]{lopez-valdivia2019}
{L{\'o}pez-Valdivia}, R., Mace, G.~N., Sokal, K.~R., {et~al.} 2019, The
  Astrophysical Journal, 879, 105, \dodoi{10.3847/1538-4357/ab2129}

\bibitem[{{L{\'o}pez-Valdivia} {et~al.}(2021){L{\'o}pez-Valdivia}, Sokal, Mace,
  Kidder, Hussaini, Nofi, Prato, {Johns-Krull}, Oh, Lee, Park, Oh, Kraus,
  Kaplan, Llama, Mann, Kim, {Gully-Santiago}, Lee, Pak, Hwang, \&
  Jaffe}]{lopez-valdivia2021}
{L{\'o}pez-Valdivia}, R., Sokal, K.~R., Mace, G.~N., {et~al.} 2021, ApJ, 921,
  53, \dodoi{10.3847/1538-4357/ac1a7b}

\bibitem[{Luck \& Heiter(2006)}]{luck2006}
Luck, R.~E., \& Heiter, U. 2006, The Astronomical Journal, 131, 3069,
  \dodoi{10.1086/504080}

\bibitem[{Luhman {et~al.}(2017)Luhman, Mamajek, Shukla, \&
  Loutrel}]{luhman2017}
Luhman, K.~L., Mamajek, E.~E., Shukla, S.~J., \& Loutrel, N.~P. 2017, The
  Astronomical Journal, 153, 46, \dodoi{10.3847/1538-3881/153/1/46}

\bibitem[{Mace {et~al.}(2016)Mace, Kim, Jaffe, Park, Lee, Kaplan, Yu, Yuk,
  Chun, Pak, Kim, Lee, Sneden, Afsar, Pavel, Lee, Oh, Jeong, Park, Kidder, Lee,
  Nguyen~Le, McLane, {Gully-Santiago}, Oh, Lee, Hwang, \& Park}]{mace2016}
Mace, G., Kim, H., Jaffe, D.~T., {et~al.} 2016, in {{SPIE Astronomical
  Telescopes}} + {{Instrumentation}}, ed. C.~J. Evans, L.~Simard, \& H.~Takami,
  Edinburgh, United Kingdom, 99080C, \dodoi{10.1117/12.2232780}

\bibitem[{Maldonado {et~al.}(2015)Maldonado, Affer, Micela, Scandariato,
  Damasso, Stelzer, Barbieri, Bedin, Biazzo, Bignamini, Borsa, Claudi, Covino,
  Desidera, Esposito, Gratton, Gonz{\'a}lez~Hern{\'a}ndez, Lanza, Maggio,
  Molinari, Pagano, Perger, Pillitteri, Piotto, Poretti, Prisinzano, Rebolo,
  Ribas, Shkolnik, Southworth, Sozzetti, \&
  Su{\'a}rez~Mascare{\~n}o}]{maldonado2015a}
Maldonado, J., Affer, L., Micela, G., {et~al.} 2015, Astronomy and
  Astrophysics, 577, A132, \dodoi{10.1051/0004-6361/201525797}

\bibitem[{Manick {et~al.}(2024)Manick, Sousa, Bouvier, Almenara, Rebull, Bayo,
  Carmona, Martioli, Venuti, Pantolmos, K{\'o}sp{\'a}l, Zanni, Bonfils, Moutou,
  Delfosse, \& {consortium}}]{manick2024}
Manick, R., Sousa, A.~P., Bouvier, J., {et~al.} 2024, Long Period Modulation of
  the Classical {{T Tauri}} Star {{CI Tau}}: Evidence for an Eccentric Close-in
  Massive Planet at 0.17 Au,  arXiv.
\newblock \doeprint{2403.03706}

\bibitem[{Mann {et~al.}(2013{\natexlab{a}})Mann, Brewer, Gaidos, L{\'e}pine, \&
  Hilton}]{mann2013a}
Mann, A.~W., Brewer, J.~M., Gaidos, E., L{\'e}pine, S., \& Hilton, E.~J.
  2013{\natexlab{a}}, The Astronomical Journal, 145, 52,
  \dodoi{10.1088/0004-6256/145/2/52}

\bibitem[{Mann {et~al.}(2014)Mann, Deacon, Gaidos, Ansdell, Brewer, Liu,
  Magnier, \& Aller}]{mann2014}
Mann, A.~W., Deacon, N.~R., Gaidos, E., {et~al.} 2014, The Astronomical
  Journal, 147, 160, \dodoi{10.1088/0004-6256/147/6/160}

\bibitem[{Mann {et~al.}(2015)Mann, Feiden, Gaidos, Boyajian, \& {von
  Braun}}]{mann2015}
Mann, A.~W., Feiden, G.~A., Gaidos, E., Boyajian, T., \& {von Braun}, K. 2015,
  The Astrophysical Journal, 804, 64, \dodoi{10.1088/0004-637X/804/1/64}

\bibitem[{Mann {et~al.}(2013{\natexlab{b}})Mann, Gaidos, \& Ansdell}]{mann2013}
Mann, A.~W., Gaidos, E., \& Ansdell, M. 2013{\natexlab{b}}, The Astrophysical
  Journal, 779, 188, \dodoi{10.1088/0004-637X/779/2/188}

\bibitem[{Nelissen {et~al.}(2023)Nelissen, McGinnis, Folsom, Ray, Vidotto,
  Alecian, Bouvier, Morin, Donati, \& Devaraj}]{nelissen2023}
Nelissen, M., McGinnis, P., Folsom, C.~P., {et~al.} 2023, Astronomy and
  Astrophysics, 670, A165, \dodoi{10.1051/0004-6361/202245194}

\bibitem[{Neves {et~al.}(2012)Neves, Bonfils, Santos, Delfosse, Forveille,
  Allard, Nat{\'a}rio, Fernandes, \& Udry}]{neves2012}
Neves, V., Bonfils, X., Santos, N.~C., {et~al.} 2012, Astronomy and
  Astrophysics, 538, A25, \dodoi{10.1051/0004-6361/201118115}

\bibitem[{O'Neal \& Neff(1997)}]{oneal1997}
O'Neal, D., \& Neff, J.~E. 1997, The Astronomical Journal, 113, 1129,
  \dodoi{10.1086/118331}

\bibitem[{O'Neal {et~al.}(2001)O'Neal, Neff, Saar, \& Mines}]{oneal2001}
O'Neal, D., Neff, J.~E., Saar, S.~H., \& Mines, J.~K. 2001, The Astronomical
  Journal, 122, 1954, \dodoi{10.1086/323093}

\bibitem[{Park {et~al.}(2014)Park, Jaffe, Yuk, Chun, Pak, Kim, Pavel, Lee, Oh,
  Jeong, Sim, Lee, Nguyen~Le, Strubhar, {Gully-Santiago}, Oh, Cha, Moon, Park,
  Brooks, Ko, Han, Nah, Hill, Lee, Barnes, Yu, Kaplan, Mace, Kim, Lee, Hwang,
  \& Park}]{park2014}
Park, C., Jaffe, D.~T., Yuk, I.-S., {et~al.} 2014, in {{SPIE Astronomical
  Telescopes}} + {{Instrumentation}}, ed. S.~K. Ramsay, I.~S. McLean, \&
  H.~Takami, Montr{\'e}al, Quebec, Canada, 91471D, \dodoi{10.1117/12.2056431}

\bibitem[{Parks {et~al.}(2014)Parks, Plavchan, White, \& Gee}]{parks2014}
Parks, J.~R., Plavchan, P., White, R.~J., \& Gee, A.~H. 2014, The Astrophysical
  Journal Supplement Series, 211, 3, \dodoi{10.1088/0067-0049/211/1/3}

\bibitem[{Passegger {et~al.}(2022)Passegger, {Bello-Garc{\'i}a},
  {Ordieres-Mer{\'e}}, {Antoniadis-Karnavas}, Marfil, {Duque-Arribas}, Amado,
  {Delgado-Mena}, Montes, {Rojas-Ayala}, Schweitzer, Tabernero, B{\'e}jar,
  Caballero, Hatzes, Henning, Pedraz, Quirrenbach, Reiners, \&
  Ribas}]{passegger2022}
Passegger, V.~M., {Bello-Garc{\'i}a}, A., {Ordieres-Mer{\'e}}, J., {et~al.}
  2022, Astronomy and Astrophysics, 658, A194,
  \dodoi{10.1051/0004-6361/202141920}

\bibitem[{Piskunov(1999)}]{piskunov1999a}
Piskunov, N. 1999, 243, 515, \dodoi{10.1007/978-94-015-9329-8_45}

\bibitem[{Polyansky {et~al.}(2018)Polyansky, Kyuberis, Zobov, Tennyson,
  Yurchenko, \& Lodi}]{polyansky2018a}
Polyansky, O.~L., Kyuberis, A.~A., Zobov, N.~F., {et~al.} 2018, Monthly Notices
  of the Royal Astronomical Society, 480, 2597, \dodoi{10.1093/mnras/sty1877}

\bibitem[{Prato {et~al.}(2008)Prato, Huerta, {Johns-Krull}, Mahmud, Jaffe, \&
  Hartigan}]{prato2008}
Prato, L., Huerta, M., {Johns-Krull}, C.~M., {et~al.} 2008, ApJ, 687, L103,
  \dodoi{10.1086/593201}

\bibitem[{Prato {et~al.}(2002)Prato, Simon, Mazeh, McLean, Norman, \&
  Zucker}]{prato2002}
Prato, L., Simon, M., Mazeh, T., {et~al.} 2002, ApJ, 569, 863,
  \dodoi{10.1086/339397}

\bibitem[{Ram{\'i}rez {et~al.}(2012)Ram{\'i}rez, Fish, Lambert, \&
  Allende~Prieto}]{ramirez2012}
Ram{\'i}rez, I., Fish, J.~R., Lambert, D.~L., \& Allende~Prieto, C. 2012, The
  Astrophysical Journal, 756, 46, \dodoi{10.1088/0004-637X/756/1/46}

\bibitem[{Rampalli {et~al.}(2021)Rampalli, Ag{\"u}eros, Curtis, Douglas,
  N{\'u}{\~n}ez, Cargile, Covey, Gosnell, Kraus, Law, \& Mann}]{rampalli2021}
Rampalli, R., Ag{\"u}eros, M.~A., Curtis, J.~L., {et~al.} 2021, The
  Astrophysical Journal, 921, 167, \dodoi{10.3847/1538-4357/ac0c1e}

\bibitem[{Rayner {et~al.}(2009)Rayner, Cushing, \& Vacca}]{rayner2009}
Rayner, J.~T., Cushing, M.~C., \& Vacca, W.~D. 2009, The Astrophysical Journal
  Supplement Series, 185, 289, \dodoi{10.1088/0067-0049/185/2/289}

\bibitem[{Rebull {et~al.}(2020)Rebull, Stauffer, Cody, Hillenbrand, Bouvier,
  Roggero, \& David}]{rebull2020}
Rebull, L.~M., Stauffer, J.~R., Cody, A.~M., {et~al.} 2020, The Astronomical
  Journal, 159, 273, \dodoi{10.3847/1538-3881/ab893c}

\bibitem[{Reiners {et~al.}(2022)Reiners, Shulyak, K{\"a}pyl{\"a}, Ribas, Nagel,
  Zechmeister, Caballero, Shan, Fuhrmeister, Quirrenbach, Amado, Montes,
  Jeffers, Azzaro, B{\'e}jar, Chaturvedi, Henning, K{\"u}rster, \&
  Pall{\'e}}]{reiners2022}
Reiners, A., Shulyak, D., K{\"a}pyl{\"a}, P.~J., {et~al.} 2022, Astronomy and
  Astrophysics, 662, A41, \dodoi{10.1051/0004-6361/202243251}

\bibitem[{Rizzuto {et~al.}(2020)Rizzuto, Dupuy, Ireland, \&
  Kraus}]{rizzuto2020a}
Rizzuto, A.~C., Dupuy, T.~J., Ireland, M.~J., \& Kraus, A.~L. 2020, The
  Astrophysical Journal, 889, 175, \dodoi{10.3847/1538-4357/ab5aed}

\bibitem[{{Rojas-Ayala} {et~al.}(2012){Rojas-Ayala}, Covey, Muirhead, \&
  Lloyd}]{rojas-ayala2012}
{Rojas-Ayala}, B., Covey, K.~R., Muirhead, P.~S., \& Lloyd, J.~P. 2012, The
  Astrophysical Journal, 748, 93, \dodoi{10.1088/0004-637X/748/2/93}

\bibitem[{Rota {et~al.}(2022)Rota, Manara, Miotello, Lodato, Facchini,
  Koutoulaki, Herczeg, Long, Tazzari, Cabrit, Harsono, M{\'e}nard, Pinilla,
  {van der Plas}, Ragusa, \& Yen}]{rota2022}
Rota, A.~A., Manara, C.~F., Miotello, A., {et~al.} 2022, Astronomy and
  Astrophysics, 662, A121, \dodoi{10.1051/0004-6361/202141035}

\bibitem[{Ryabchikova {et~al.}(2015)Ryabchikova, Piskunov, Kurucz, Stempels,
  Heiter, Pakhomov, \& Barklem}]{ryabchikova2015a}
Ryabchikova, T., Piskunov, N., Kurucz, R.~L., {et~al.} 2015, Phys. Scr., 90,
  054005, \dodoi{10.1088/0031-8949/90/5/054005}

\bibitem[{Saar \& Donahue(1997)}]{saar1997}
Saar, S.~H., \& Donahue, R.~A. 1997, The Astrophysical Journal, 485, 319,
  \dodoi{10.1086/304392}

\bibitem[{Sawczynec {et~al.}(2022)Sawczynec, Mace, {Gully-Santiago}, \&
  Jaffe}]{sawczynec2022}
Sawczynec, E., Mace, G., {Gully-Santiago}, M., \& Jaffe, D. 2022, 54, 203.06.
\newblock \url{https://ui.adsabs.harvard.edu/abs/2022AAS...24020306S}

\bibitem[{Sawczynec {et~al.}(2023)Sawczynec, Mace, {Gully-Santiago}, \&
  Jaffe}]{sawczynec2023}
---. 2023, 55, 207.14.
\newblock \url{https://ui.adsabs.harvard.edu/abs/2023AAS...24120714S}

\bibitem[{Scargle(1982)}]{scargle1982}
Scargle, J.~D. 1982, The Astrophysical Journal, 263, 835,
  \dodoi{10.1086/160554}

\bibitem[{Sheehan {et~al.}(2019)Sheehan, Wu, Eisner, \& Tobin}]{sheehan2019}
Sheehan, P.~D., Wu, Y.-L., Eisner, J.~A., \& Tobin, J.~J. 2019, The
  Astrophysical Journal, 874, 136, \dodoi{10.3847/1538-4357/ab09f9}

\bibitem[{Sikora {et~al.}(2023)Sikora, Rowe, Barat, Bean, Brady, D{\'e}sert,
  Feinstein, Gilbert, Henry, Kasper, Lizotte, Matesic, Panwar, Seifahrt,
  Shivkumar, Stef{\'a}nsson, \& St{\"u}rmer}]{sikora2023}
Sikora, J., Rowe, J., Barat, S., {et~al.} 2023, The Astronomical Journal, 165,
  250, \dodoi{10.3847/1538-3881/acc865}

\bibitem[{Skrutskie {et~al.}(2006)Skrutskie, Cutri, Stiening, Weinberg,
  Schneider, Carpenter, Beichman, Capps, Chester, Elias, Huchra, Liebert,
  Lonsdale, Monet, Price, Seitzer, Jarrett, Kirkpatrick, Gizis, Howard, Evans,
  Fowler, Fullmer, Hurt, Light, Kopan, Marsh, McCallon, Tam, Van~Dyk, \&
  Wheelock}]{skrutskie2006}
Skrutskie, M.~F., Cutri, R.~M., Stiening, R., {et~al.} 2006, The Astronomical
  Journal, 131, 1163, \dodoi{10.1086/498708}

\bibitem[{Sokal {et~al.}(2020)Sokal, {Johns-Krull}, Mace, Nofi, Prato, Lee, \&
  Jaffe}]{sokal2020}
Sokal, K.~R., {Johns-Krull}, C.~M., Mace, G.~N., {et~al.} 2020, ApJ, 888, 116,
  \dodoi{10.3847/1538-4357/ab59d8}

\bibitem[{Stahl {et~al.}(2021)Stahl, Tang, {Johns-Krull}, Prato, Llama, Mace,
  Joon~Lee, Oh, Luna, \& Jaffe}]{stahl2021}
Stahl, A.~G., Tang, S.-Y., {Johns-Krull}, C.~M., {et~al.} 2021, AJ, 161, 283,
  \dodoi{10.3847/1538-3881/abf5e7}

\bibitem[{Stellingwerf(1978)}]{stellingwerf1978}
Stellingwerf, R.~F. 1978, The Astrophysical Journal, 224, 953,
  \dodoi{10.1086/156444}

\bibitem[{{STScI Development Team}(2013)}]{stscidevelopmentteam2013}
{STScI Development Team}. 2013, Astrophysics Source Code Library,
  ascl:1303.023.
\newblock \url{https://ui.adsabs.harvard.edu/abs/2013ascl.soft03023S}

\bibitem[{Tang {et~al.}(2021)Tang, Stahl, {Johns-Krull}, Prato, \&
  Llama}]{tang2021a}
Tang, S.-Y., Stahl, A., {Johns-Krull}, C., Prato, L., \& Llama, J. 2021, JOSS,
  6, 3095, \dodoi{10.21105/joss.03095}

\bibitem[{Tang {et~al.}(2023)Tang, Stahl, Prato, Schaefer, {Johns-Krull},
  Skiff, Beichman, \& Uyama}]{tang2023}
Tang, S.-Y., Stahl, A.~G., Prato, L., {et~al.} 2023, The Astrophysical Journal,
  950, 92, \dodoi{10.3847/1538-4357/acc58b}

\bibitem[{Taniguchi {et~al.}(2018)Taniguchi, Matsunaga, Kobayashi, Fukue,
  Hamano, Ikeda, Kawakita, Kondo, Sameshima, \& Yasui}]{taniguchi2018}
Taniguchi, D., Matsunaga, N., Kobayashi, N., {et~al.} 2018, Monthly Notices of
  the Royal Astronomical Society, 473, 4993, \dodoi{10.1093/mnras/stx2691}

\bibitem[{Tayar {et~al.}(2022)Tayar, Claytor, Huber, \& {van
  Saders}}]{tayar2022}
Tayar, J., Claytor, Z.~R., Huber, D., \& {van Saders}, J. 2022, The
  Astrophysical Journal, 927, 31, \dodoi{10.3847/1538-4357/ac4bbc}

\bibitem[{Tennyson \& Yurchenko(2012)}]{tennyson2012}
Tennyson, J., \& Yurchenko, S.~N. 2012, Monthly Notices of the Royal
  Astronomical Society, 425, 21, \dodoi{10.1111/j.1365-2966.2012.21440.x}

\bibitem[{Terrien {et~al.}(2015)Terrien, Mahadevan, Bender, Deshpande, \&
  Robertson}]{terrien2015}
Terrien, R.~C., Mahadevan, S., Bender, C.~F., Deshpande, R., \& Robertson, P.
  2015, The Astrophysical Journal, 802, L10,
  \dodoi{10.1088/2041-8205/802/1/L10}

\bibitem[{Tran {et~al.}(2023)Tran, Bedell, {Foreman-Mackey}, \&
  Luger}]{tran2023a}
Tran, Q.~H., Bedell, M., {Foreman-Mackey}, D., \& Luger, R. 2023, The
  Astrophysical Journal, 950, 162, \dodoi{10.3847/1538-4357/acd05c}

\bibitem[{Valenti \& Fischer(2005)}]{valenti2005}
Valenti, J.~A., \& Fischer, D.~A. 2005, The Astrophysical Journal Supplement
  Series, 159, 141, \dodoi{10.1086/430500}

\bibitem[{Valenti \& Piskunov(1996)}]{valenti1996a}
Valenti, J.~A., \& Piskunov, N. 1996, Astronomy and Astrophysics Supplement
  Series, 118, 595.
\newblock \url{https://ui.adsabs.harvard.edu/abs/1996A&AS..118..595V}

\bibitem[{VanderPlas(2018)}]{vanderplas2018}
VanderPlas, J.~T. 2018, The Astrophysical Journal Supplement Series, 236, 16,
  \dodoi{10.3847/1538-4365/aab766}

\bibitem[{{von Braun} {et~al.}(2011){von Braun}, Boyajian, Kane, {van Belle},
  Ciardi, {L{\'o}pez-Morales}, McAlister, Henry, Jao, Riedel, Subasavage,
  Schaefer, {ten Brummelaar}, Ridgway, Sturmann, Sturmann, Mazingue, Turner,
  Farrington, Goldfinger, \& Boden}]{vonbraun2011}
{von Braun}, K., Boyajian, T.~S., Kane, S.~R., {et~al.} 2011, The Astrophysical
  Journal, 729, L26, \dodoi{10.1088/2041-8205/729/2/L26}

\bibitem[{Wallace {et~al.}(1996)Wallace, Livingston, Hinkle, \&
  Bernath}]{wallace1996}
Wallace, L., Livingston, W., Hinkle, K., \& Bernath, P. 1996, The Astrophysical
  Journal Supplement Series, 106, 165, \dodoi{10.1086/192333}

\bibitem[{Watson {et~al.}(2006)Watson, Henden, \& Price}]{watson2006}
Watson, C.~L., Henden, A.~A., \& Price, A. 2006, Society for Astronomical
  Sciences Annual Symposium, 25, 47.
\newblock \url{https://ui.adsabs.harvard.edu/abs/2006SASS...25...47W}

\bibitem[{Wehrhahn {et~al.}(2023)Wehrhahn, Piskunov, \&
  Ryabchikova}]{wehrhahn2023}
Wehrhahn, A., Piskunov, N., \& Ryabchikova, T. 2023, Astronomy and
  Astrophysics, 671, A171, \dodoi{10.1051/0004-6361/202244482}

\bibitem[{White {et~al.}(2018)White, Huber, Mann, Casagrande, Grunblatt,
  Justesen, Silva~Aguirre, Bedding, Ireland, Schaefer, \& Tuthill}]{white2018}
White, T.~R., Huber, D., Mann, A.~W., {et~al.} 2018, Monthly Notices of the
  Royal Astronomical Society, 477, 4403, \dodoi{10.1093/mnras/sty898}

\bibitem[{Yu {et~al.}(2019)Yu, Donati, Grankin, Collier~Cameron, Moutou,
  Hussain, Baruteau, Jouve, \& {MaTYSSE Collaboration}}]{yu2019}
Yu, L., Donati, J.~F., Grankin, K., {et~al.} 2019, Monthly Notices of the Royal
  Astronomical Society, 489, 5556, \dodoi{10.1093/mnras/stz2481}

\bibitem[{Yuk {et~al.}(2010)Yuk, Jaffe, Barnes, Chun, Park, Lee, Lee, Wang,
  Park, Pak, Strubhar, Deen, Oh, Seo, Pyo, Park, Lacy, Goertz, Rand, \&
  {Gully-Santiago}}]{yuk2010}
Yuk, I.-S., Jaffe, D.~T., Barnes, S., {et~al.} 2010, in {{SPIE Astronomical
  Telescopes}} + {{Instrumentation}}, ed. I.~S. McLean, S.~K. Ramsay, \&
  H.~Takami, San Diego, California, USA, 77351M, \dodoi{10.1117/12.856864}

\end{thebibliography}
\bibliographystyle{aasjournal}

\end{document}